%% file: sk_atm_ext_2017.tex
\documentclass[twocolumn,showpacs,prd,preprintnumbers,superscriptaddress]{revtex4-1}

\usepackage{graphicx}
\usepackage{epsfig}
\usepackage{subfigure}
\usepackage{xspace}
\usepackage{dcolumn}
\usepackage{multirow}
\usepackage{longtable}
\usepackage{amsmath}
\usepackage{mathtools}
\usepackage{xcolor}
\usepackage{lineno}

\newcommand{\val}[2]{\ensuremath{#1 \; \mathrm{#2}\xspace}}

\newcommand{\UP}{UP-$\mu$\xspace}

\newcommand{\numu}{\ensuremath{\nu_{\mu}}\xspace}
\newcommand{\nubar}{\ensuremath{\bar{\nu}_{\mu}}\xspace}
\newcommand{\numubar}{\nubar}
\newcommand{\nue}{\ensuremath{\nu_{e}}\xspace}
\newcommand{\nuebar}{\ensuremath{\bar{\nu}_{e}}\xspace}
\newcommand{\nutau}{\ensuremath{\nu_{\tau}}\xspace}

\newcommand{\pizero}{\ensuremath{\pi^0}\xspace}

\newcommand{\dm}{\ensuremath{\Delta m^{2}}\xspace}
\newcommand{\dmsq}[1]{\ensuremath{\dm_{#1}}\xspace}

\newcommand{\sn}[1]{\ensuremath{ \sin^{2}(\theta_{#1}) }\xspace }

\newcommand{\superk}      {Super-Kamiokande\xspace}

\newcolumntype{d}[1]{D{.}{\cdot}{#1}}
\newcolumntype{.}{D{.}{.}{-1}}
\newcolumntype{,}{D{,}{,}{2}}

\newcommand{\ket}[1]{\lvert #1\rangle}

\begin{document}

\title{ Atmospheric neutrino oscillation analysis 
        with external constraints in Super-Kamiokande I-IV }

\input{authors}

\begin{abstract}

An analysis of atmospheric neutrino data from all four run
periods of \superk optimized for sensitivity to the neutrino mass
hierarchy is presented. Confidence intervals for $\Delta
m^2_{32}$, $\sin^2 \theta_{23}$, $\sin^2 \theta_{13}$ and
$\delta_{CP}$ are presented for normal neutrino mass hierarchy and
inverted neutrino mass hierarchy hypotheses, based on atmospheric neutrino data
alone. Additional constraints from reactor data on $\theta_{13}$ and
from published binned T2K data on muon neutrino disappearance and
electron neutrino appearance are added to the atmospheric neutrino fit
to give enhanced constraints on the above parameters. Over the range
of parameters allowed at 90\% confidence level, the normal mass
hierarchy is favored by between 91.9\% and 94.5\% based on the combined
Super-Kamiokande plus T2K result.

\end{abstract}


\maketitle


\section{Introduction}
\label{sec:intro}

The principal goal of contemporary neutrino oscillation experiments is
to fully test the three-neutrino mixing paradigm based on the
Pontecorvo-Maki-Nakagawa-Sakata (PMNS)
matrix~\cite{Pontecorvo:1967fh,Maki:1962mu}. This paradigm is
characterized by three mixing angles, two mass splittings, and one 
CP-violating phase. Some neutrino mixing parameters have been
experimentally determined, such as the magnitude of the two mass
splittings, the ordering of the mass states with the smallest splitting,
and the values of the mixing angles. In particular, measurements by
reactor antineutrino ~\cite{An:2012bu,Ahn:2012nd,Abe:2011fz} 
experiments and T2K~\cite{Abe:2013hdq}
have established
that the mixing angle $\theta_{13}$ is small but non-zero and they
have precisely measured its value. There remain unknown parameters in
the PMNS formalism, most notably the ordering of the mass states with
the largest splitting, which is mathematically expressed as the sign of
$\Delta m^2_{31}$, and is commonly referred to as the neutrino mass
hierarchy. Although it is known that muon and tau neutrino mixing is
nearly maximal, i.e. $\theta_{23}$ is near $\pi/4$, it is not known if
$\theta_{23}$ takes exactly that value, or is slightly larger or
slightly smaller~\cite{Adamson:2017qqn,Abe:2017uxa}. With all three
neutrino flavors and mass states mixing, it is possible to measure the
unknown CP-violating phase $\delta_{CP}$ and perhaps conclude that
neutrinos and antineutrinos have different oscillation probabilities,
if it is found that $\delta_{CP}$ is neither 0 nor $\pi$. The value of
$\delta_{CP}$ is considered to be unknown, although the T2K
and NOvA long-baseline experiments, and the results published in this
paper, are beginning to constrain it~\cite{Abe:2017uxa,Adamson:2017gxd}.

Due to the presence of neutrinos and antineutrinos, the effects of
matter on neutrino oscillations, and the wide variety of energies and
pathlengths spanned, atmospheric neutrinos are sensitive to the
unknown parameters of the PMNS formalism. The measurement of the mass
hierarchy is driven by an expected hierarchy-dependent, upward-going
excess of either electron neutrino or antineutrino interactions driven
by $\theta_{13}$-induced matter effects between 2 and 10 GeV. In order
to take advantage of this phenomenon, sign selection of neutrino
interactions and sufficient statistics are necessary. It should be
noted that the size of this event excess is a function of $\theta_{23}$,
and as will be discussed below, constraints on this parameter improve
sensitivity to the hierarchy. Determining the mass hierarchy and
measuring $\theta_{23}$ play an important role in interpreting any
neutrino versus antineutrino oscillation difference and thereby
establishing CP violation.

In this paper we analyze 328 kiloton$\cdot$years of Super-Kamiokande
(Super-K) atmospheric data. The sensitivity of our experiment is
not sufficient to definitively resolve the unknown parameters. In
particular we are limited by low statistics and difficult event
classification in the high-energy hierarchy-sensitive
sample. Nevertheless, we analyze the atmospheric neutrino data
in a manner optimized for sensitivity to the mass hierarchy and report our best
estimates and confidence intervals. 
We present results with and without constraints from
external experiments. In Section~\ref{sec:oscillations} atmospheric
neutrino oscillations are reviewed before discussing the Super-K
detector and data set in Section~\ref{sec:detector}. An analysis of
the atmospheric neutrino data by themselves is then presented in
Section~\ref{sec:atm_only} and followed by an analysis incorporating
constraints from external measurements in Section~\ref{sec:gaibu}.
These results are interpreted in Section~\ref{sec:interp} 
before concluding in Section~\ref{sec:conclusion}.

\section{Oscillations}
\label{sec:oscillations}
\subsection{In Vacuum}
Neutrinos oscillate because the neutrino eigenstates of the weak interaction are different from the neutrino mass eigenstates.   The flavor eigenstates $\nu_\alpha$ are related to the mass eigenstates $\nu_i$ by
\begin{equation}
\ket{\nu_{\alpha}}=\sum_i^3 U_{\alpha,i}^* \ket{\nu_i},
\label{eqn:flav-mass}
\end{equation}
where U is the 3x3 Pontecorvo-Maki-Nakagawa-Sakata (PMNS) matrix~\cite{Pontecorvo:1967fh,Maki:1962mu}
\begin{multline}
\textbf{U}=\begin{pmatrix}
1 & 0 & 0\\
0 & c_{23} & s_{23} \\
0 & -s_{23} & c_{23} \end{pmatrix} 
\begin{pmatrix}
c_{13} & 0 & s_{13}e^{-i\delta_{CP}}\\
0 & 1 & 0 \\
-s_{13}e^{i\delta_{CP}} & 0 & c_{13} \end{pmatrix} \\
\times\begin{pmatrix}
c_{12} & s_{12} & 0\\
-s_{12} & c_{12} & 0 \\
0 & 0 & 1 \end{pmatrix}.
\label{eqn:pmns}
\end{multline}
Here $c_{ij}=\cos{\theta_{ij}} , s_{ij}=\sin{\theta_{ij}}$.  Propagation of these states according to their vacuum Hamiltonians leads to the standard oscillation formula for relativistic neutrinos in vacuum
\begin{multline}
P(\nu_{\alpha} \rightarrow \nu_{\beta})= \delta_{\alpha\beta}-4\sum_{i>j} \Re(U_{\alpha i}^*U_{\beta i}U_{\alpha j}U_{\beta j}^*)\sin^2 \Delta_{ij} \\ \pm 2\sum_{i>j} \Im(U_{\alpha i}^*U_{\beta i}U_{\alpha j}U_{\beta j}^*)\sin 2\Delta_{ij},
\label{eqn:oscprob}
\end{multline}
where $$ \Delta_{ij}=\frac{1.27 \Delta m_{ij}^2 (\textrm{eV}^2) L(\textrm{km})}{E (\textrm{GeV})} $$ and the sign before the second summation is positive for neutrinos and negative for anti-neutrinos.
Neutrino oscillations in vacuum are thus fully described by 6 parameters: the 3 mixing angles $\theta_{13}, \theta_{12}, \theta_{23}$, the two mass splittings $\Delta m_{21}^2, \Delta m_{31}^2$, and the CP-violating phase $\delta_{CP}$.  
Data from reactor, atmospheric, solar, and long-baseline neutrino experiments indicate that nearly all of these parameters have non-zero values~\cite{Olive:2016xmw}.
Currently the sign of $\Delta m_{31}^2$ and the value of $\delta_{CP}$ are unknown.
Note that throughout this paper the indices of the mass splittings present the neutrino mass states in descending order from left to right regardless of the hierarchy assumption. 

The unoscillated atmospheric neutrino flux consists of electron- and muon-flavored neutrinos and antineutrinos.
Since $\nu_{\tau}$ charged current interactions are either kinematically disallowed or suppressed compared to $\nu_{\mu}$ and $\nu_{e}$ charged current (CC) interactions over the energy range considered in the analysis below, 
the atmospheric data are predominantly described by the $\nu_{\mu}$ and $\nu_{e}$ survival probabilities and the $\nu_{\mu} \leftrightarrow \nu_e$ oscillation probability.
For sufficiently small $L/E$, $\sin^2 \left(\frac{1.27 \Delta m_{12}^2 L}{E} \right) \ll1$ and so the $\Delta m_{12}^2$ terms in Equation~\ref{eqn:oscprob} can be ignored and the approximation $\Delta m_{31}^2 \approx \Delta m_{32}^2$ applied.  Under these assumptions, the dominant $\nu_e$ and $\nu_{\mu}$ oscillation probabilities become:
\begin{multline}
P(\nu_e \rightarrow \nu_e)\cong1-\sin^2 2\theta_{13} \sin^2\left( \frac{1.27 \Delta m_{31}^2 L}{E} \right)\\
P(\nu_{\mu} \rightarrow \nu_{\mu})\cong1-4\cos^2 \theta_{13} \sin^2 \theta_{23} (1-\cos^2 \theta_{13}\sin^2\theta_{23}) \\ \times \sin^2 \left(\frac{1.27 \Delta m_{31}^2 L}{E} \right) \\
P(\nu_{\mu} \leftrightarrow \nu_e) \cong \sin^2 \theta_{23} \sin^2 2\theta_{13} \sin^2\left(\frac{1.27 \Delta m_{31}^2 L}{E} \right).
\label{eqn:oscprob_atm_vac}
\end{multline}

\subsection{In Matter}    
When neutrinos travel through matter, the effective Hamiltonian is modified from its vacuum form due to the difference in the forward scattering amplitudes of $\nu_e$ and $\nu_{\mu,\tau}$  (presented here in the mass eigenstate basis):
\begin{equation}
H_{\textrm{matter}}=\begin{pmatrix}\frac{m_1^2}{2E}&0&0\\0&\frac{m_2^2}{2E}&0\\0&0&\frac{m_3^2}{2E} \end{pmatrix}  + U^\dagger\begin{pmatrix}a&0&0\\0&0&0\\0&0&0\end{pmatrix}U,
\label{eqn:hamiltonian_matter}
\end{equation}
where $a=\pm\sqrt{2}G_FN_e$, $G_f$ is the Fermi constant, $N_e$ is the electron density, $U$ is the PMNS matrix, and the plus (minus) sign is for neutrinos (antineutrinos). 
For constant density matter the resulting oscillation probabilities can be written using effective mixing parameters.
In particular, the $P(\nu_{\mu} \leftrightarrow \nu_{e})$ probability from Eq.~\ref{eqn:oscprob_atm_vac} can be rewritten by replacing 
$\Delta m_{31}^{2}$ and
$\theta_{13}$  by their matter-effective parameters
%
\begin{eqnarray}
\Delta m^{2}_{31,M}                 & = &\Delta m^{2}_{31} \sqrt{ \sin^{2} 2\theta_{13}
                                     + ( \Gamma - \cos 2\theta_{13} )^{2} } \\ 
\sin^{2} 2 \theta_{13}^{M} & = & \frac{ \sin^{2} 2 \theta_{13}  }
                                       { \sin^{2}2\theta_{13} + 
                                       ( \Gamma - \cos 2\theta_{13} )^{2} }, 
\label{eq:MatterVariables}
\end{eqnarray}
\noindent where $\Gamma = a E / \Delta m^{2}_{31}$.
In this form it can be seen that for neutrino energies, matter densities, and $\Delta m^{2}_{31}$ such that $\Gamma \sim \cos 2\theta_{13}$ 
the effective mixing angle becomes maximal. 
This resonant enhancement of the oscillation probability depends on the sign of the mass hierarchy and occurs for either neutrinos 
or antineutrinos through the sign of the matter potential $a$. 

In general atmospheric neutrinos do not traverse constant density matter as they travel through the earth, but such resonant oscillations are nonetheless 
present. 
The analyses presented below use exact three-flavor oscillation probabilities computed including matter effects for varying matter profiles. 
Following Ref. \cite{Barger:1980dd}, the matrix $X$, whose row vectors are the propagated mass eigenvectors, can be written as: 
\begin{equation}
\mathbf{X} =\sum_k \left[\prod_{j\neq k}\frac{2E\mathbf{H}_\textrm{matter}-M_j^2\mathbf{I}}{M_k^2-M_j^2}\right]\exp\left(-i\frac{M_k^2L}{2E}\right),
\label{eqn:Barger_X}
\end{equation}
where the $M_i^2/2E$ are the eigenvalues of the constant-density matter Hamiltonian $H_\textrm{matter}$, and $\mathbf{I}$ is the identity matrix.  The oscillation probability can then be written as:
\begin{equation}
P(\nu_\alpha \rightarrow \nu_\beta)=| (\mathbf{UXU^\dagger})_{\alpha \beta} |^2.
\label{eqn:oscprob_matter_constdensity}
\end{equation}
The eigenvalues $M_i^2/2E$ have been found 
as Equations~21 and 22 of Ref.~\cite{Barger:1980dd,*[{A sign error affecting the $\sqrt{2} G_{f} N_{e}$ term in Eq. 21 of Barger et al. was pointed out in }]Langacker:1982ih}. 

An atmospheric neutrino can pass through various densities of matter on its way to the detector.  The Earth's atmosphere is modeled as vacuum, and the Earth as a sphere of radius 6371 km, with a spherical density profile which is a simplified version of the preliminary reference Earth model (PREM)~\cite{Anonymous:pe0hLoUI}, as shown in Table \ref{tab:PREM}.

\begin{table}
\begin{tabular}{lccc}
Region&$R_\textrm{min}$ (km) & $R_\textrm{max}$ (km) & density (g/cm$^3$) \\
\hline
\hline
inner core & 0 & 1220 & 13.0 \\
outer core & 1220 & 3480 & 11.3 \\
mantle & 3480 & 5701 & 5.0\\
crust & 5701 & 6371 & 3.3\\
\hline
\hline
\end{tabular}
\caption{Model of the Earth used in the analysis, a simplified version of the PREM.}
\label{tab:PREM}
\end{table}

\begin{figure}[htbp]
\includegraphics[width=0.45\textwidth,keepaspectratio=true]{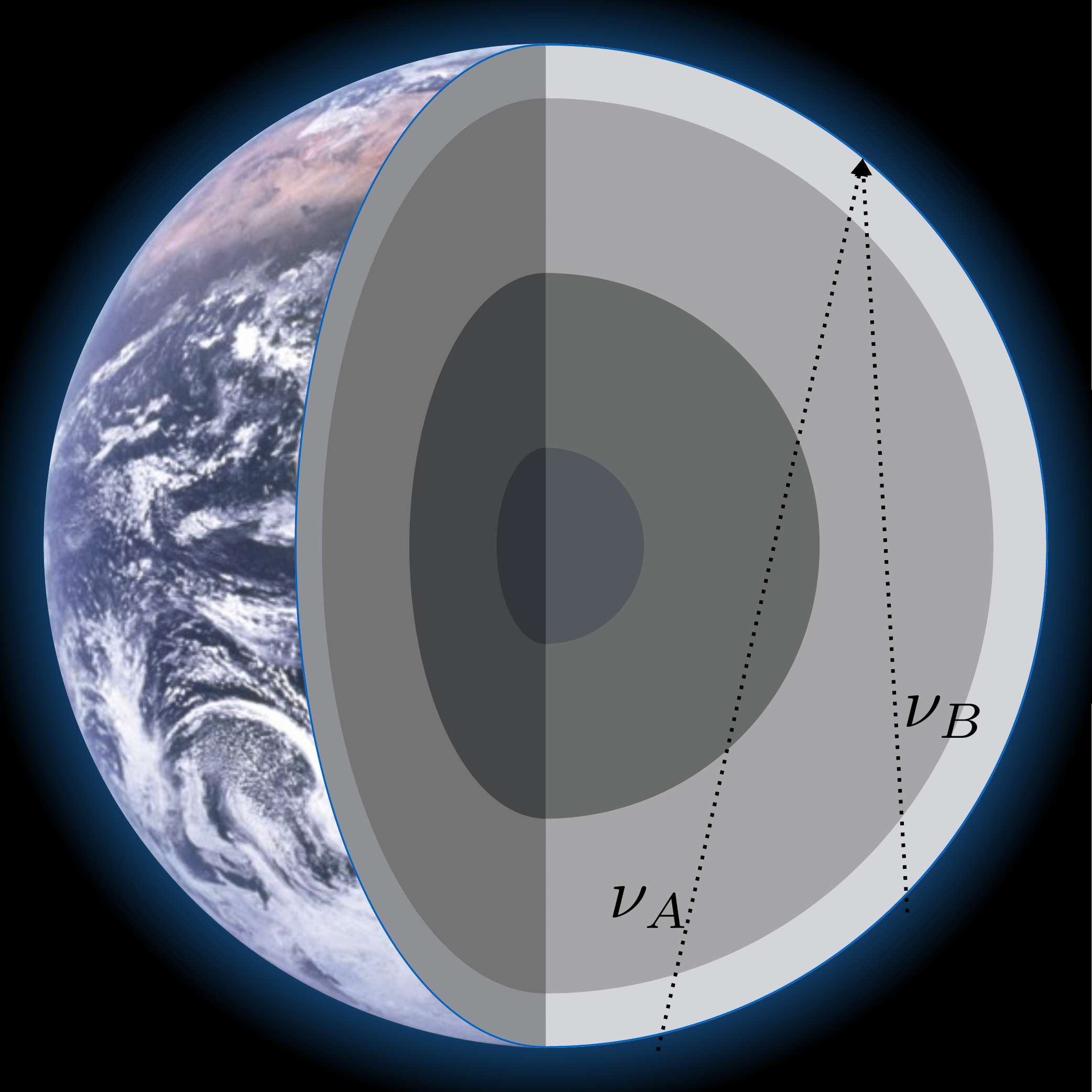}
\caption{The propagation of two neutrinos through the simplified model of the Earth used in the analysis below.  Both $\nu_A$ and $\nu_B$ are produced in the atmosphere.  $\nu_A$ then experiences 6 oscillation steps (air $\rightarrow$ crust $\rightarrow$ mantle $\rightarrow$ outer core $\rightarrow$ mantle $\rightarrow$ crust), while $\nu_B$ experiences 4 oscillation steps (air $\rightarrow$ crust $\rightarrow$ mantle $\rightarrow$ crust).}
\label{fig:prem_cartoon}
\end{figure}

\begin{figure*}[htbp]
 \subfigure[ $P( \nu_{\mu} \rightarrow \nu_{\mu} )$] {
  \includegraphics[width=0.45\textwidth]{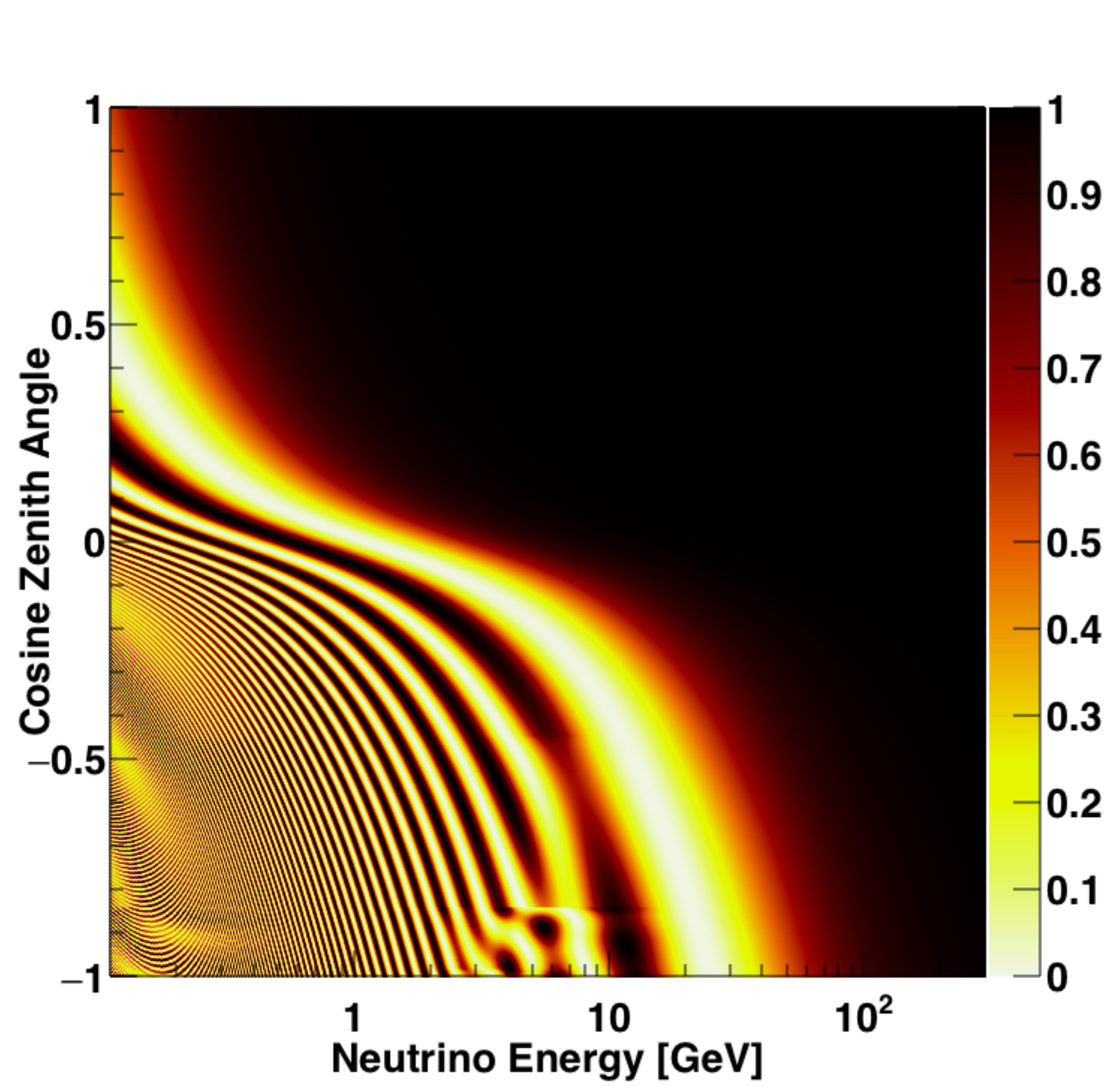}
 }
 \subfigure[ $P( \nu_{\mu} \rightarrow \nu_{e} )$] {
  \includegraphics[width=0.45\textwidth]{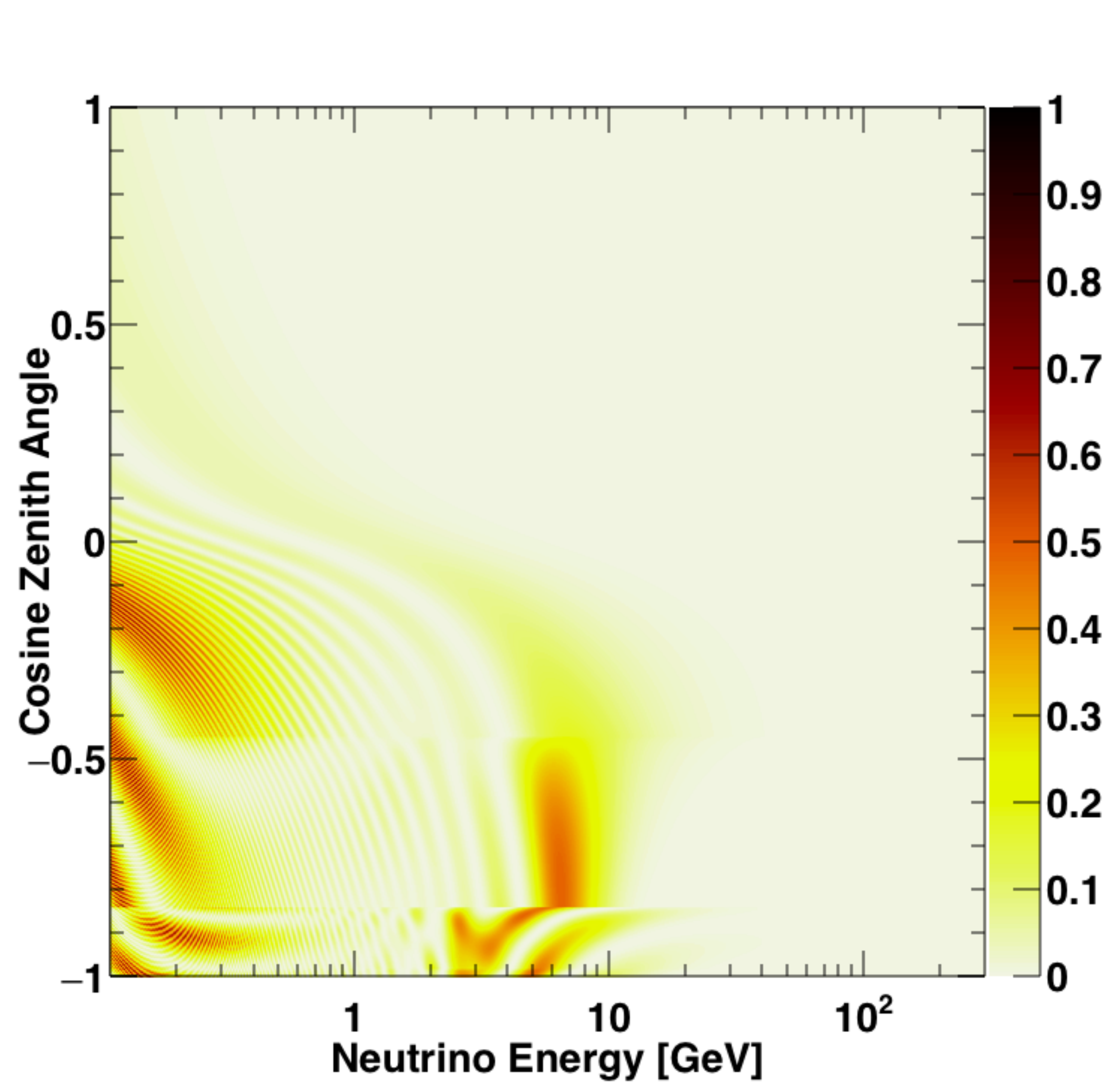}
 }
 \subfigure[ $P( \bar \nu_{\mu} \rightarrow \bar \nu_{\mu} )$] {
  \includegraphics[width=0.45\textwidth]{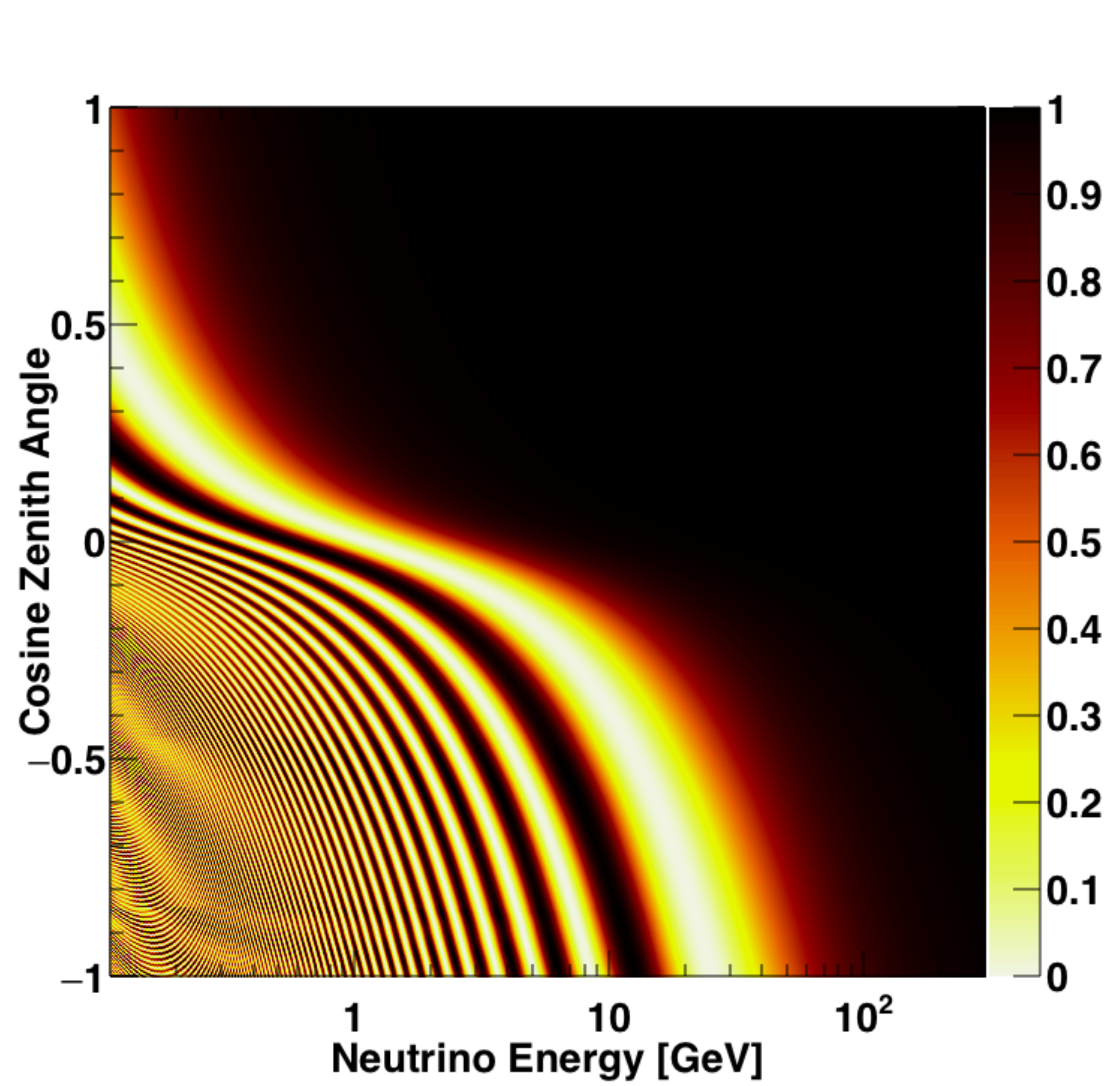}
 }
 \subfigure[ $P( \bar \nu_{\mu} \rightarrow \bar \nu_{e} )$] {
  \includegraphics[width=0.45\textwidth]{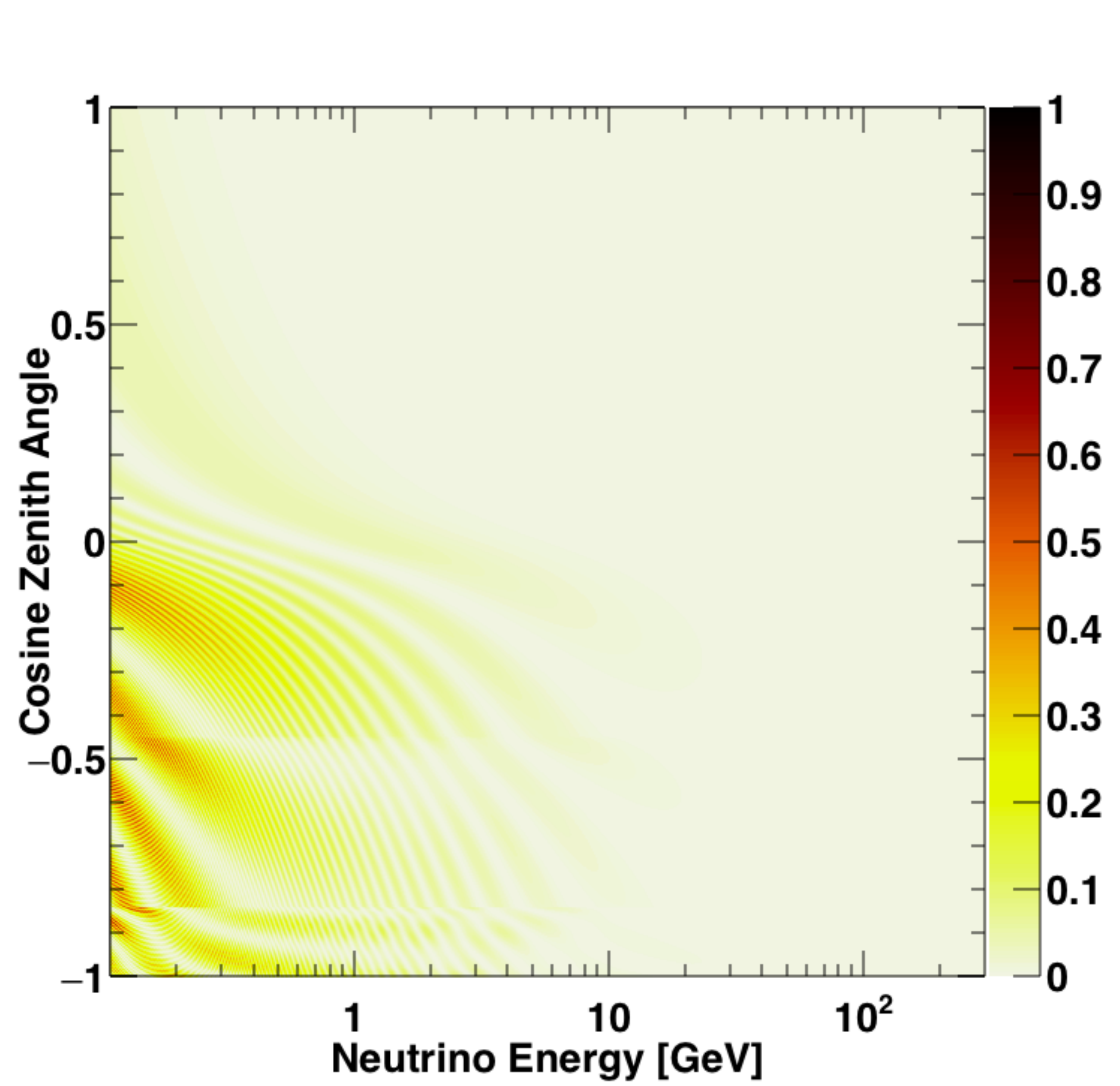}
 }
  \caption{ Oscillation probabilities for neutrinos (upper panels) and antineutrinos (lower panels) as a function 
            of energy and zenith angle assuming a normal mass hierarchy. 
            Matter effects in the Earth produce the distortions in the neutrino 
            figures between two and ten~GeV, which are not present in the antineutrino figures.
            Distortions in the $\nu_{\mu}$ survival probability and enhancements in the $\nu_{e}$ appearance
            probability occur primarily in angular regions corresponding to neutrino propagation 
            across both the outer core and mantle regions (cosine zenith $ <-0.9$)
            and propagation through the mantle and crust ($-0.9< $  cosine zenith $<-0.45$ ). 
            For an inverted hierarchy the matter effects appear in the antineutrino figures instead. 
            Here the oscillation parameters are taken to be 
            $\Delta m_{32}^{2} = 2.5 \times 10^{-3} \mbox{eV}^{2}$, 
            $\mbox{sin}^{2} \theta_{23} = 0.5$, 
            $\mbox{sin}^{2} \theta_{13} = 0.0219$, and
            $\delta_{CP} = 0 $.} 
  \label{fig:osc_prob}
\end{figure*}

The use of the full PREM model with 82 layers provides no perceptible 
change in the sensitivity of the Super-Kamiokande analysis, so the simplified matter profile is adopted to reduce computation times.
To calculate the oscillation probability of a neutrino with energy E produced at a height $h$ above the surface of the Earth, the path from the detector to the neutrino production location is 
traced through $N$ steps across the atmosphere and different regions of the Earth's interior (Fig.~\ref{fig:prem_cartoon}).  Note that because the Earth is modeled as spherically symmetric, this path is a function of only the production height and  zenith angle; it is independent of azimuthal angle.  
The oscillation probability for a given neutrino is calculated by stepping along its path:
\begin{equation}
P_{\nu_\alpha \rightarrow \nu_\beta}(E,h,\cos \theta_\textrm{zenith})=|(\mathbf{U} \prod_i^N\mathbf{X}(L_i,\rho_i,E)\mathbf{U^\dagger})_{\alpha \beta}|^2,
\end{equation}
where $L_{i}$ and $\rho_{i}$ are the length and density of the $i^{th}$ step.
Figure~\ref{fig:osc_prob} shows the $\nu_\mu$ survival and $\nu_\mu \rightarrow \nu_e$ transition probabilities for neutrinos and antineutrinos assuming the normal mass hierarchy. 
Resonant oscillation effects are clear in both 
channels for upward-going neutrinos with energies between two and ten~GeV.
In this region matter effects suppress the disappearance of $\nu_\mu$ while enhancing the 
appearance of $\nu_{e}$.
The discontinuity in the oscillation probabilities for cosine zenith angles steeper than 
 $-0.9$ corresponds to neutrinos crossing both the outer core and mantle regions of the Earth.
For shallower zenith angles the distortion in the $\nu_{\mu}$ survival probability and the 
resonant feature in the $\nu_{e}$ appearance probability are caused by matter effects in the mantle region.
Note that none of these features appear in the antineutrino plots.
If the inverted hierarchy were assumed instead, the roles of neutrinos and antineutrinos switch completely and the 
discontinuities and resonance effects appear with nearly the same magnitude but in the antinuetrino plots.

\section{The Super-Kamiokande Detector}
\label{sec:detector}

\begin{table*}
\begin{center}
\begin{tabular}{lcc@{\hspace{1em}}ccccc@{\hspace{1em}}cc}
Sample    & Energy bins & cos $\theta_{z}$ bins & CC $\nu_{e}$  &  CC $\bar{\nu_{e}}$ & CC $\nu_{\mu} + \bar{\nu_{\mu}} $ & CC $\nu_{\tau}$ & NC & Data & MC \\ 
\hline
\hline
\\
\multicolumn{10}{l}{\textbf{Fully Contained (FC) Sub-GeV}} \\
\hspace*{4pt} e-like, Single-ring  \\
\hspace*{8pt} 0 decay-e  & 5 $e^\pm$ momentum   & 10 in $[-1, 1]$  & 0.717 &  0.248 &  0.002 &  0.000 &  0.033  & 10294 & 10266.1  \\
\hspace*{8pt} 1 decay-e  & 5 $e^\pm$ momentum   & single bin               & 0.805 &  0.019 &  0.108 &  0.001 &  0.067  &  1174 & 1150.7  \\
\hspace*{4pt} $\mu$-like, Single-ring  \\
\hspace*{8pt} 0 decay-e   & 5 $\mu^\pm$ momentum & 10 in $[-1, 1]$  & 0.041 &  0.013 &  0.759 &  0.001 &  0.186 & 2843 & 2824.3  \\
\hspace*{8pt} 1 decay-e   & 5 $\mu^\pm$ momentum & 10 in $[-1, 1]$  & 0.001 &  0.000 &  0.972 &  0.000 &  0.027 & 8011 & 8008.7  \\
\hspace*{8pt} 2 decay-e   & 5 $\mu^\pm$ momentum & single bin       & 0.000 &  0.000 &  0.979 &  0.001 &  0.020 & 687  & 687.0  \\
\hspace*{4pt} $\pi^{0}$-like  \\
\hspace*{8pt} Single-ring & 5 $e^\pm$   momentum & single bin       & 0.096 &  0.033 &  0.015 &  0.000 &  0.856 & 578  & 571.8  \\
\hspace*{8pt} Two-ring    & 5 $\pi^0$   momentum & single bin       & 0.067 &  0.025 &  0.011 &  0.000 &  0.897 & 1720 & 1728.4  \\
\hspace*{4pt} Multi-ring         &  &                  & 0.294 &  0.047 &  0.342 &  0.000 &  0.318 & (1682)  & (1624.2)  \\
\\
\multicolumn{10}{l}{\textbf{Fully Contained (FC) Multi-GeV}} \\
\hspace*{4pt} Single-ring  \\
\hspace*{8pt} $\nu_{e}$-like   & 4 $e^\pm$ momentum   & 10 in $[-1, 1]$  & 0.621 &  0.090 &  0.100 &  0.033 &  0.156 & 705  & 671.3  \\
\hspace*{8pt} \nuebar-like     & 4 $e^\pm$ momentum   & 10 in $[-1, 1]$  & 0.546 &  0.372 &  0.009 &  0.010 &  0.063 & 2142 & 2193.7  \\
\hspace*{8pt} $\mu$-like       & 2 $\mu^\pm$ momentum & 10 in $[-1, 1]$  & 0.003 &  0.001 &  0.992 &  0.002 &  0.002 & 2565 & 2573.8  \\
\hspace*{4pt} Multi-ring  \\
\hspace*{8pt} \nue-like        & 3 visible energy     & 10 in $[-1, 1]$  & 0.557 &  0.102 &  0.117 &  0.040 &  0.184 & 907  & 915.5   \\
\hspace*{8pt} \nuebar-like     & 3 visible energy     & 10 in $[-1, 1]$  & 0.531 &  0.270 &  0.041 &  0.022 &  0.136 & 745  & 773.8   \\
\hspace*{8pt} $\mu$-like       & 4 visible energy     & 10 in $[-1, 1]$  & 0.027 &  0.004 &  0.913 &  0.005 &  0.051 & 2310 & 2294.0  \\
\hspace*{8pt} Other            & 4 visible energy     & 10 in $[-1, 1]$  & 0.275 &  0.029 &  0.348 &  0.049 &  0.299 & 1808 & 1772.6  \\
\\
\multicolumn{10}{l}{\textbf{Partially Contained (PC)}} \\
\hspace*{4pt} Stopping        & 2 visible energy      & 10 in $[-1, 1]$  & 0.084 &  0.032 &  0.829 &  0.010 &  0.045 & 566  & 570.0  \\
\hspace*{4pt} Through-going   & 4 visible energy      & 10 in $[-1, 1]$  & 0.006 &  0.003 &  0.978 &  0.007 &  0.006 & 2801 & 2889.9  \\
\\
\multicolumn{10}{l}{\textbf{Upward-going Muons (Up-$\mu$\xspace})} \\
\hspace*{4pt} Stopping        & 3 visible energy      & 10 in $[-1, 0]$  & 0.008 &  0.003 &  0.986 &  0.000 &  0.003 & 1456.4 & 1448.9  \\
\hspace*{4pt} Through-going \\
\hspace*{8pt} Non-showering   & single bin            & 10 in $[-1, 0]$  & 0.002 &  0.001 &  0.996 &  0.000 &  0.001 & 5035.3 & 4900.4  \\
\hspace*{8pt} Showering       & single bin            & 10 in $[-1, 0]$  & 0.001 &  0.000 &  0.998 &  0.000 &  0.001 & 1231.0 & 1305.0  \\
\hline
\hline
\end{tabular}
\caption{Sample purity broken down by neutrino flavor assuming neutrino oscillations with $\Delta m^{2}_{32} = 2.4\times 10^{-3}\mbox{eV}^{2}$ and $\sin^{2}\theta_{23} = 0.5$. 
         The data and MC columns refer to the total number of observed and expected events, respectively, 
         including oscillations but before fitting, for the full 328~kiloton-year exposure. 
         Sub-GeV multi-ring interactions are not used in the present analysis. 
         The numbers of observed and expected events in this sample are enclosed in parenthesis.} 
\label{tbl:full_samples}
\end{center}
\end{table*}

Super-Kamiokande is a cylindrical 50-kiloton water Cherenkov detector, located inside the Kamioka mine in Gifu, Japan. 
An inner detector (ID) volume is viewed by more than 11,000 inward-facing 20-inch photomultiplier tubes (PMTs) and contains a 32-kiloton target 
volume. 
The outer detector, which is defined by the two meter-thick cylindrical shell surrounding the ID, is lined with reflective Tyvek to increase light collection to 
1,885 outward-facing eight-inch PMTs mounted on the shell's inner surface.
Since the start of operations in 1996, Super-Kamiokande has gone through four data taking periods, SK-I, -II, -III, and -IV.

Though the basic configuration the detector is similar across the phases there are a few important differences.
At the start of the SK-IV period in 2008 the front-end electronics were upgraded to 
a system with an ASIC based on a high-speed charge-to-time converter~\cite{Nishino:2009zu}.
The new system allows for the loss-less data acquisition of all PMT hits above threshold 
and has improved the tagging efficiency of delayed Michel electrons from muon decay 
from 73\% in SK-III to 88\%. 

Further, following a period of detector maintenance and upgrades at the end of SK-I (1996-2001), the implosion of a single PMT at 
the bottom of the detector on November 12, 2001, created a shock wave and chain reaction 
that went on to destroy 6,665 ID and 1,027 OD PMTs.
The detector was rebuilt the following year with nearly half of the photocathode coverage (19\%) in the ID (5,137 PMTs) and the full 
complement of OD PMTs for the SK-II period (2002-2005).
Since that time all ID PMTs have been encased in fiber-reinforced plastic shells with 1.0~cm thick acrylic covers 
to prevent further chain reactions.
This resulted in an increased threshold of 7.0 MeV in SK-II compared to 5.0 MeV in SK-I. 
In 2006 the detector underwent a second upgrade in which the remaining ID PMTs were replaced and additional optical barriers
were added to the top and bottom portions of the OD to improve separation with its barrel region.
Both SK-III (2006-2008) and SK-IV (2008-present) were operated with the full 40\% photocathode coverage in the ID.

Neutrino interactions which produce charged particles above the Cherenkov threshold in water are reconstructed
based on the observed ring patterns projected on the detector walls.
Photomultiplier timing information is used to reconstruct the initial interaction vertex after correcting 
for the photon time of flight.
Particles are divided into two broad categories based upon their Cherenkov ring pattern and opening angle.
Rings from particles which produce electromagnetic showers, such as electrons and photons, tend to have
rough edges due to the many overlapping rings from particles in the shower and are labeled $e$-like or showering.  
Muons and charged pions on the other hand, which do not form showers, produce Cherenkov rings with 
crisp edges. 
Such rings are labeled $\mu$-like or non-showering.
The event reconstruction assigns momenta to each reconstructed ring in an event based on the observed number
of photons in the ring.
Particles with higher momenta produce brighter Cherenkov rings.
Similarly, particle directions are inferred based on the shape of their ring pattern.
Since the neutrino itself is unobserved, energy and direction variables for use in the 
oscillation analysis described below are based on the properties of their 
daughter particles. 

More detailed descriptions of the detector and its electronics can be found in~\cite{Fukuda:2002uc,Nishino:2009zu,Abe:2013gga}.

\subsection{Detector Calibration}

Over the 20 year history of the experiment changes in the run conditions have been unavoidable.
Seasonal changes in precipitation and the expansion of underground activities at the 
Kamioka site have variable impact on the quality and quantity of underground water available 
to fill the detector and maintain its temperature.
These changes impact the water transparency and subsequent performance of the detector and 
therefore must be corrected through calibrations.
Since neutrino oscillations are a function of the neutrino energy, a thorough understanding of the 
detector energy scale is important for precision measurements.

\begin{figure*}[htbp]
  \includegraphics[width=0.99\textwidth]{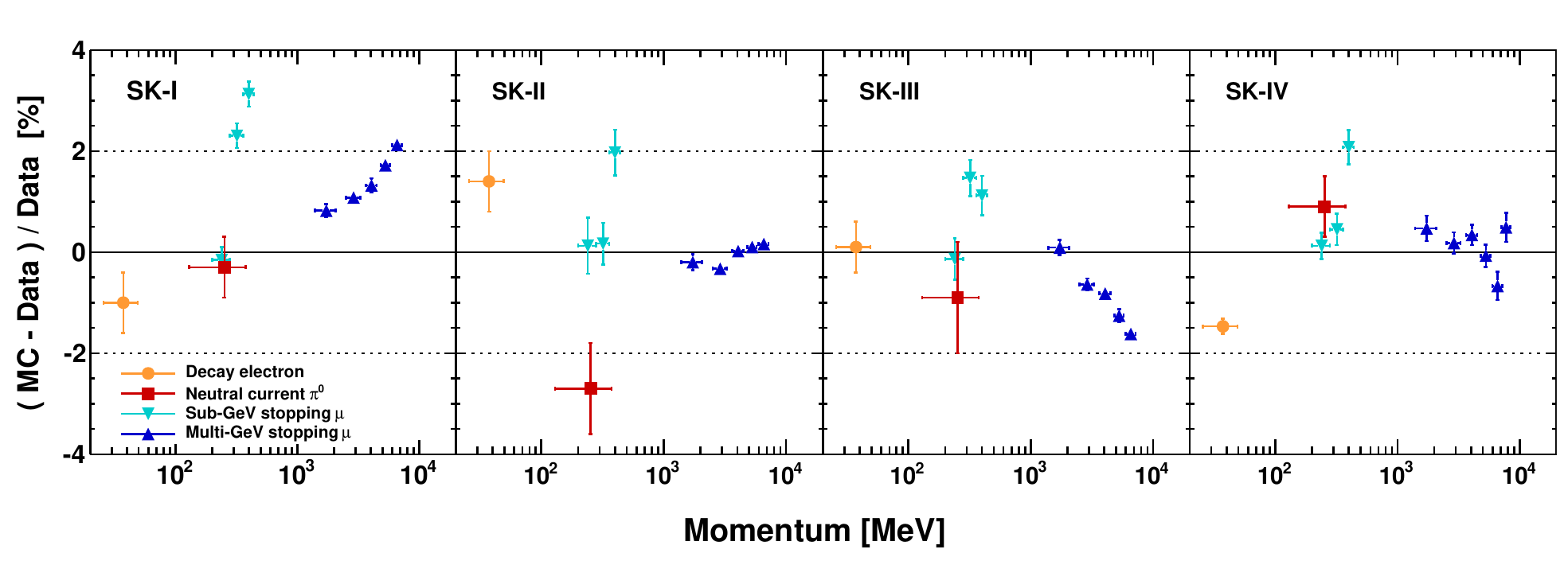}
  \caption{ Absolute energy scale measurements for each SK period. Vertical error bars denote the statistical uncertainty 
            and horizontal error bars the momentum range spanned by each analysis. }
\label{fig:abs_escale}
\end{figure*}

At the same time the range of energies of interest to atmospheric neutrino analysis spans 
from tens of MeV to tens of TeV, eliminating the possibility of calibration through radioactive 
isotopes.
Accordingly, the energy scale is calibrated using natural sidebands covering a variety of energies. 
Neutral pions reconstructed from atmospheric neutrino interactions provide a calibration point 
via the $\pi^{0}$ momentum and stopping cosmic ray muons of various momenta are used to 
measure photoelectron production as a function of 
muon track length (Cherenkov angle) for multi-GeV (sub-GeV) energies.
Here the muon track length is estimated using the distance between the 
entering vertex and the position of the electron produced in its subsequent decay. 
The energy spectrum of these Michel electrons additionally serves as a low energy calibration point.
Figure~\ref{fig:abs_escale} shows the absolute energy scale measurement using each of these
samples. 

\begin{figure}[htbp]
  \includegraphics[width=0.5\textwidth]{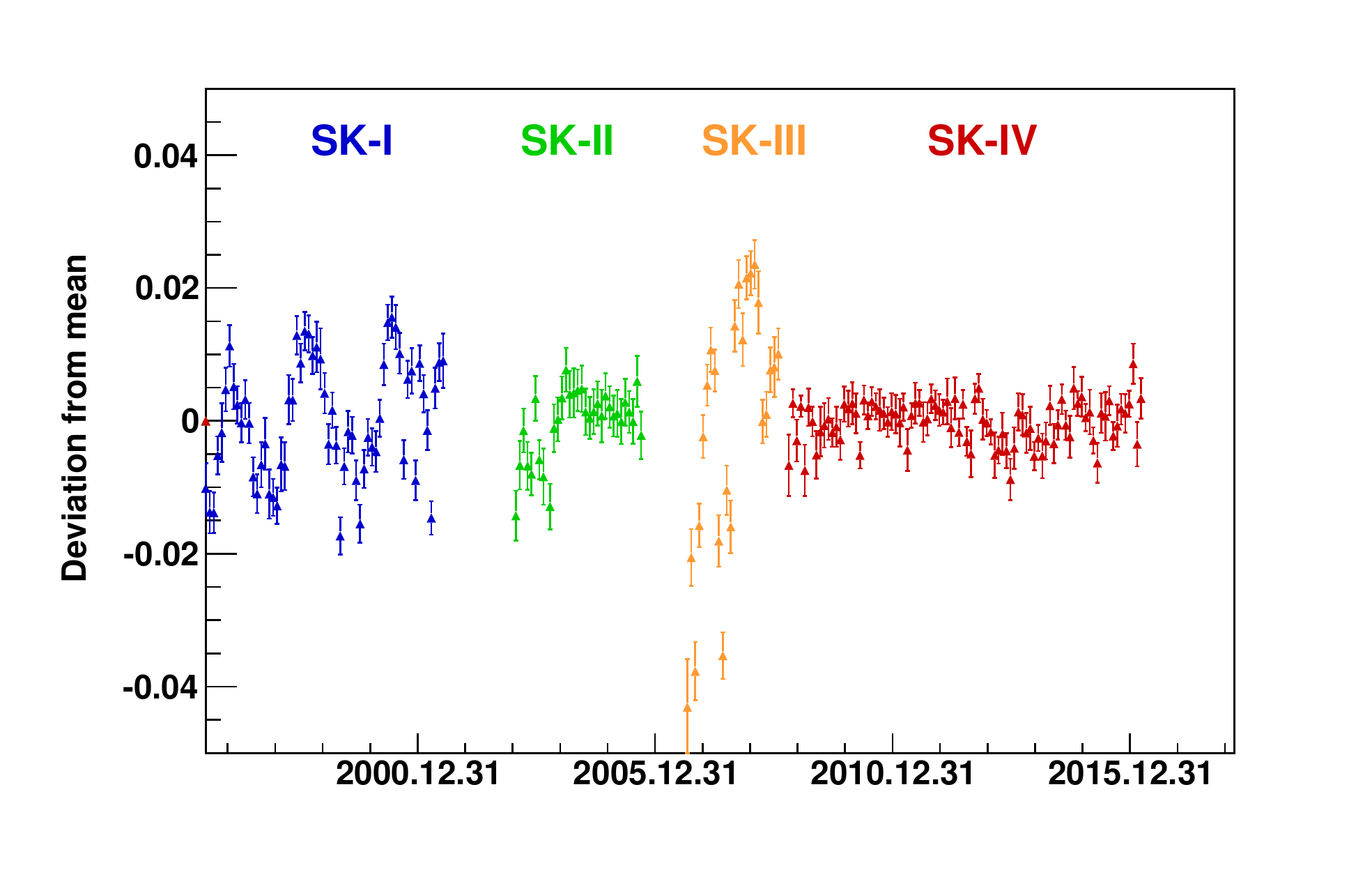}
  \caption{ Energy scale stability measured as a function of date since the start of SK operations.
            The energy scale is taken as the average of the reconstructed momentum divided by 
            range of stopping cosmic ray muon data in each bin. 
            The vertical axis shows the deviation of this parameter from the mean value 
            for each SK period separately. Error bars are statistical.}
\label{fig:escale_time}
\end{figure}

In the oscillation analysis the absolute energy scale uncertainty 
is conservatively taken to be the value of the most discrepant sample 
from this study in each run period. 
The total systematic error is assigned taking this value summed in quadrature 
with the time variation of the energy scale, which is measured 
using the variation in the average reconstructed momentum of Michel electrons 
and the variation in the stopping muon momentum divided by range.
An example of the latter showing the energy scale stability since SK-I appears in 
Fig.~\ref{fig:escale_time}.
Note the SK-III period was subject to poor and volatile water transparency 
conditions, resulting in a comparatively turbulent energy scale. 
The stability seen in the SK-IV period is a result of 
improvements in the water purification system and in corrections
for the time variation of the PMT response.
The total energy scale uncertainty in each period has been estimated as 
3.3\% in SK-I, 2.8\% in SK-II, 2.4\% in SK-III, and 2.1\% in SK-IV.

%
%

\subsection{Sample Selection}

The current analysis utilizes atmospheric neutrino data collected during each of the SK run periods and corresponds to a 
total livetime of 5,326 days, 2,519 of which are from SK-IV. 
Super-Kamiokande's atmospheric neutrino data are separated into three broad categories, fully contained (FC), 
partially contained (PC) and upward-going muons (Up-$\mu$) that are further sub-divided into the final analysis 
samples. 
Fully contained events have a reconstructed vertex within the 22.5~kton fiducial volume, defined as the region located 
more than 2~m from the ID wall, and with no activity in the OD. 
The FC data are sub-divided based upon the number of observed Cherenkov rings, the particle ID (PID) of the most energetic ring,
and visible energy or momentum into combinations of single- or multi-ring, electron-like ($e$-like) or muon-like ($\mu$-like), and 
sub-GeV ($E < 1330.0$ MeV) or multi-GeV ($E > 1330.0$ MeV). 
Additional selections are made based on the number of observed electrons from muon decays and the likelihood of containing a $\pi^{0}$.
For the SK-I, -II, and -III data periods the latter selection is based on~\cite{Wendell:2010md} and 
for SK-IV it is performed using the improved algorithm presented in~\cite{Abe:2013hdq}.
After all selections there are a total of 14 FC analysis samples. 
Events with a fiducial vertex but with energy deposition in the OD are classified as PC. 
Based on the energy deposition within the OD, PC events are further classified into ``stopping'' and ``through-going'' 
subsamples.

\begin{figure*}[phtb]
\includegraphics[width=1.0\textwidth]{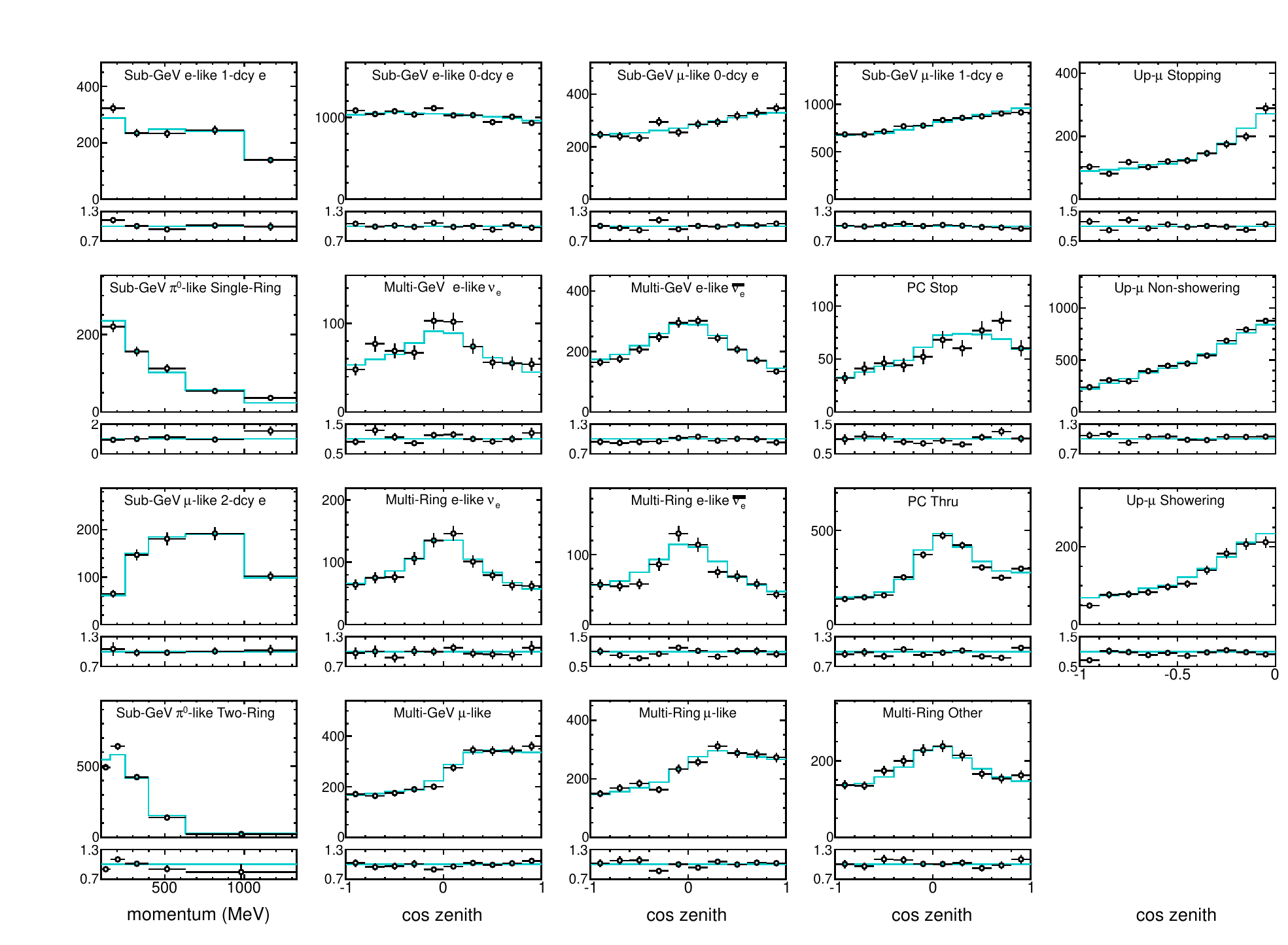}
\caption{ Data and MC comparisons for the entire Super-K data divided into 19 analysis samples. 
          Samples with more than one zenith angle bin (c.f. Table~\ref{tbl:full_samples}) are shown as zenith angle distributions 
          (second through fifth column) and other samples are shown as reconstructed momentum 
          distributions (first column). Lines denote the best fit MC assuming the normal hierarchy.
          Narrow panels below each distribution show the ratio of the data to this MC. 
          In all panels the error bars represent the statistical uncertainty.  
          In this projection each bin contains events of all energies, which 
          obscures the difference between the hierarchies. If the inverted hierarchy MC were also drawn it 
          would lie on top of the normal hierarchy line and for this reason it is not shown here.
          Figure~\ref{fig:ud_ratio} provides a better projection for comparing the hierarchies. 
        }
\label{fig:sk_zenith}
\end{figure*}

\begin{figure*}
    \includegraphics[width=\textwidth,keepaspectratio=true,type=pdf,ext=.pdf,read=.pdf]{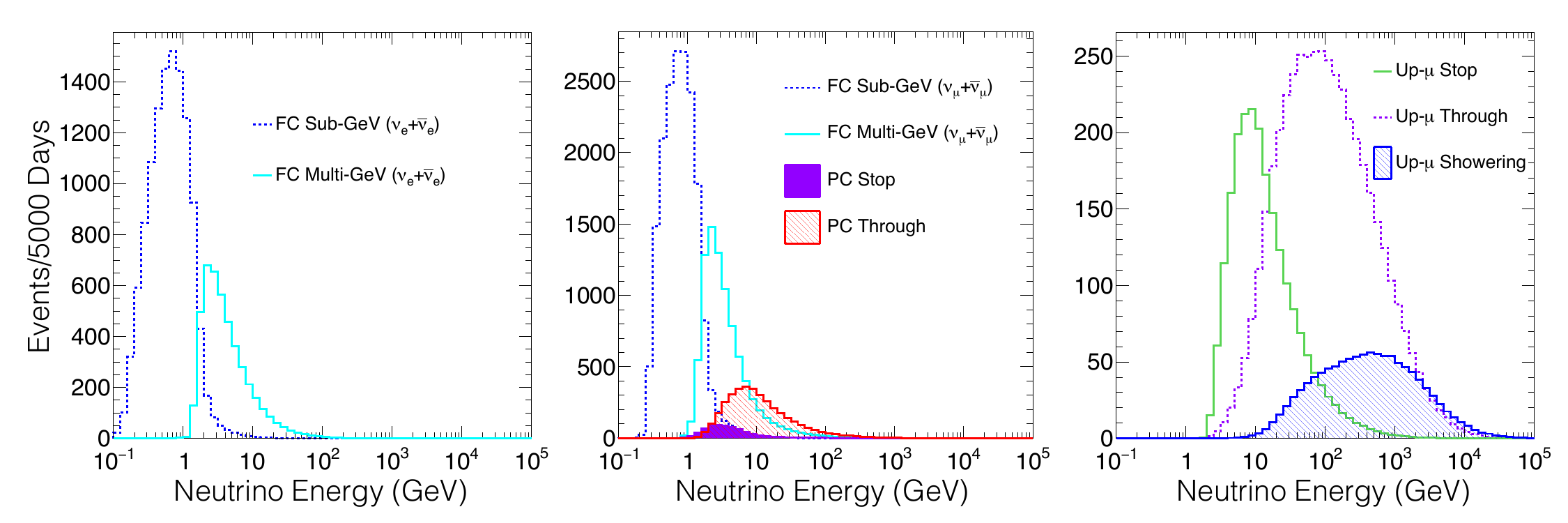}
    \caption{ True Super-K atmospheric neutrino energy spectra from simulation without oscillations.}
  \label{fig:true_spectrum}
\end{figure*}

\begin{figure}
    \includegraphics[width=0.50\textwidth]{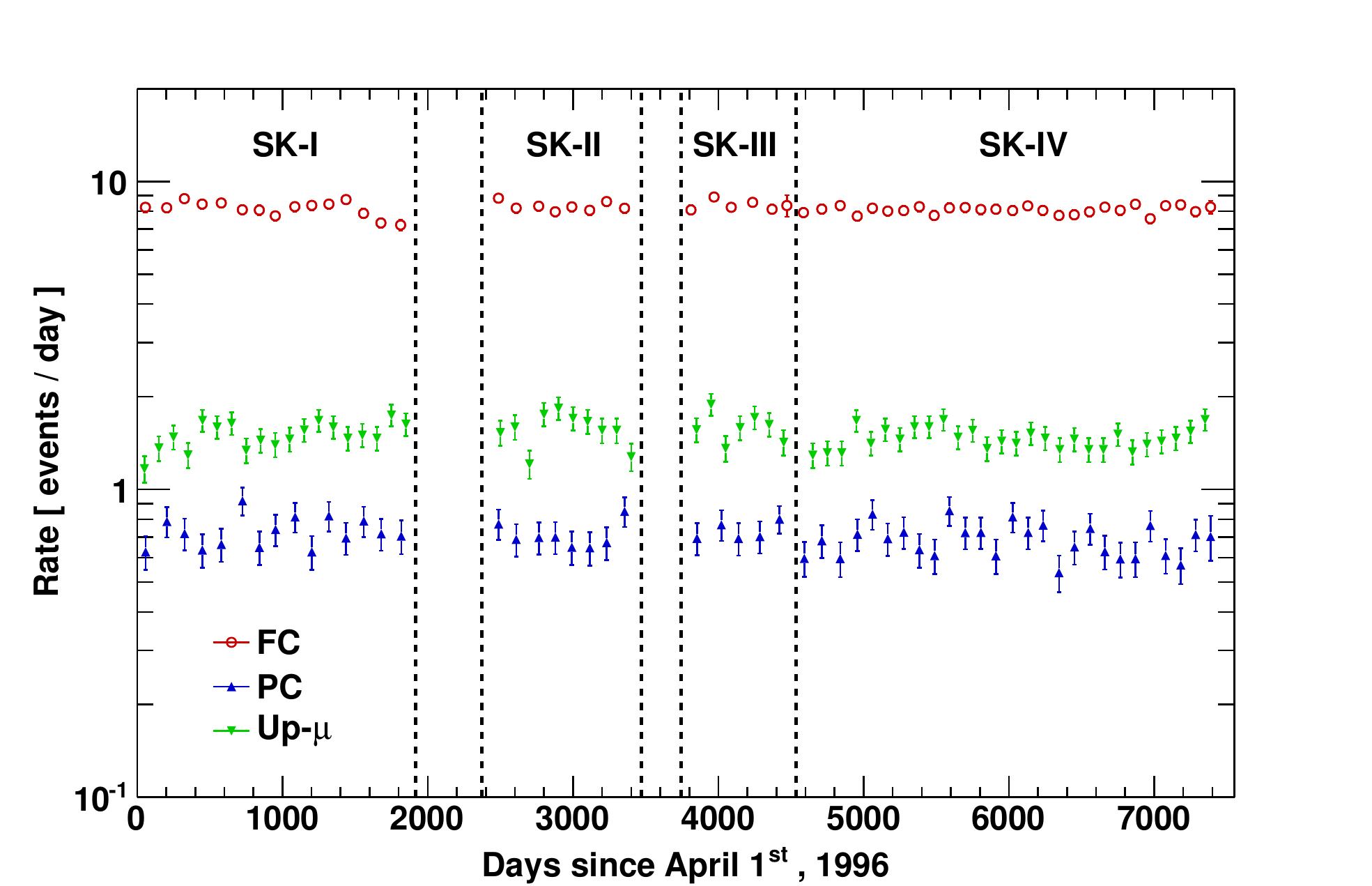}
    \caption{ Final event rates as a function of time since the start of SK operations. 
              The error bars are statistical.
              Circles denote the fully contained event rate and upward-facing (downward-facing) triangles 
              show the partially contained (upward-going muon) event rates.}
  \label{fig:reduction}
\end{figure}

The Up-$\mu$ sample is composed of upward-going muon events produced by neutrino interactions in the rock surrounding SK or in the OD water. 
Accordingly, light deposition in both the OD and ID is expected and the sample is divided into ``through-going'' and ``stopping'' 
subsamples for events that cross or stop within the ID, respectively. 
Through-going events with energy deposition consistent with radiative losses are separated into a ``showering'' subsample.
The 19 analysis samples defined for each of the SK run periods are summarized in Table~\ref{tbl:full_samples}.
Zenith angle distributions of each sample are shown in Fig.~\ref{fig:sk_zenith}.
Distributions of the true neutrino energy for the FC, PC, and Up-$\mu$ event categories appear in Fig.~\ref{fig:true_spectrum}.
Their event rates over the lifetime of the experiment have been stable at 8.3 FC events per day, 0.73 PC events per day, 
and 1.49 Up-$\mu$ events per day, as shown in Fig.~\ref{fig:reduction}.
In total 27,505 $\mu$-like and 
20,946 $e$-like data events are used in the analysis.
Though events classified as sub-GeV multi-ring interactions are present in the data, they are a small fraction of the 
available events and provide little additional oscillation sensitivity. 
As a result they are excluded from the present analysis. 

As outlined in Section~\ref{sec:oscillations} the primary handle for distinguishing the normal from the inverted 
mass hierarchy is whether neutrinos or antineutrinos undergo resonant oscillations as they traverse 
the earth. 
The effect of resonant oscillations would manifest most prominently as an excess of 
upward-going $e$-like events at 
$\mathcal{O}$(GeV) energies driven by $\nu_{\mu} \rightarrow \nu_{e}$ oscillations, so 
extracting the mass hierarchy requires separation of $\nu_{e}$ from $\bar \nu_{e}$ interactions.
As the SK detector is insensitive to the charge sign of particles traversing the detector, 
charged-current (CC) neutrino interactions and antineutrino interactions cannot be differentiated on an event-by-event basis.
Instead this separation is done statistically. 
It should be noted that due to their larger cross section and higher flux, more than twice as many neutrino interactions are expected in the data. 
Further, while hierarchy-sensitive matter effects are also present in the $\nu_{\mu} \rightarrow \nu_{\mu}$
channel, attempts to similarly separate the $\mu$-like data yielded no significant change in sensitivity and are not considered here.
 
Between two and ten GeV, in addition to charged-current quasi-elastic interactions, single-pion ($1\pi$) production via 
$\Delta$ resonance excitation and deep inelastic scattering (DIS) processes are significant.  
In the case of the former, the outgoing $\pi^{-}$ in antineutrino reactions, such as $\bar \nu_{e} + n \rightarrow  e^{+} n \pi^{-}$,
will often capture on a $^{16}\mbox{O}$ nucleus leaving the positron as the only Cherenkov light-emitting particle. 
Neutrino interactions, on the other hand, are accompanied by a $\pi^{+}$, such as in $\nu_{e} + n \rightarrow  e^{-} n \pi^{+}$,
where the $\pi^{+}$ does not capture in this manner and can therefore survive long enough to produce a delayed electron through its decay chain. 
For CC $\bar \nu_{e}$ interactions in which the $\pi^{-}$ has captured there will be no such decay electrons.
Accordingly, an antineutrino enriched subsample is extracted from the single-ring multi-GeV $e$-like sample by additionally 
requiring there are no decay electrons present. 
This cut defines the single-ring multi-GeV $\bar \nu_{e}$-like sample and its rejected events 
form the single-ring multi-GeV $\nu_{e}$-like sample. 
After this selection the fractions of charged-current electron neutrino and antineutrino events
in the $\nu_{e}$-like sample are 62.1\% and 9.0\%, respectively.
For the $\bar \nu_{e}$-like sample the fractions are 54.6\% and 37.2\%.

At these energies, events with more than one reconstructed ring are often DIS interactions, which produce both multiple charged pions 
and nuclear fragments. 
In order to purify the neutrino and antineutrino components of the multi-ring samples a two-stage likelihood method has been developed. 
Due to the presence of several light-producing particles the Cherenkov ring produced by the leading lepton is often obscured, 
resulting in degraded PID performance and accordingly, significant NC and $\nu_{\mu}$-induced backgrounds in multi-ring events whose 
most energetic ring is $e$-like. 
The first stage of the separation is designed to extract and purify CC $\nu_{e} + \bar \nu_{e}$ interactions from 
this base sample.  
To perform this selection a likelihood function, detailed in a previous publication~\cite{Wendell:2010md}, is built from 
the PID variable of the event's most energetic ring, the fraction of the event's total momentum it carries, the number of 
decay electrons, and the largest distance to a decay electron vertex from the primary event vertex. 
The efficiency of this method for selecting true CC $\nu_{e} + \bar \nu_{e}$ events is 72.7\% and results in a sample 
that is 73.0\% pure in these interactions.
Separate likelihoods are prepared for each of the run periods and yield similar efficiencies and purities. 
Events that pass this selection are classified as ``multi-ring $e$-like'' while those that fail are termed 
``multi-ring other.''
Though the multi-ring other sample has not been used in previous Super-K oscillation analyses it is introduced here 
since its $\nu_{e}$ component offers some hierarchy sensitivity and its oscillation-induced $\nu_{\tau}$ 
and NC components provide additional constraints on related systematic uncertainties.

\begin{figure}
    \includegraphics[width=0.5\textwidth]{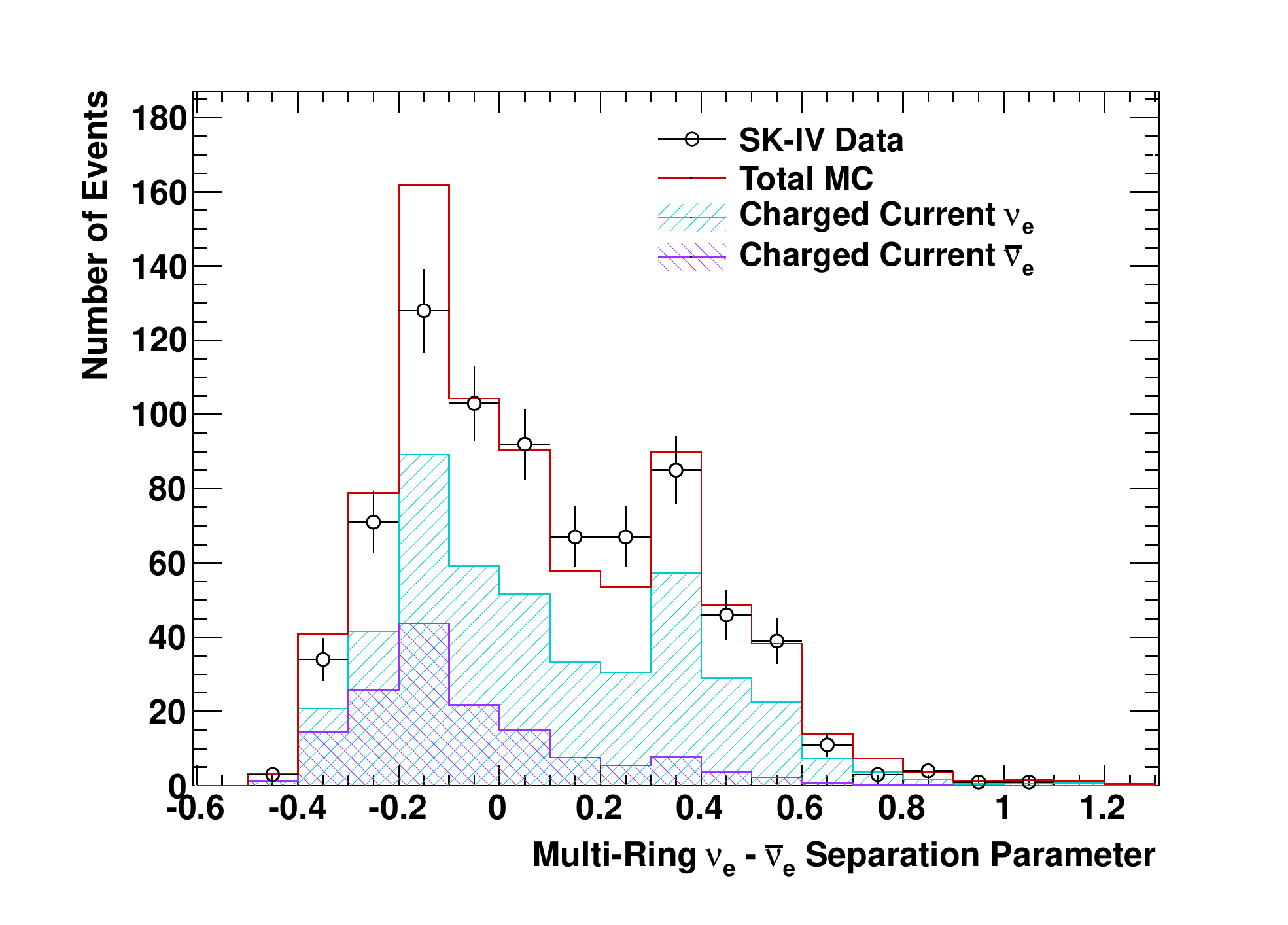}
    \caption{ Likelihood distribution used to separate SK-IV Multi-ring $e$-like events 
              into the neutrino-like and antineutrino-like samples. 
              Error bars represent the statistical uncertainty of the data.
              Events with negative (positive) likelihood values are designated \nuebar-like (\nue-like). }
  \label{fig:mmebar}
\end{figure}

The second stage of the separation process focuses on separating samples enriched in neutrino and antineutrino interactions 
from the multi-ring $e$-like data.
A second likelihood method is introduced based on three variables, the number of reconstructed rings, 
the number of decay electrons, and the event's transverse momentum.  
For charged-current interactions the conservation of charge implies the total charge of the recoiling hadronic system 
must be positive to balance the negative charge of the out-going lepton. 
The total charge carried by hadrons emerging from antineutrino interactions, on the other hand, will be zero or negative.
As a result, the charged pion multiplicity, and hence number of visible Cherenkov rings, in neutrino-induced events is expected 
to exceed that from antineutrino events.
This difference is enhanced by the propensity for $\pi^{-}$ to capture in water. 
In combination these two effects suggest that more electrons from the $\pi$ decay chain are expected in $\nu$ interactions. 
Due to the V-A structure of the weak interaction, the angular distribution of the leading lepton from $\bar \nu$ 
interactions is more forward than those from $\nu$ processes. 
As a result, the transverse momentum of the system is expected to be smaller for the former. 
Since there is no direct knowledge of an incoming atmospheric neutrino's direction the transverse momentum of each event 
is defined relative to the direction of the most energetic ring.
The final likelihood is defined over five visible energy bins, 1.33-2.5 GeV , 2.5-5.0 GeV, 5.0-10.0 GeV, 10.0-20 GeV and 
$>$ 20 GeV for each SK run period.
Figure~\ref{fig:mmebar} shows the combined likelihood distribution used in SK-IV.  
The efficiency for identifying true CC $\bar{\nu}_{e}$ ($\nu_e$) events as $\bar{\nu}_{e}$-like is 71.5\% (47.1\%).

\subsection{Simulation}

The simulation of atmospheric neutrinos is performed following the 
flux calculation of Honda et. al~\cite{Honda:2011nf} and using the NEUT~\cite{Hayato:2002sd} 
simulation software (version 5.3.6) to generate neutrino interactions for tracking 
in a GEANT3~\cite{Brun:1994aa}-based simulation of the Super-K detector~\cite{Abe:2013gga}.
Several improvements to NEUT have been made since the previous version used for 
atmospheric neutrino analysis (c.f.~\cite{Abe:2014gda}).
Charged-current quasi-elastic interactions are simulated using the 
Llewellyn-Smith formalism~\cite{LlewellynSmith:1971uhs} with nucleons distributed 
according to the Smith-Moniz relativistic Fermi gas~\cite{Smith:1972xh} assuming 
an axial mass $M_{A} = 1.21 \mbox{GeV/c}^{2}$ and form factors from~\cite{Bradford:2006yz}.
Interactions on correlated pairs of nucleons, so-called meson exchange currents (MEC), 
have been included following the model of Nieves~\cite{Nieves:2004wx}.
Pion-production processes are simulated using the Rein-Sehgal model~\cite{Rein:1980wg}
with Graczyk form factors~\cite{Graczyk:2007bc}. 
Since the MEC simulation includes delta absorption processes, the pionless 
$\Delta$ decay process, $\Delta + N' \rightarrow N'' + N'$, in NEUT's previous pion production 
model has been removed in the present version.

NEUT's cascade model is used in the detector simulation to treat 
the hadronic interactions of pions with nuclei in the detector.
The cross sections underlying the model, including charge exchange, absorption, inelastic scattering, and hadron production processes, 
have been tuned using a fit to external pion scattering data as 
described in Ref.~\cite{Abe:2015awa} (c.f. Table~IV).
Uncertainties from that fit have been propagated as systematic uncertainties in the present analysis.
Differences in the expected number of pions in the final state between the NEUT prediction and
measurements from the CHORUS experiment~\cite{KayisTopaksu:2007pe} are considered as an additional source 
of systematic uncertainty affecting the event selection presented above. 

\section{Atmospheric Neutrino Analysis}
\label{sec:atm_only}

Three fits, each incorporating a different degree of external
information, are performed to estimate oscillation parameters.
In the first and least-constrained fit, the Super-K atmospheric
neutrino data are fit allowing $\theta_{13}$ to vary as a free
parameter.  The second fit similarly uses only atmospheric neutrino
data, but assumes $\theta_{13}$ to be the average of several reactor
neutrino disappearance experiment measurements, $\mbox{sin}^{2}
\theta_{13} = 0.0219 \pm 0.0012$~\cite{Agashe:2014kda}.  Finally, the T2K samples
discussed in section~\ref{sec:gaibu} are fit alongside the atmospheric
neutrino data under the same assumption.  In each of these fits the
data are fit against both the normal and inverted hierarchy
hypotheses.

%
%
%
%
\begin{figure*}[htbp]
 \subfigure{
  \includegraphics[width=0.40\textwidth]{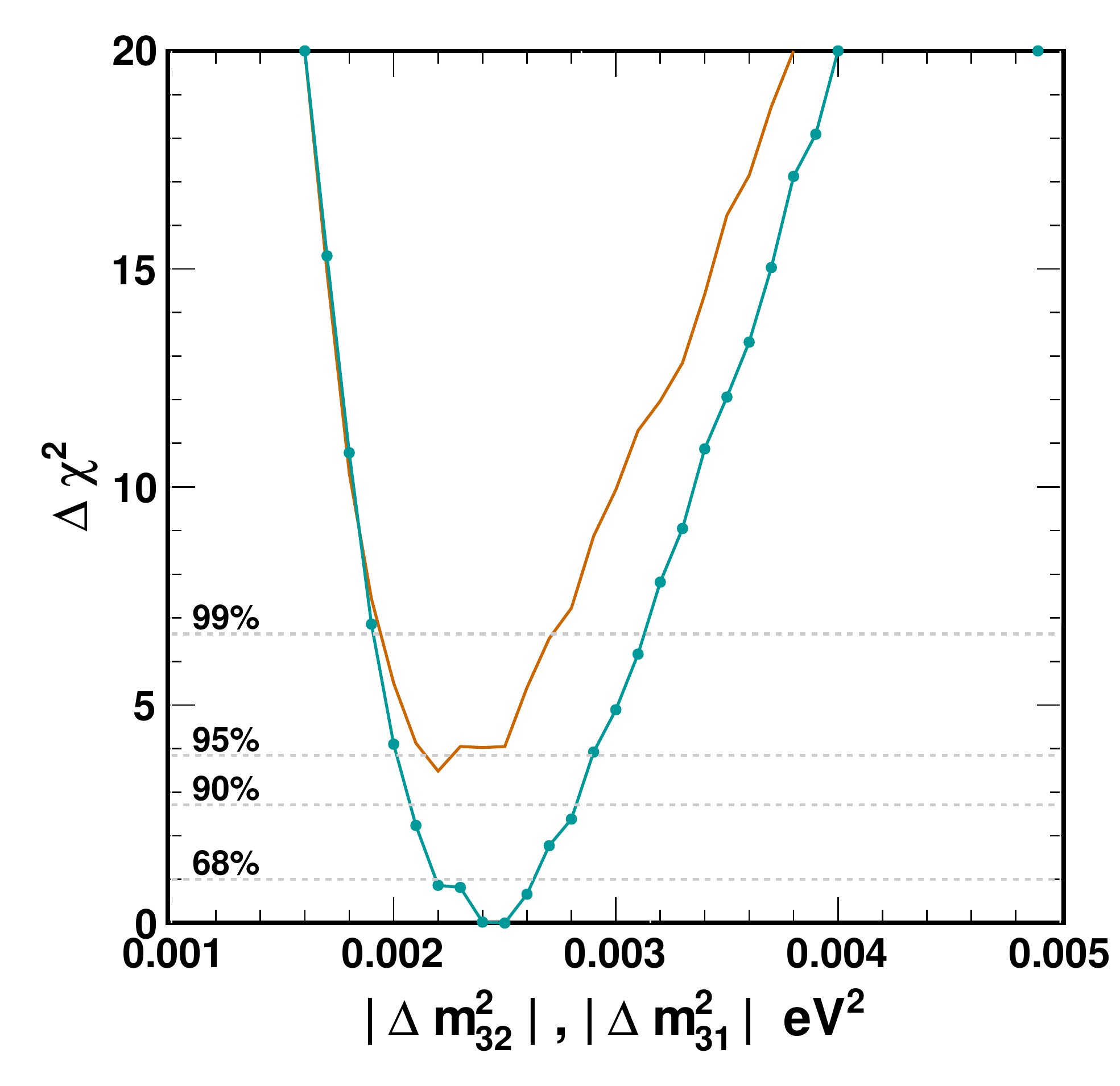}
 }
 \subfigure{
  \includegraphics[width=0.40\textwidth]{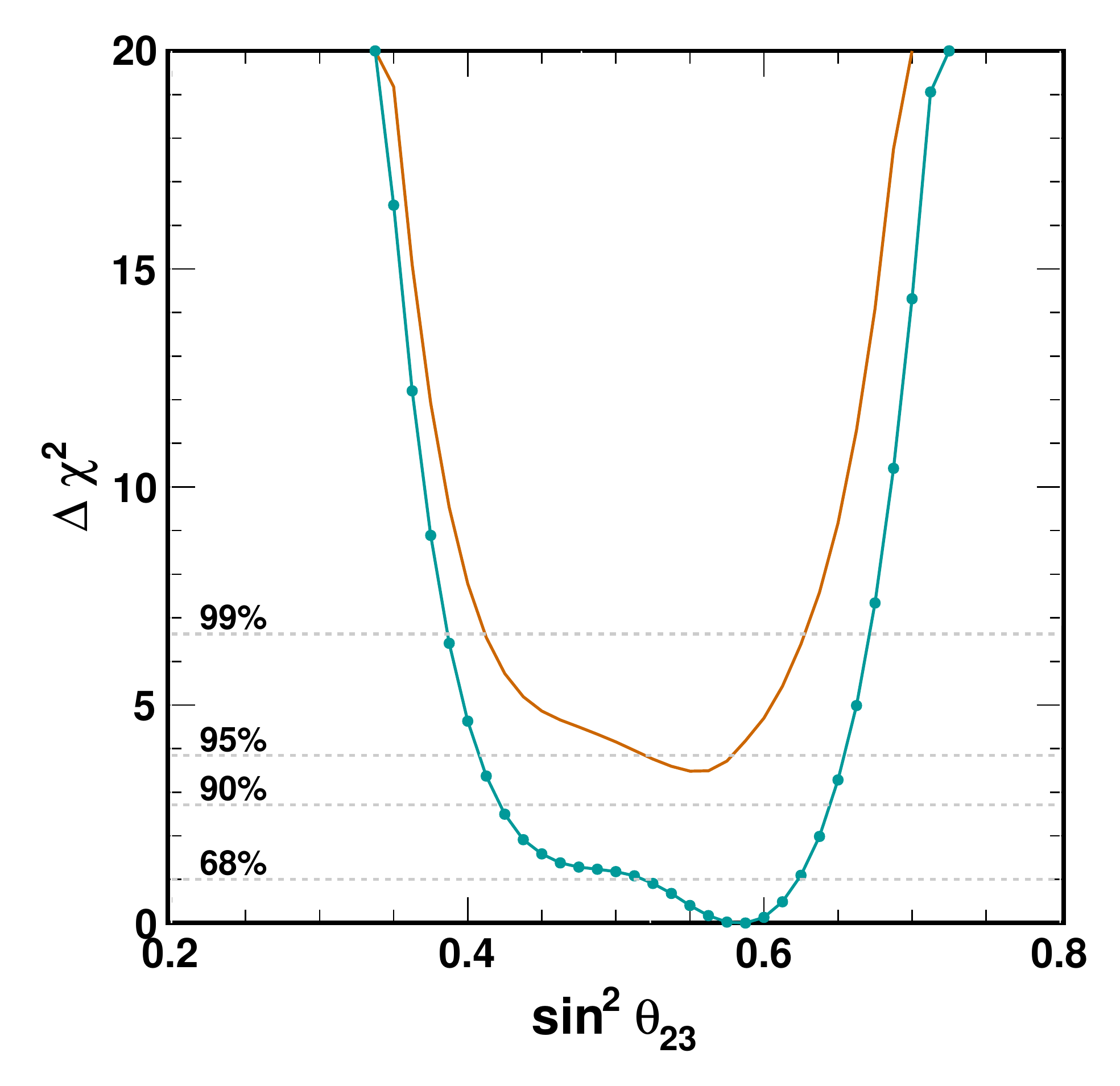}
 }
 \subfigure{
  \includegraphics[width=0.40\textwidth]{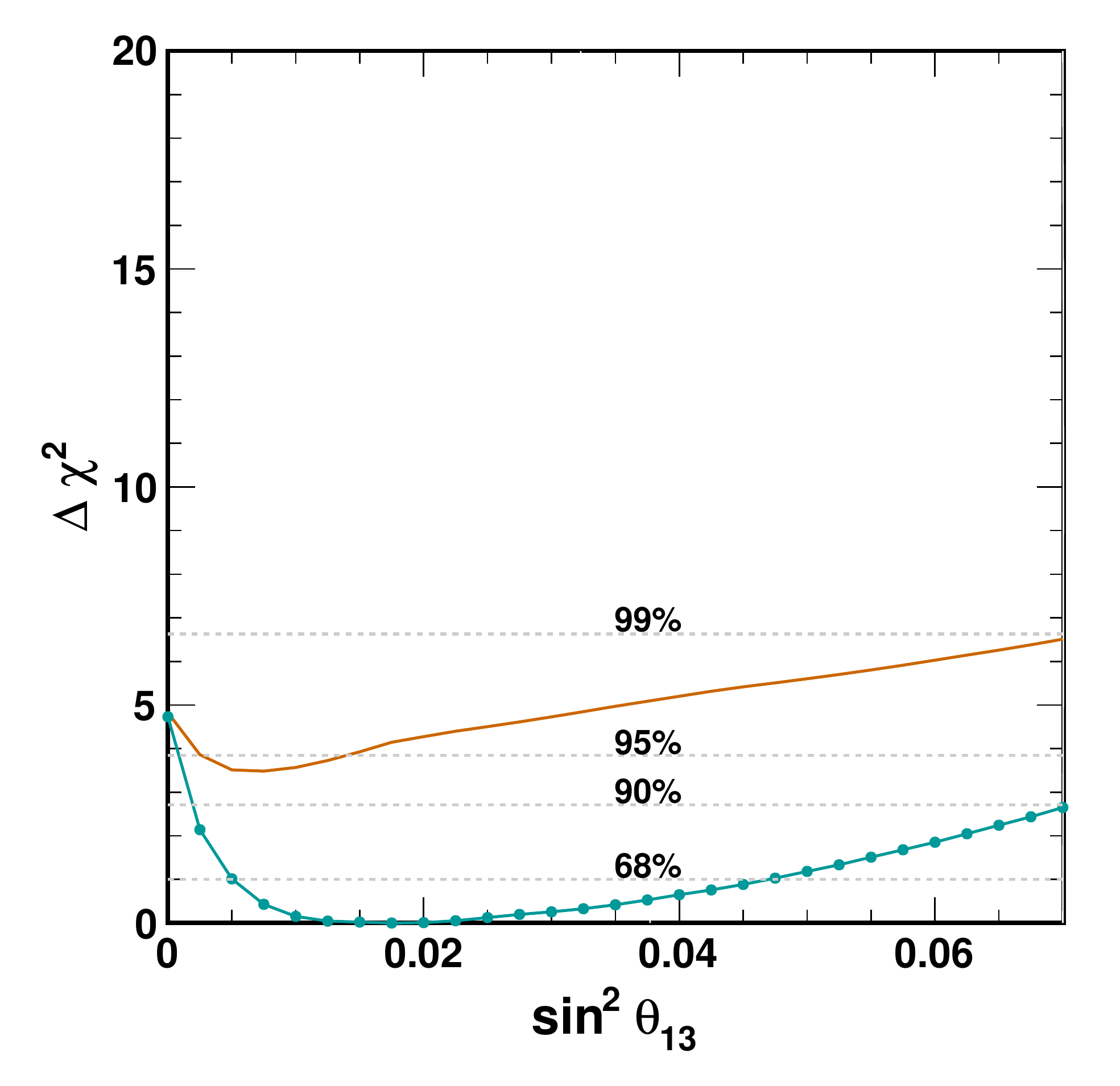}
 }
 \subfigure{
  \includegraphics[width=0.40\textwidth]{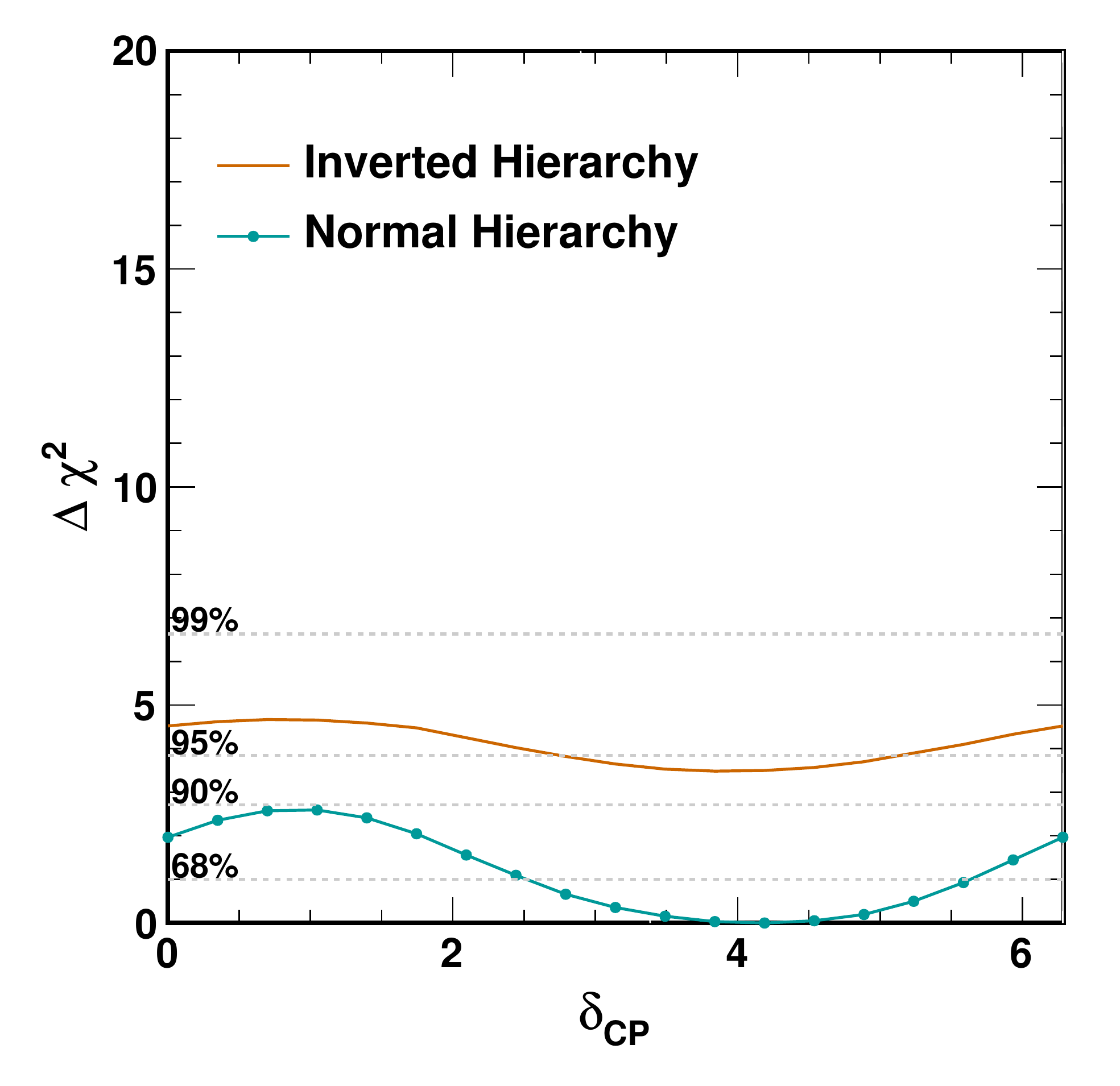}
 }
  \caption{ Constraints on neutrino oscillation parameters from
            the Super-K atmospheric neutrino data fit with no external constraints.
            Orange lines denote the inverted hierarchy result, which
            has been offset from the normal hierarchy result, shown in
            blue, by the difference in their minimum $\chi^{2}$ values.}
  \label{fig:q13free_sk_cont}
\end{figure*}

Data are fit to the MC using a binned $\chi^{2}$ method built assuming
Poisson statistics and incorporating systematic errors as scaling
factors on the MC in each bin~\cite{Fogli:2002pt}:
%
\begin{eqnarray}
\chi^{2}  = 2 \displaystyle \sum_{n} \left(  E_{n} -\mathcal{O}_{n}  + \mathcal{O}_{n} \ln \frac{ \mathcal{O}_{n} }{ E_{n} } \right)
             + \displaystyle \sum_{i} \left( \frac{ \epsilon_{i} }{ \sigma_{i} } \right)^{2}, 
\label{eq:fullchi}
\end{eqnarray}
\noindent where, 
%
\begin{eqnarray}
E_{n} &=& \displaystyle \sum_{j} E_{n,j}(1 + \displaystyle \sum_{i}
f^{i}_{n,j} \epsilon_{i} ) \\ \mathcal{O}_{n} &=& \displaystyle
\sum_{j} \mathcal{O}_{n,j}.
\end{eqnarray}
\noindent In this equation $E_{n,j}$ represents the MC expectation in
the $n^{th}$ analysis bin for the $j^{th}$ SK period.  Similarly,
$\mathcal{O}_{n,j}$ is the corresponding data in that bin and
$f^{i}_{n,j}$ is a coefficient describing the fractional change in the
bin's MC under a $1{\sigma_{i}}$ variation of the $i^{th}$
systematic error source.  Systematic errors penalize the $\chi^{2}$
based on their corresponding fitting parameters, $\epsilon_{i}$.
Solving the system of equations defined by the requirement $\partial
\chi^{2} / \partial \epsilon_{i} = 0 $ for each systematic error
brings the data and MC into the best agreement allowed by the
systematic errors.  This minimization in the systematic error
parameters is repeated over a grid of oscillation parameters and the
parameter set returning the smallest value of $\chi^{2}$ is taken as the
best fit.

The fit is performed over 520 analysis bins for each of the SK periods
and a total of 155 systematic error sources.  In
addition, a systematic error on the presence of meson exchange
currents has been added to the analysis where the difference
between the NEUT model with and without MEC is taken as the $1{\sigma}$
uncertainty.  Further, the single pion production error of
previous analyses has been broken down into three parts following the
parameterization of Ref.~\cite{Graczyk:2007bc}.
Systematic errors and their sizes at the best fit point of the analysis 
are presented in Tables~\ref{tab:sysa}, ~\ref{tab:sysb}, and ~\ref{tab:sysc}.

%
%
\begin{table}
\begin{center}
\begin{tabular}{lc}
Parameter & Value  \\
\hline 
\hline 
$\Delta m^{2}_{21}$             & $(7.53 \pm 0.18) \times 10^{-5} \mbox{eV}^{2}$  \\  
$\mbox{sin}^{2}    \theta_{12}$ & $0.304  \pm 0.014$  \\  
$\mbox{sin}^{2}    \theta_{13}$ & $0.0219 \pm 0.0012$  \\  
\hline 
\hline 
\end{tabular}
\end{center}
\caption{Values of oscillation parameters fixed in the analysis and
  their systematic errors.  Note that $\mbox{sin}^{2} \theta_{13}$ is
  only fixed in the ``$\theta_{13}$ constrained''analyses described in
  Section~\ref{sec:gaibu}.}
\label{tbl:oscparm}
\end{table}

When the atmospheric data are studied without external constraints the
fit is performed over four parameters. The agreement between the data
and MC is evaluated using Equation~\ref{eq:fullchi} at each point in
the grid spanned by $ 0.0 \leq \mbox{sin}^{2} \theta_{13} \leq 0.10$
(15 points), $ 0.3 \leq \mbox{sin}^{2} \theta_{23} \leq 0.7 $ (25
points), $ 1.0 \times 10^{-3} \leq |\Delta m^{2}_{32,31}| \leq 5.0
\times 10^{-3} \mbox{eV}^{2}$ (51 points), and $ 0.0 \leq \delta_{CP}
\leq 2 \pi $ (19 points).  The solar mixing parameters are set to the
values in Table~\ref{tbl:oscparm} but their uncertainties are treated
as a source of systematic error in the analysis.  For the normal
(inverted) hierarchy fit the fitting parameter is $\Delta m^{2}_{32}$
($\Delta m^{2}_{31}$).  Independent fits are performed for the
normal and inverted hierarchies and the grid point returning the
smallest value of $\chi^{2}$ is termed the best fit for each.
The smallest of these is taken as the global best fit.

Further, the compatibility of the atmospheric neutrino data with
oscillations subject to matter effects in the Earth is evaluated by
performing the same fits with $\mbox{sin}^{2} \theta_{13}$ constrained
to $0.0219 \pm 0.0012$ (discussed below) and introducing an additional
scaling parameter on the electron density in
Equation~\ref{eqn:hamiltonian_matter}, $\alpha$.  This parameter is
allowed to range in 20 steps from 0.0 to 1.9, with $\alpha = 1.0$
corresponding to the standard electron density in the Earth.

\subsection*{Results and Discussion }

Figure~\ref{fig:q13free_sk_cont} shows one-dimensional allowed regions
for $| \Delta m^{2}_{32,31} |$, $\mbox{sin}^{2} \theta_{23}$,
$\theta_{13}$ and $\delta_{CP}$.  In each plot the curve is drawn such
that the $\chi^{2}$ for each point on the horizontal axis is the
smallest value among all parameter sets including that point.  When
the atmospheric neutrino data are fit by themselves with no constraint
on $\theta_{13}$, the normal hierarchy hypothesis yields better data-MC
agreement than the inverted hierarchy hypothesis with 
$\chi^{2}_{NH,min} - \chi^{2}_{IH,min} = -3.48$.
The preferred value of $\mbox{sin}^{2}
\theta_{13}$ is $0.018 (0.008)$ assuming the former (latter).  Though
both differ from the globally preferred value of $0.0219$ the
constraints are weak and include this value at the $1{\sigma}$ level.
In the normal hierarchy fit the point at $\mbox{sin}^{2} \theta_{13} =
0.0$ is disfavored at approximately $2{\sigma}$ indicating the data have
a weak preference for non-zero values.  A summary of the best fit
information and parameter constraints is presented in
Table~\ref{tbl:bestfits}.

\begin{figure*}[htbp]
\includegraphics[width=0.90\textwidth]{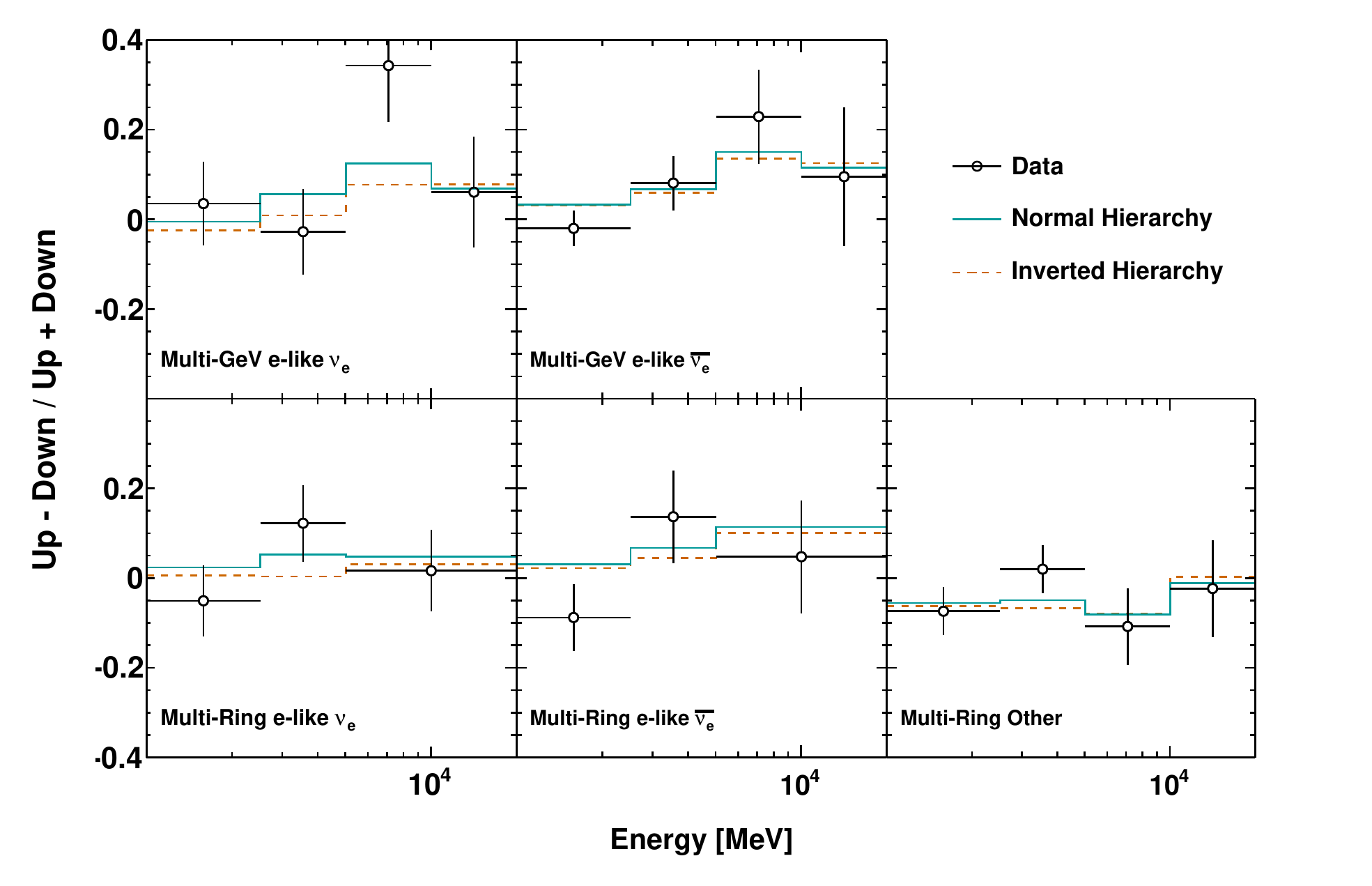}
\caption{ Upward- ($\mbox{cos}\theta < -0.4$) to downward-going
  ($\mbox{cos}\theta > 0.4$) event ratio as a function of
  energy. The error bars are statistical. For the single-ring samples the energy is taken to be the visible energy 
  assuming the light-producing particle was an electron. For the 
  multi-ring samples the total energy is used after accounting for 
  the particle type (electron or muon) of each reconstructed ring. 
  The cyan line denotes the best fit from the normal hierarchy hypothesis, and the orange dashed line 
  the best fit from the inverted hierarchy hypothesis.
  The error on the prediction is dominated by the uncertainty in the $\nu_{\tau}$ cross section
  and is not more than 3\% (absolute) in any bin of the figure. }
\label{fig:ud_ratio}
\end{figure*}

The data's preference for both non-zero $\mbox{sin}^{2} \theta_{13}$
and the normal mass hierarchy suggest the presence of upward-going
electron neutrino appearance at multi-GeV energies driven by matter
effects in the Earth (c.f. Fig.~\ref{fig:osc_prob}).
Figure~\ref{fig:ud_ratio} shows the up-down asymmetry of the multi-GeV
single- and multi-ring electron-like analysis samples.  Here the
asymmetry is defined as $N_{U}-N_{D}/N_{U}+N_{D}$, where
$N_{U}$($N_{D}$) are the number of events whose zenith angle satisfy
$\mbox{cos}\theta_{z} < -0.4$ ($\mbox{cos}\theta_{z} > 0.4$).
Small excesses seen between a few and ten GeV in the
Multi-GeV e-like $\nu_{e}$ and the Multi-Ring e-like $\nu_{e}$ and
$\bar \nu_{e}$ samples drive these preferences.

The normal hierarchy fits to the atmospheric mixing parameters yield
$\Delta m^{2}_{32} = 2.50^{+0.13}_{-0.31}\times 10^{-3} \mbox{eV}^{2}$ and $\mbox{sin}^{2}
\theta_{23} = 0.587^{+0.036}_{-0.069}$.
However, the Super-K data show a weak preference for the second octant
of $\theta_{23}$, disfavoring maximal mixing ($\mbox{sin}^{2}
\theta_{23} = 0.5$) at around $1{\sigma}$ significance.  This preference
is driven by data excesses (deficits) at multi-GeV energies in the
upward-going regions of the single-ring e-like $\nu_{e}$ ($\mu$-like)
and multi-ring other samples.  These features are consistent with
expectations from $\nu_{\mu}\rightarrow \nu_{e}$ oscillations driven
by non-zero $\theta_{13}$.

The best fit value of $\delta_{CP}$ is found to be 4.18 (3.84) radians
in the normal (inverted) fit, with the least preferred parameter value
near 0.8 radians disfavored by $\Delta \chi^{2} $ = 2.7 (1.0).  This
preference is driven predominantly by the sub-GeV e-like samples, via
$\nu_{\mu} \rightarrow \nu_{e}$ oscillations.  
Though the effect of this parameter
at these energies is a complicated function of both energy and the
neutrino path length, the point at 4.18 radians generally induces more
electron neutrino appearance in the sub-GeV e-like samples.  At higher
energies the effect of $\delta_{CP}$ modulates the
$\theta_{13}$-driven $\nu_{\mu} \rightarrow \nu_{e}$ probability in
the resonance region, but is secondary in size and induces more (less)
appearance at 4.18 (0.8) radians.  As there are fewer antineutrino
events relative to neutrino events in the atmospheric sample there is
accordingly more freedom to adjust $\theta_{13}$ to bring the MC
prediction into agreement with data in the inverted hierarchy fit.  As
a result a weaker constraint on $\delta_{CP}$ is obtained.

\begin{figure}[htbp]
  \includegraphics[width=0.47\textwidth,keepaspectratio=true]{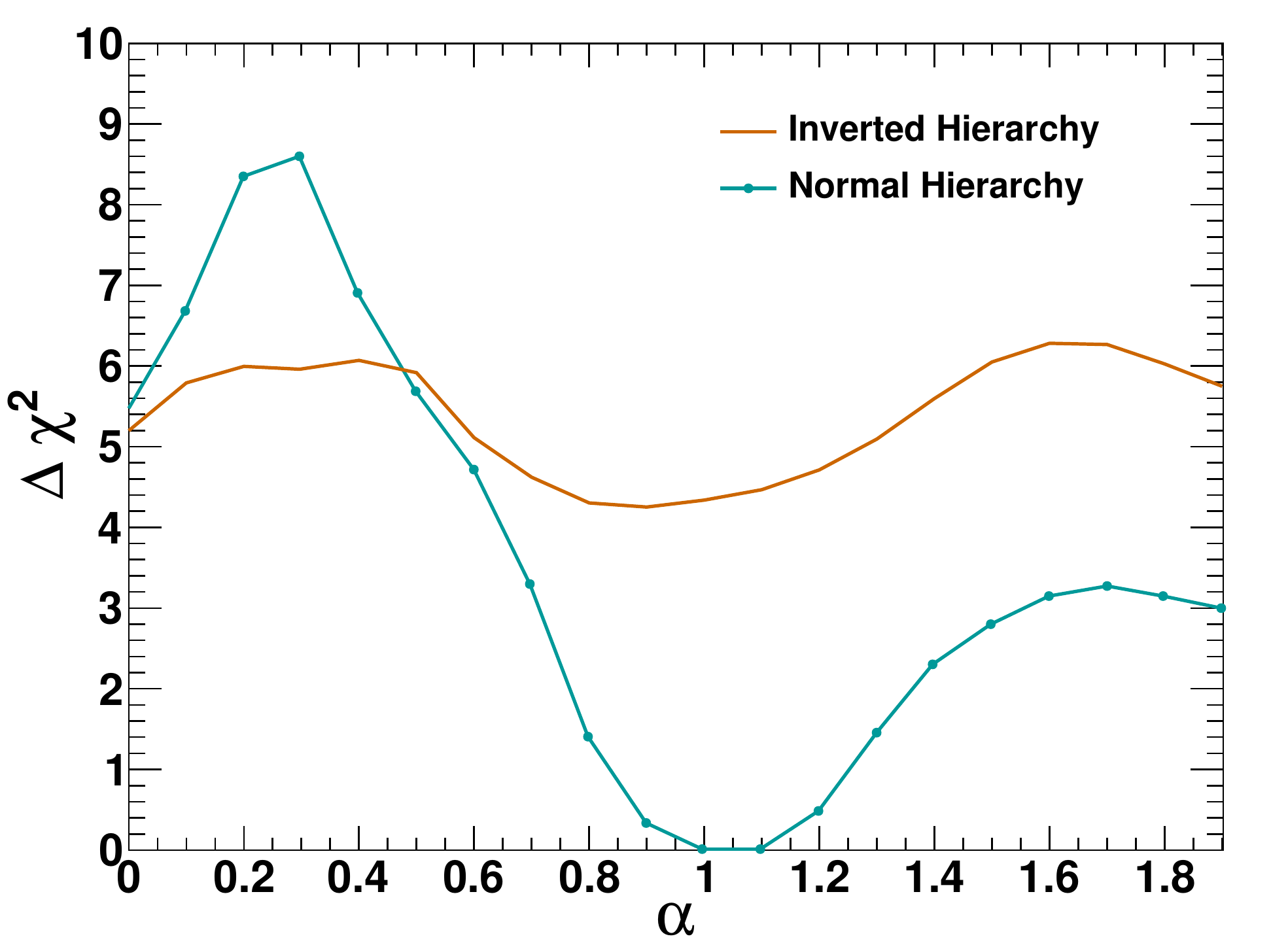}
  \caption{ Constraints on the matter effect parameter $\alpha$ from
    the Super-K atmospheric neutrino data fit assuming $\mbox{sin}^{2}
    \theta_{13} = 0.0219\pm 0.0012$ .  Orange lines denote the inverted hierarchy
    result, which has been offset from the normal hierarchy result,
    shown in blue, by the difference in their minimum $\chi^{2}$
    values.  Vacuum corresponds to $\alpha=0$, while the standard
    matter profile used in the rest of the analyses presented here
    corresponds to $\alpha=1$. }
  \label{fig:sk_matter}
\end{figure}

The consistency of these data with the presence of matter effects is
illustrated in Fig.~\ref{fig:sk_matter}.  With $\mbox{sin}^{2}
\theta_{13}$ set to $0.0219\pm0.0012$, the data prefer the normal hierarchy with
an electron density consistent with that of standard matter
($\alpha=1.0$).  Purely vacuum oscillations, represented by
$\alpha=0.0$, are disfavored by the fit by 
$\chi^{2}_{\alpha = 0}- \chi^{2}_{min} = 5.2$ after accounting for the
hierarchy uncertainty.  Based on toy Monte Carlo studies, this
corresponds to a significance of excluding vacuum oscillations at $1.6{\sigma}$.

\section{Atmospheric Neutrinos With External Constraints}
\label{sec:gaibu}

\renewcommand{\arraystretch}{1.0}

Though the atmospheric neutrino data are sensitive to the values of
$\theta_{13}$, $\theta_{23}$, and $|\Delta m^{2}_{32}|$, the size of
the mass hierarchy signal is a function of these parameters.  As such,
larger uncertainties translate directly into reduced hierarchy
sensitivity.  
Indeed, toy MC data sets which were generated with a particular hierarchy 
but were best fit to the alternative hierarchy often preferred values of 
the atmospheric mixing parameters different from the input values.
For example, a true normal hierarchy MC generated with 
$\theta_{23}$ in the lower octant can be reasonably fit by the inverted 
hierarchy hypothesis and the second octant of this parameter.
Since there is relatively poor separation between neutrino 
and antineutrino interactions, the expected increase in the event rates 
in both scenarios is roughly equal.
Restricting the allowed
regions of the atmospheric mixing parameters therefore provides increased hierarchy
sensitivity by effectively removing such degenerate combinations.  The
constraints adopted in the present analysis are based exclusively on
information available in the literature and are described below.

\subsection{Reactor Constraint on $\theta_{13}$ }

Currently the most precise measurements of $\sin^2 2\theta_{13}$ come
from the Daya Bay, RENO, and Double Chooz
experiments
In the analysis
described below the central value of this parameter is taken to be
$\sin^2 \theta_{13} = 0.0219 \pm 0.0012$ based on the average of these
measurements presented in~\cite{Agashe:2014kda}.  A systematic error
representing the size of the uncertainty from this average is
incorporated in the analysis.

\subsection{Constraints from T2K}

The T2K (Tokai-to-Kamioka) long-baseline neutrino experiment sends a
beam composed primarily of $\nu_{\mu}$ from Tokai-village, Japan,
$2.5^{\circ}$ off-axis toward the Super-Kamiokande detector 295~km
away.  A complex of detectors (the near detectors) located 280~m
downstream of the neutrino production point and at the same off-axis
angle is used to measure the unoscillated beam spectrum and to thereby
constrain the expected spectrum at Super-K (the far detector).  A
sharp beam profile peaking at $600~\mbox{MeV}$ is expected at the far
detector and provides for sensitive measurements of $\theta_{23}$ and
$\Delta m^{2}_{32}$.  Currently T2K's measurements~\cite{Abe:2015awa,Abe:2017uxa}
of these parameters are more constraining than the Super-K atmospheric
neutrino measurement and provide a statistically-independent
constraint.  These together with inherent correlations in some
systematic error sources, such as the detector response and cross
section model, make T2K a powerful input to the Super-K hierarchy
analysis.  A more detailed description of the T2K experiment is
presented elsewhere~\cite{Abe:2011ks}.

\begin{table}[htp]
\begin{center}
  \begin{tabular}{p{1cm}p{1cm}c} 
  \hline 
  \hline 
  $\Phi$ & $\sigma$ &
  Int./22.5~kton\\ 
  \hline 
  \hline 
  $\nu_{\mu}$ & $\nu_{e}$ & 1722.3 \\
  $\nu_{\mu}$ & $\nu_{\mu}$ & 1643.3 \\ 
  $\bar \nu_{\mu}$ & $\bar \nu_{\mu}$ & 53.3 \\ 
  $\nu_{e}$ & $\nu_{e}$ & 29.3 \\
  $\bar \nu_{e}$ & $\bar \nu_{e}$ & 4.3 \\ 
  \hline 
  \hline 
 \end{tabular} 
\end{center}
  \caption{Expected interaction
  rates within the SK 22.5~kton fiducial volume for the T2K beam
  fluxes ($\Phi$) and cross section type ($\sigma$) 
  presented in ~\cite{Abe:2012av}.
  Rates correspond to the number of interactions
  per $1.0\times 10^{21}$ protons on target.  }
\label{tblrate}
\end{table}

Since Super-K serves as the far detector for T2K many aspects of the
experiments are shared.  Notably the detector simulation as well as
the neutrino interaction generator, NEUT~\cite{Hayato:2009zz}, and the
event reconstruction tools at Super-K are common between the two.
From the standpoint of Super-K then, only the neutrino source and
associated systematics differ between the beam and atmospheric
neutrino measurements.  For this reason it is possible to create a
reliable simulation of the T2K experiment using software and methods
specific to atmospheric neutrino measurements, provided only
information about the beam flux and systematic errors.  Accordingly,
in addition to the $19 \times 4$ data samples presented in
Section~\ref{sec:detector}, simulated T2K $\nu_{e}$ appearance and
$\nu_{\mu}$ disappearance samples are introduced into the atmospheric
analysis in order to directly incorporate T2K's measurements.  Monte
Carlo corresponding to these samples is constructed from reweighted
atmospheric neutrino MC and data are taken from the literature.  This
scheme allows various oscillation hypotheses to be tested against the
published T2K data and in conjunction with the Super-K data.  Provided
the model samples reproduce T2K's results when fit without
the atmospheric neutrino data, the results of a combined analysis can
be taken as reliable.

Neutrino MC samples at Super-K are generated according to the Honda 2011 flux
calculation~\cite{Honda:2011nf} and a sample equivalent to a 500~year
exposure of the SK-IV detector, the run period which contains the T2K
beam data, is reweighted according to the beam flux prediction
presented in~\cite{Abe:2012av}.  Detailed predictions assuming no
oscillations are available for the
$\nu_{\mu}, \bar \nu_{\mu}, \nu_{e},$ and $\bar \nu_{e}$ components of
both the beam and atmospheric fluxes at Super-K.  Atmospheric neutrino
interactions are reweighted according to neutrino flavor, arrival
direction, and energy to match the beam spectrum.
Though the T2K beam enters the Super-K tank from one direction 
and atmospheric neutrinos enter from all directions, the uniformity of 
the detector's response is such that this reweighting results in negligible 
biases in the model samples.
Both T2K analysis samples considered here are fully contained
interactions based on the same fiducial volume as the atmospheric
neutrino sample.  The normalization of the reweighted MC (hereafter
beam MC) is computed based on the total neutrino interaction cross
section on 22.5~kton of water convolved with the beam flux.
Table~\ref{tblrate} lists the interaction rate for $1.0\times
10^{21}$ protons on the T2K target for several combinations of
neutrino flux and cross section.

Separate T2K $e$-like and $\mu$-like samples are constructed from the
beam MC using the selection criteria presented in~\cite{Abe:2013hdq}
and~\cite{Abe:2014ugx}, respectively.
Both samples are composed of fully contained fiducial volume events
with more than 30~MeV of visible energy and a single reconstructed
Cherenkov ring.  To be included in the $e$-like sample the PID of the ring is
required to be $e$-like and must have more than 100~MeV of visible energy.  
Additionally, there must not be any activity
consistent with the electron from a decayed muon and the reconstructed
neutrino energy (described below) must be less than 1250 MeV.  A final
cut designed to reduce backgrounds from NC $\pi^{0}$ interactions is
applied according to~\cite{Abe:2013hdq}.  Events whose Cherenkov ring
has $\mu$-like PID with a momentum greater than 200~MeV/c and at most
one decay electron comprise the $\mu$-like sample.

During the analysis, both samples are binned using the reconstructed
neutrino energy calculated assuming charged-current quasi-elastic 
(CCQE) interactions in water:
\begin{widetext}
\begin{equation}
E_{\nu}^{\rm rec} = \frac{
(M_{n} - V_{\rm nuc})\cdot E_{l} - m_{l}^2/2 + M_{n}\cdot V_{\rm nuc}-V_{\rm nuc}^2/2+ \left(M_{p}^2-M_{n}^2\right)/2
}
{M_{n}-V_{nuc}-E_{l}+P_{l} \cos \theta}.
\label{eqn:erec}
\end{equation}
\end{widetext}
\noindent Here $M_{n}$($M_{p}$) is the neutron (proton) mass 
and $V_{\rm nuc}$ is the average nucleon binding energy in $^{16}O$,
27~MeV.  The charged lepton mass, $m_{l}$, is assumed to be that of an
electron for the $e$-like sample and that of a muon for the $\mu$-like
sample.  Similarly, the total energy, $E_{l}$, is computed for each
sample using the corresponding $m_{l}$ and the reconstructed momentum,
$P_{l}$.  The $\cos \theta$ term represents the opening angle
between the neutrino and lepton directions, which is computed using MC
truth information for the parent neutrino and the reconstructed
direction of the charged lepton ring.  Though the official T2K
analyses use maximum likelihood methods, without detailed information
of each data event, reproducing the analyses exactly using only
published information is infeasible.  
Instead the data are binned as specified in the T2K publications. 
 The $e$-like sample uses 50~MeV wide
bins evenly spaced from 100 to 1250~MeV and the $\mu$-like sample uses
50 MeV bins from 0.2 to 3.0 GeV, 100~MeV wide bins from 3.0 to 5.0~GeV, 
and a single bin for more energetic events.

%
%
%
\begin{figure}[htbp]
\includegraphics[width=0.48\textwidth]{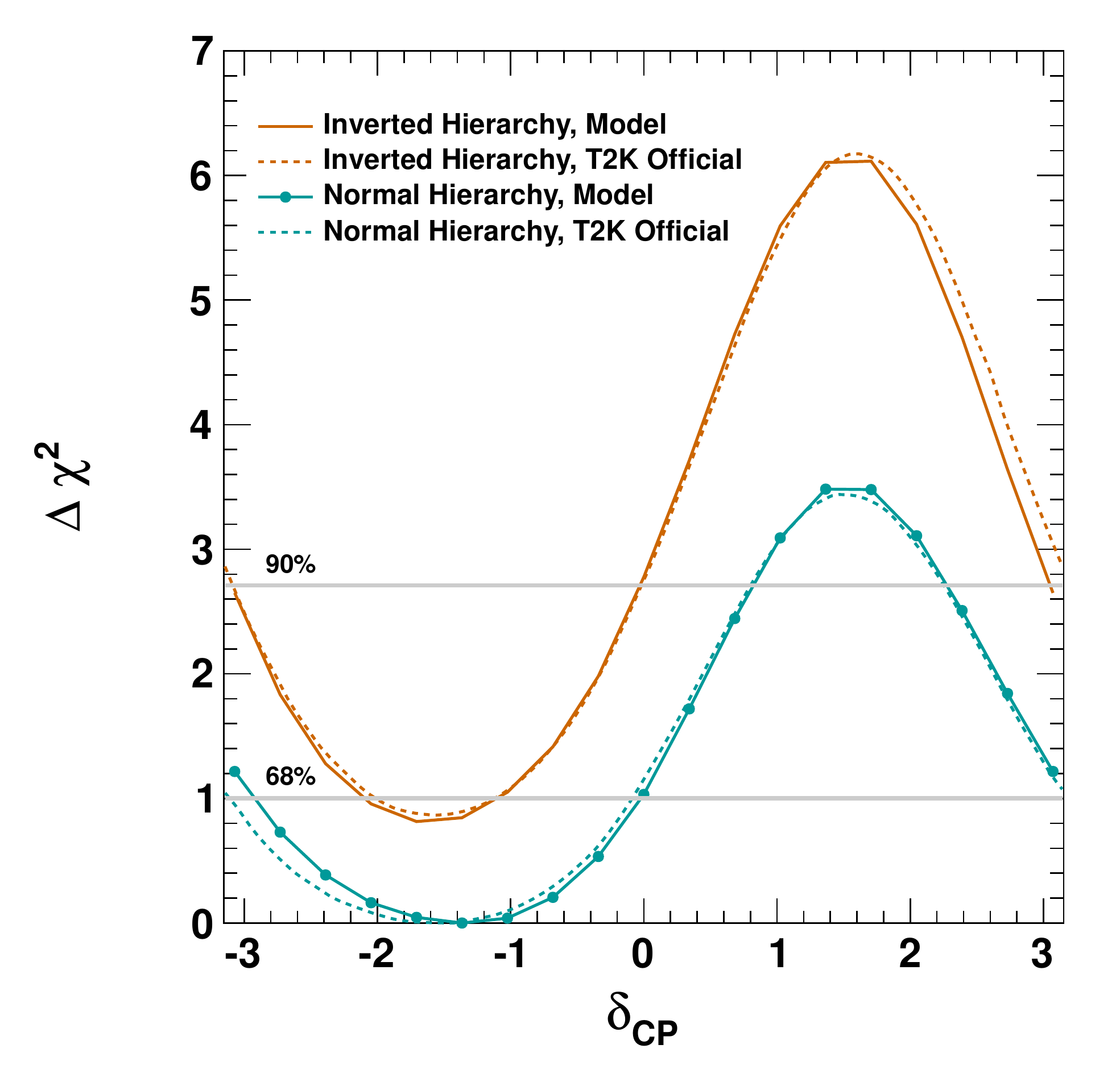}
\caption{ Comparison of the T2K model (solid) with T2K results 
digitized from Fig.~33 of Ref.~\cite{Abe:2015awa} (dashed).  Cyan and
orange lines correspond to the normal hierarchy and inverted hierarchy
fits, respectively, with the offset in the two representing the
difference in their minimum $\chi^{2}$ values.  }
\label{fig:t2k_compare}
\end{figure}

A critical component of the T2K analysis is the constraint coming from
measurements of the unoscillated neutrino flux and interactions at its
near detector complex.  Measurements of the CC $\nu_{\mu}$ interaction
rate adjust the central values and uncertainties on parameters
describing the flux and cross section models underlying the simulation
at Super-K.  Incorporation of these constraints alters the shape and
composition of the expected spectrum at Super-K and is therefore
essential for an accurate reproduction of the T2K results.  Energy
dependent normalization parameters for the beam's
$\nu_{\mu}, \bar \nu_{\mu}, \nu_{e},$ and $\bar \nu_{e}$ flux
components from~\cite{Abe:2015awa} are applied as additional weighting
factors for the beam MC.  Constraints on the interaction model, such
as the value of axial mass parameters for quasi-elastic processes and
pion production interactions via the $\Delta$ resonance, as well as
the CCQE, CC single pion, and NC $\pi^{0}$ cross section
normalizations are similarly incorporated as multiplicative weighting
factors.  For example, the T2K-measured change in the CCQE axial mass
parameter, $M_{A}^{QE}$ from the default value of $1.21 \pm 0.45$ to
$1.33 \pm 0.20$ is incorporated into the present analysis by computing
the ratio of the CCQE cross section for each MC event based on its
generated lepton and hadron kinematics.  Errors assigned to the flux
and cross section parameters in~\cite{Abe:2015awa} are used in the
construction of systematic error response coefficients discussed below.
It should be noted however, that the complete spectral response of the
T2K error model is not publicly available, and the influence of
systematic errors is often expressed as the expected change in each
sample's event rate.  In these cases the error model used in the
atmospheric neutrino analysis is adapted to produce the same event
rate change in the T2K samples.  In the combined analysis of
atmospheric data and the T2K model, detector and cross section
systematic errors are considered completely correlated between the two
data sets, while the flux errors are uncorrelated.

\begin{figure}[htbp]
\includegraphics[width=0.50\textwidth]{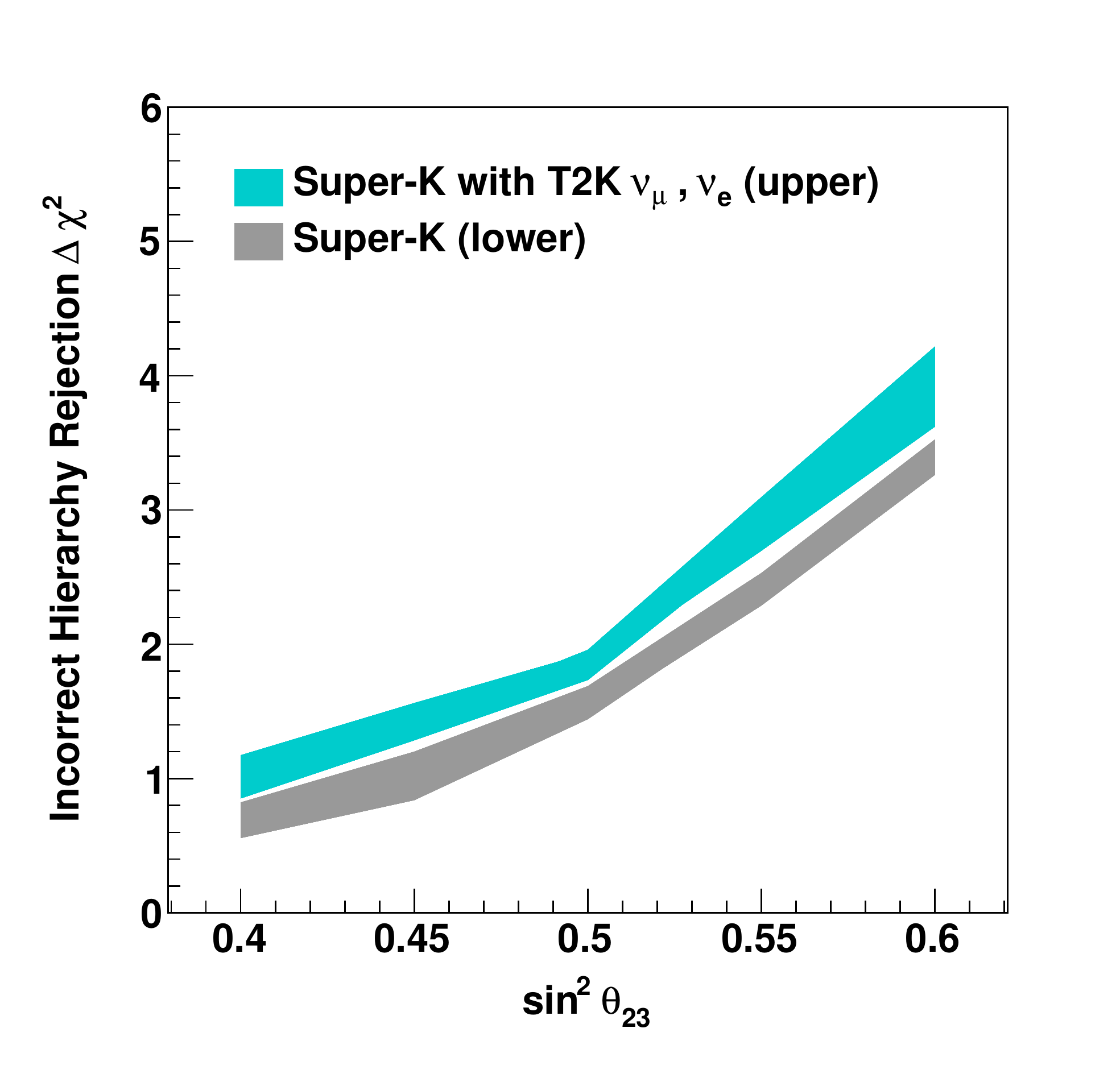}
\caption{ Expected sensitivity to the normal mass hierarchy as a function 
of the true value of $\sin^2 \theta_{23}$. Grey (lower) and cyan (upper) bands
show the sensitivity of the atmospheric neutrino sample alone and when
combined with the T2K model, respectively. The width of the bands corresponds to the
uncertainty in $\delta_{CP}$.  The inverted hierarchy shows a
qualitatively similar improvement in sensitivity and is not shown.  }
\label{fig:hier_sens}
\end{figure}

The model constructed here is based on $6.57\times 10^{20}$ protons on
target taken with T2K's neutrino-enhanced beam.  Though antineutrino
data and contours are available in the literature
(c.f.~\cite{Abe:2017bay}), the statistics are too low to impact the
sensitivity of the present analysis and are not included in the model.
Figure~\ref{fig:t2k_compare} shows a comparison of the model with T2K's
constraints on $\delta_{CP}$ and the mass hierarchy after removing
(profiling out) the effect of other oscillation parameters.
The expected impact of the T2K model on the atmospheric neutrino
sensitivity to the mass hierarchy is illustrated in
Figure~\ref{fig:hier_sens}.  For all assumed values of
$\sin^2 \theta_{23}$, the T2K model's constraint on the
atmospheric mixing parameters strengthens the sensitivity.

It should be noted that other long-baseline neutrino experiments have
made precision measurements of atmospheric mixing parameters, which,
when adapted as external constraints in this analysis, could improve
the expected sensitivity in the same manner as T2K. For example, as
seen in Fig.~\ref{fig:sk_world}, MINOS~\cite{Adamson:2014vgd}
constrains $\Delta m^2_{32}$ roughly as precisely as T2K, although T2K
constrains $\sin^2 \theta_{23}$ better. Moreover, the neutrino
interactions in MINOS are on iron nuclei, not water, introducing an
uncancelled systematic uncertainty. Measurements by
NOvA~\cite{Adamson:2016tbq,Adamson:2016xxw} of muon neutrino
disappearance and electron neutrino appearance should benefit the
present analysis; their inclusion is anticipated in a future effort.

%
%
%

\subsection*{Analysis}

After the introduction of external constraints the atmospheric neutrino data 
are analyzed in two ways using modified versions of the fitting scheme outlined 
in Section~\ref{sec:atm_only}.
In the first analysis the same atmospheric neutrino data samples and binning 
are fit over a restricted parameter space, with $\mbox{sin}^{2}\theta_{13}$ 
constrained to 0.0219 as described above and other parameter ranges 
unchanged. 
An additional systematic error parameter representing the effect of the 
uncertainty in external measurements of $\theta_{13}$ on the SK analysis 
samples is included in the fit. 

The second analysis imposes the same constraint but introduces additional 
analysis bins and systematic errors to accommodate the T2K analysis samples 
described above. 
Using this model of the T2K samples the analysis is performed over 
the same oscillation parameter grid and does not rely on knowledge of 
T2K's published likelihood surface.
Systematic error parameters for the T2K samples are fit simultaneously with 
those for the atmospheric neutrino samples.

\subsection*{Results and Discussion }


%
%
%
%
\begin{figure*}[htbp]
  \includegraphics[width=0.32\textwidth]{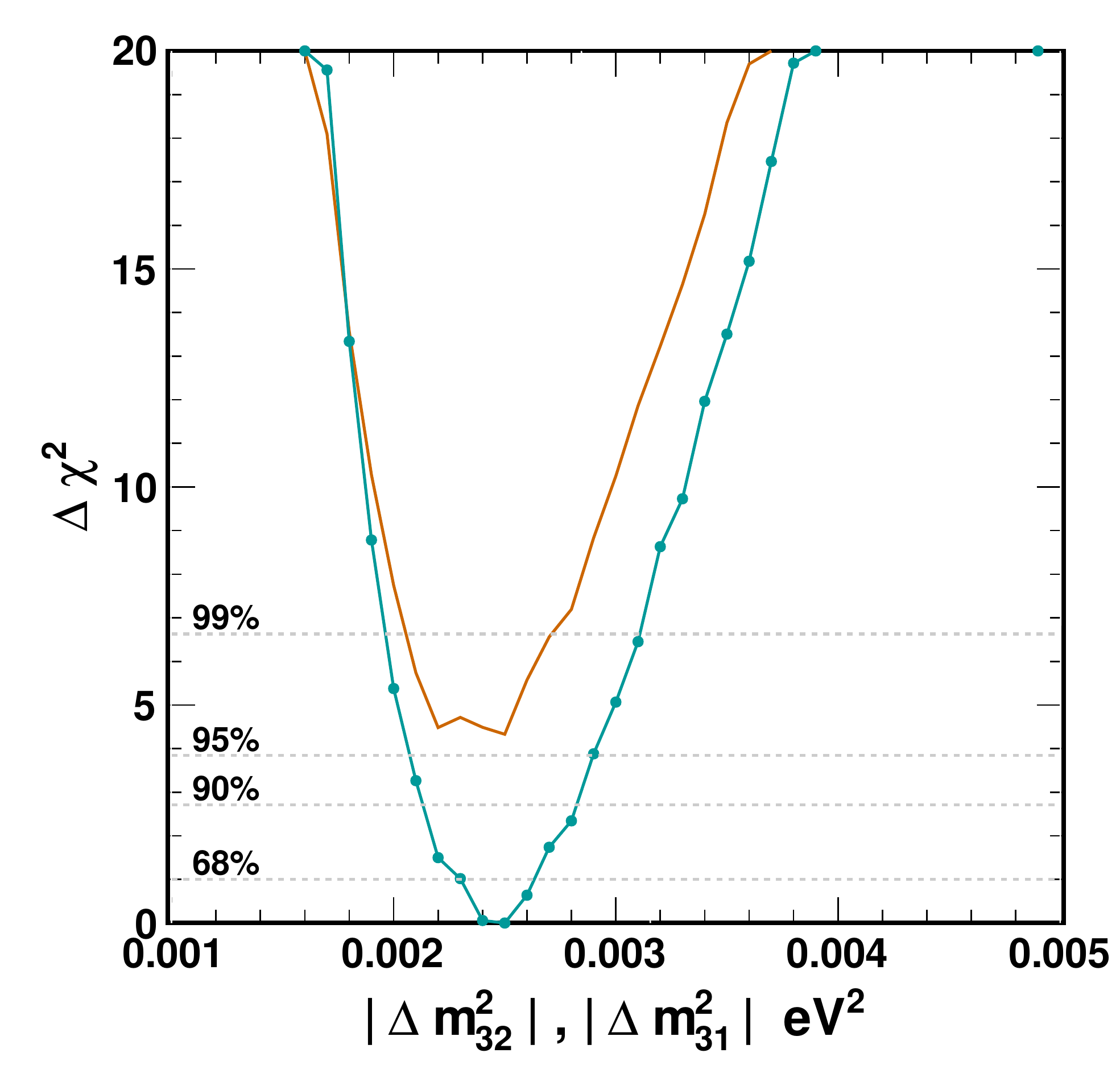}
  \includegraphics[width=0.32\textwidth]{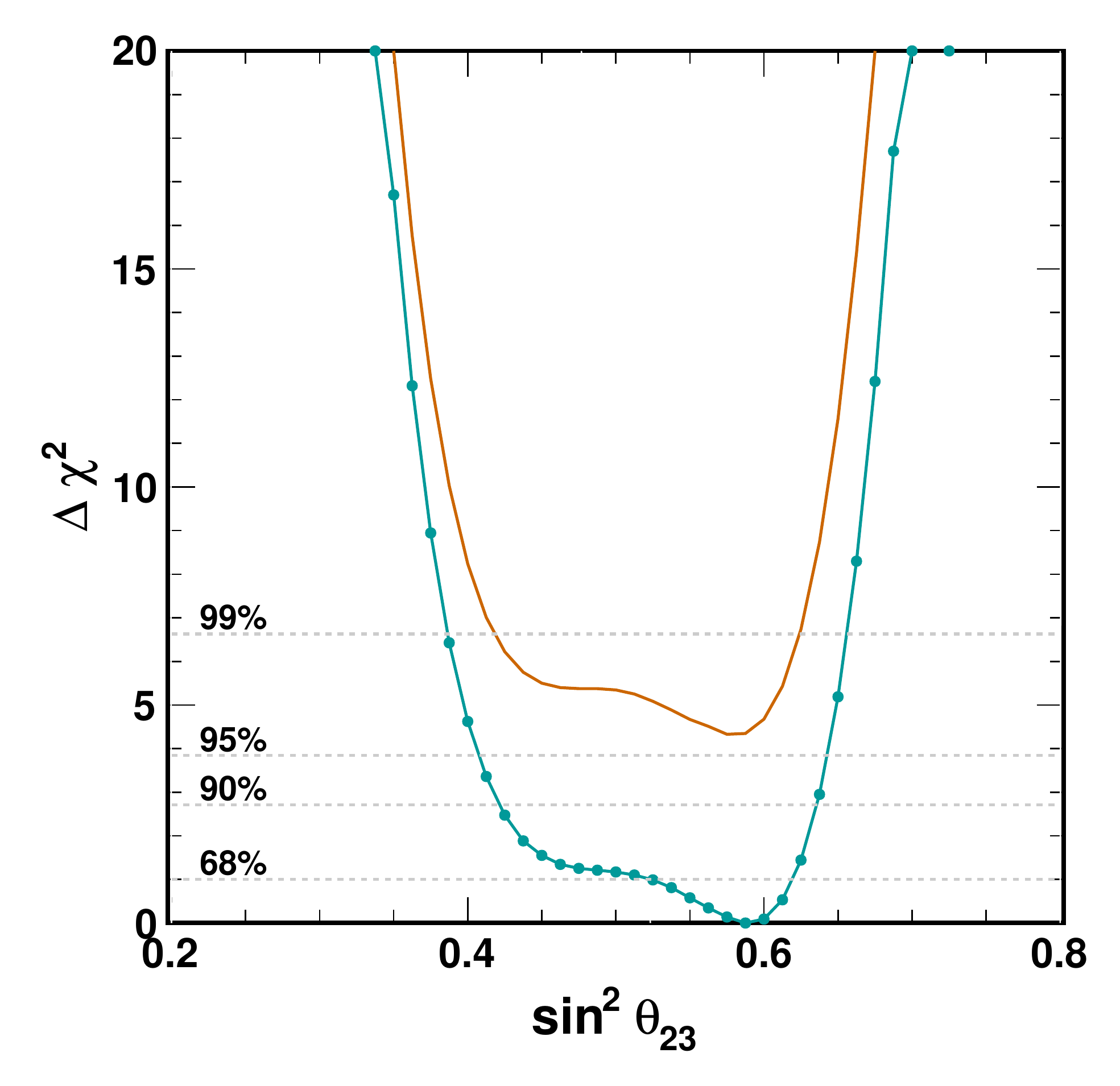}
  \includegraphics[width=0.32\textwidth]{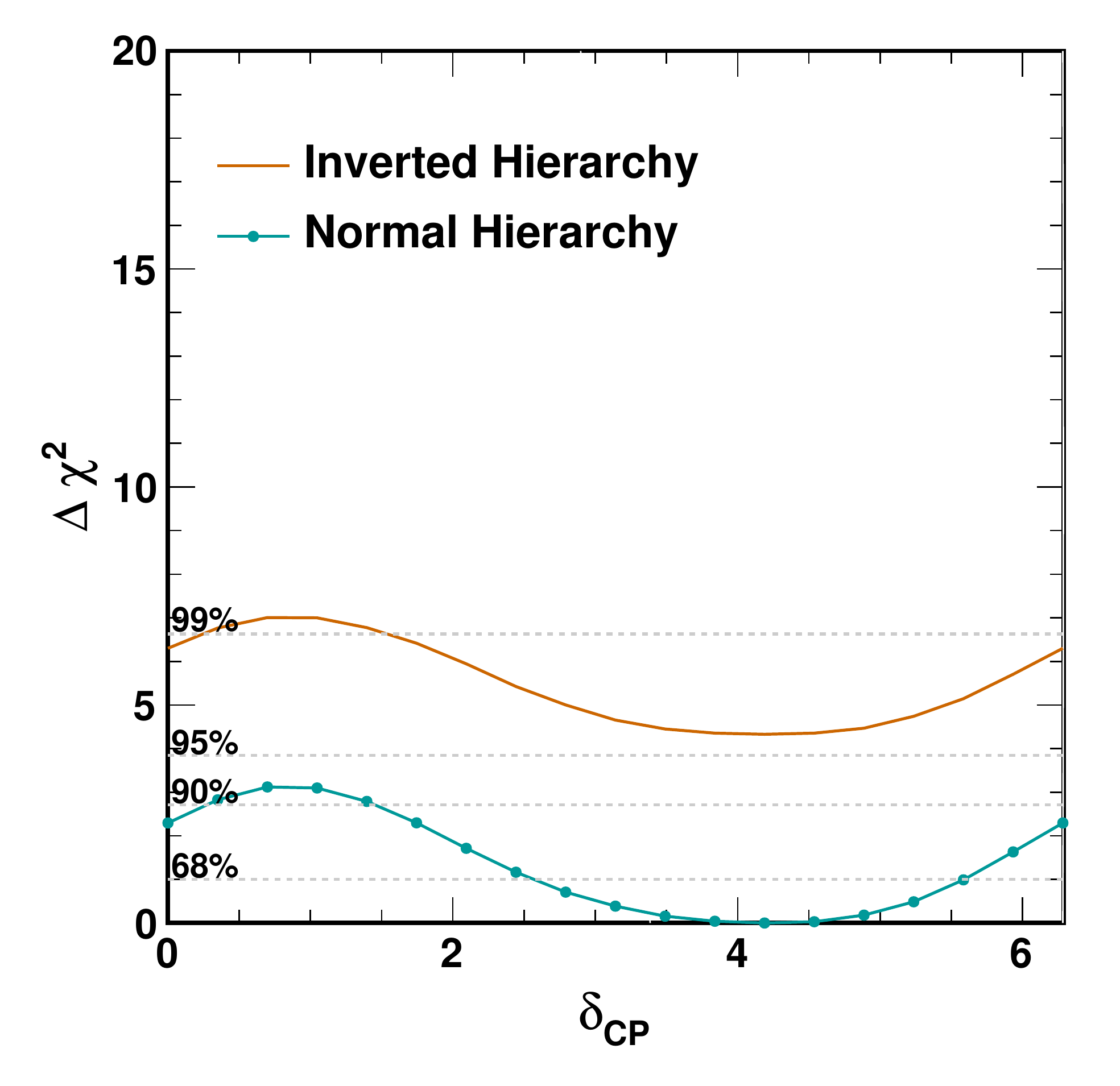}
  \caption{ Constraints on neutrino oscillation parameters from
            the Super-K atmospheric neutrino data fit assuming $\mbox{sin}^{2} \theta_{13} = 0.0219\pm 0.0012$ .
            Orange lines denote the inverted hierarchy result, which
            has been offset from the normal hierarchy result, shown in
            cyan, by the difference in their minimum $\chi^{2}$ values.}
  \label{fig:q13fixed_sk_cont}
\end{figure*}

%
%
%
\begin{figure}[htpb]
  \includegraphics[width=0.50\textwidth]{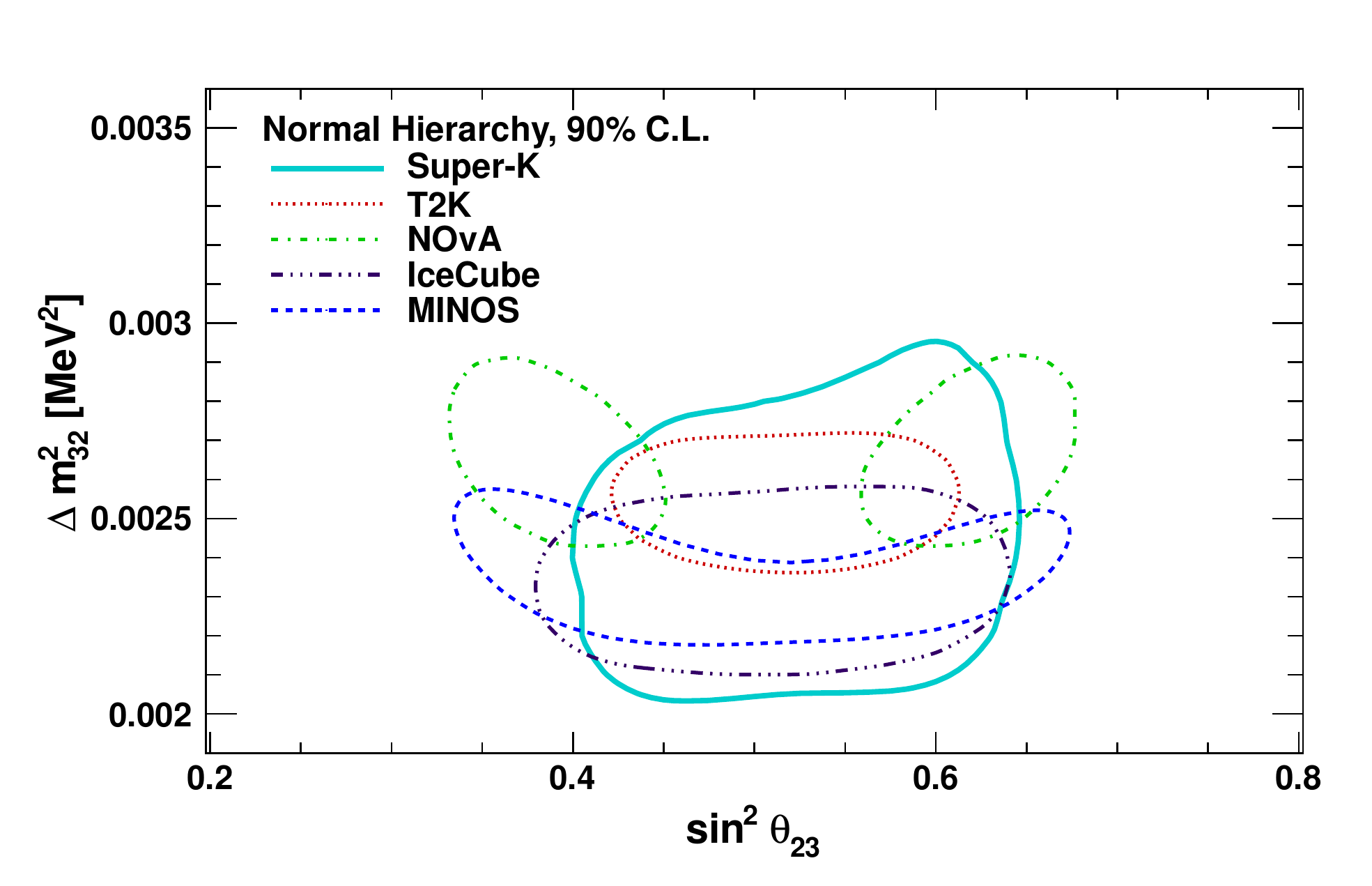}
  \caption{ Constraints on neutrino oscillation contours at the 90\%
    C.L. from analyses assuming the normal mass hierarchy.  The
    Super-K contour (cyan) is taken from the analysis with
    $\mbox{sin}^{2} \theta_{13}$ assumed to be $0.0219 \pm 0.0012$.  Contours
    from the T2K (violet)~\cite{Abe:2017uxa}, NOvA (dashed
    green)~\cite{Adamson:2017qqn}, MINOS$+$ (dashed
    blue)~\cite{Adamson:2014vgd}, and IceCube
    (red)~\cite{Aartsen:2016psd} experiments are also shown.}
  \label{fig:sk_world}
\end{figure}

Constraints on the atmospheric neutrino mixing parameters and $\delta_{CP}$ 
in the $\theta_{13}$-constrained fit without the T2K samples are shown in 
Figure~\ref{fig:q13fixed_sk_cont}.
As in the unconstrained fit the data prefer the normal hierarchy 
over the inverted hierarchy with
$\Delta \chi^{2} \equiv \chi^{2}_{NH,min} - \chi^{2}_{IH,min} = -4.33$.
While the best fit value of $|\Delta m^{2}_{32}|$ has shifted slightly, 
it is within errors of the unconstrained fit and in good agreement
with other measurements (c.f. Fig.~\ref{fig:sk_world}).
Similarly, the preference for the second octant of $\theta_{23}$ 
remains unchanged and no significant change is seen in the 
width of the parameter's allowed region at $1{\sigma}$.
The best fit value of $\delta_{CP}$ is 4.18 for both hierarchies,
with a tighter constraint on other values relative to the unconstrained fit. 
Parameter values and their $1{\sigma}$ errors are summarized in Table~\ref{tbl:bestfits}.

%
%
%
%
\begin{figure*}[htpb]
  \includegraphics[width=0.32\textwidth]{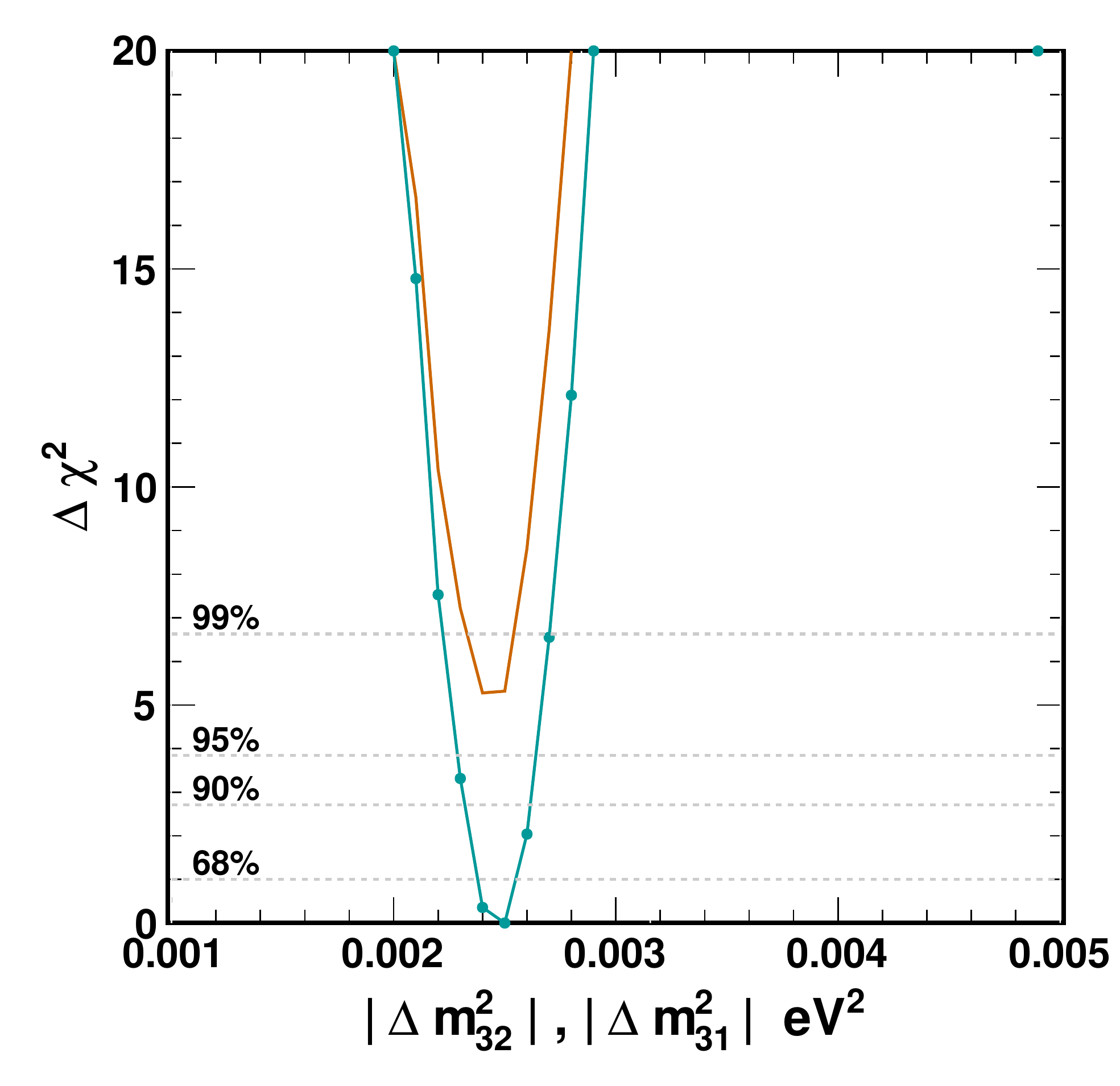}
  \includegraphics[width=0.32\textwidth]{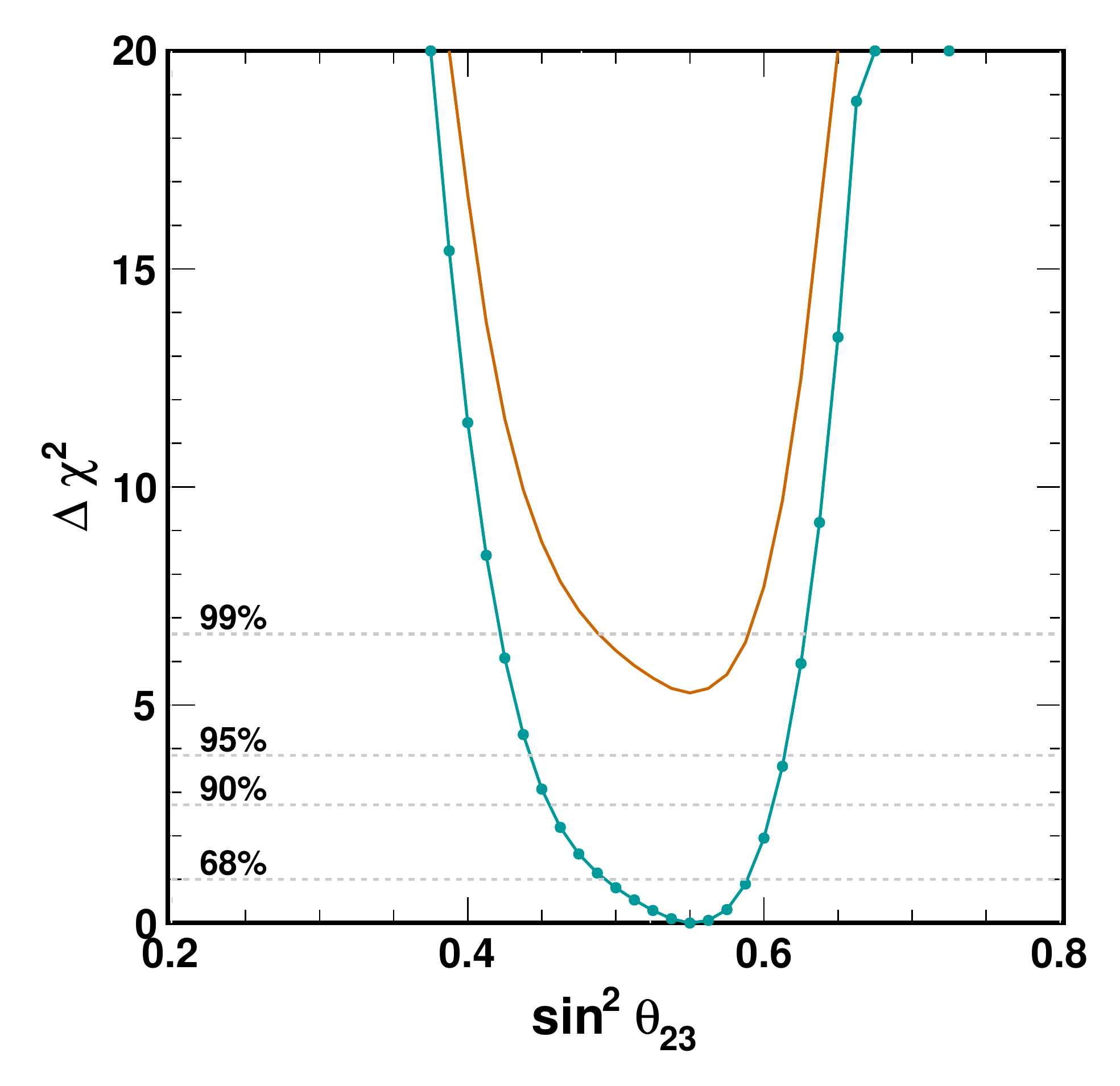}
  \includegraphics[width=0.32\textwidth]{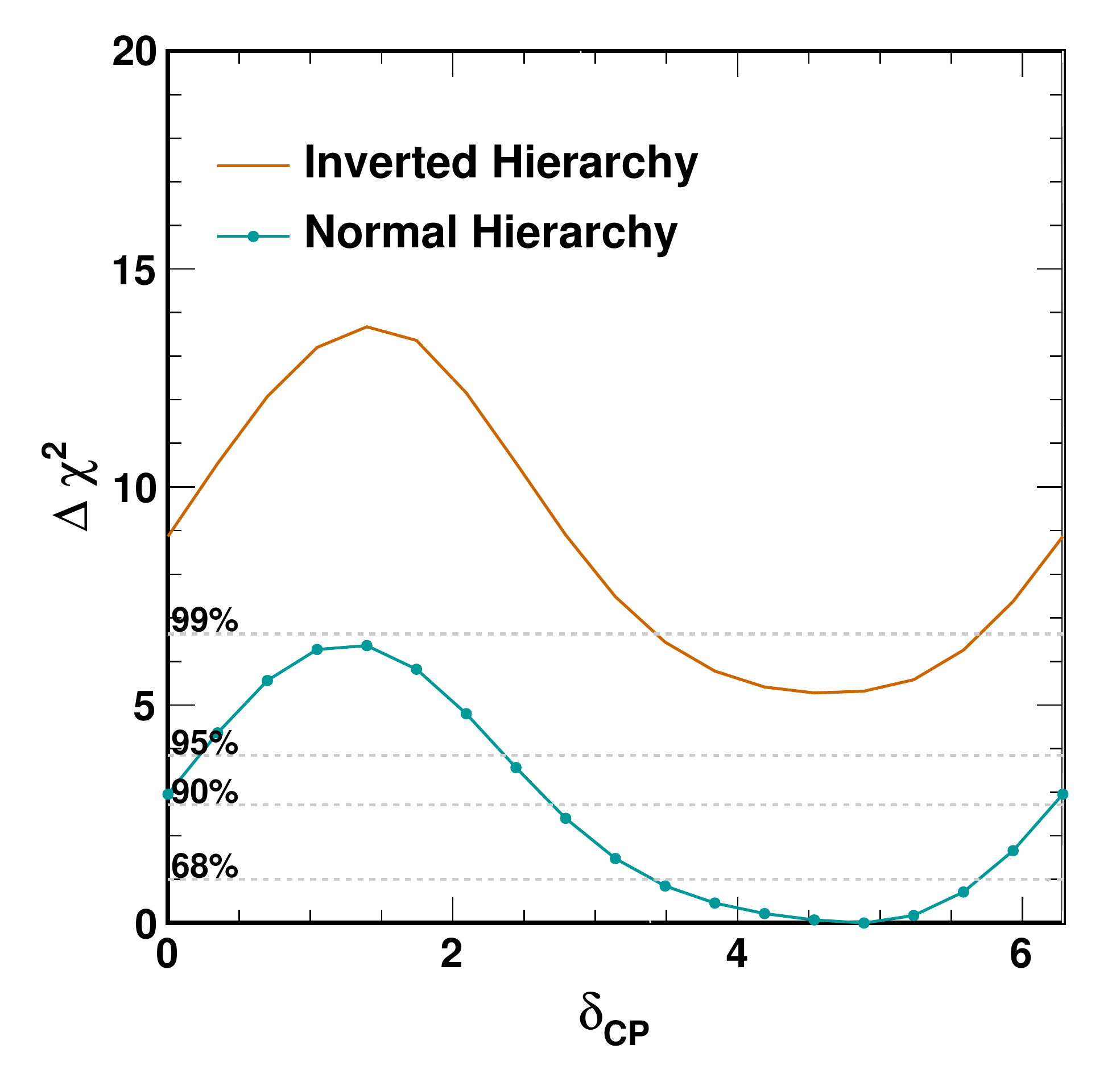}
  \caption{ Constraints on neutrino oscillation contours from a combined
            fit of Super-K atmospheric neutrino data and a model of the
            T2K experiment assuming $\mbox{sin}^{2} \theta_{13} = 0.0219 \pm 0.0012 $ .
            Orange lines denote the inverted hierarchy result, which
            has been offset from the normal hierarchy result, shown in
            cyan, by the difference in their minimum $\chi^{2}$ values.
           }
  \label{fig:skt2k_cont}
\end{figure*}

%
%
%
\begin{figure*}[htpb]
  \includegraphics[width=0.45\textwidth]{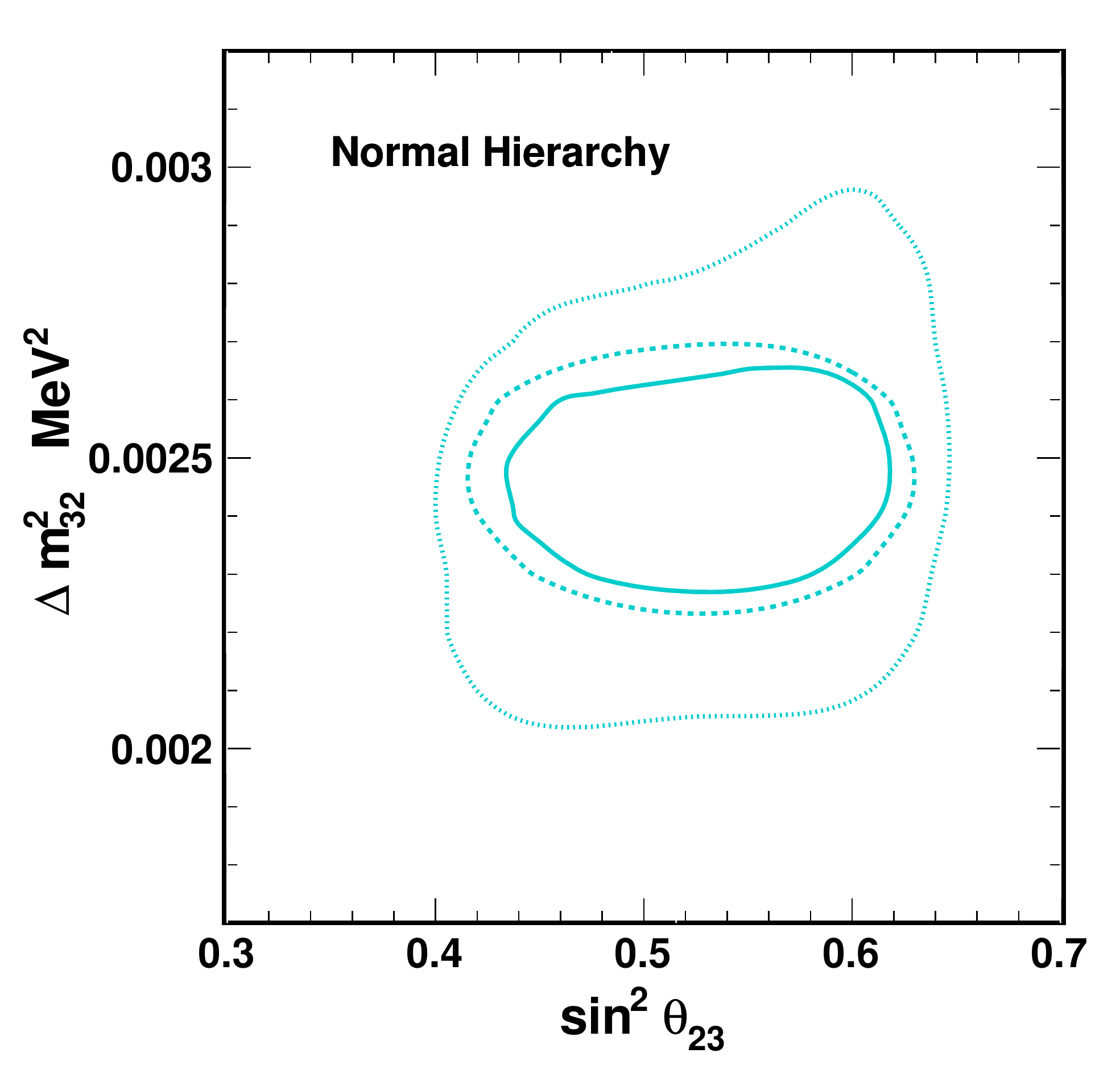}
  \includegraphics[width=0.45\textwidth]{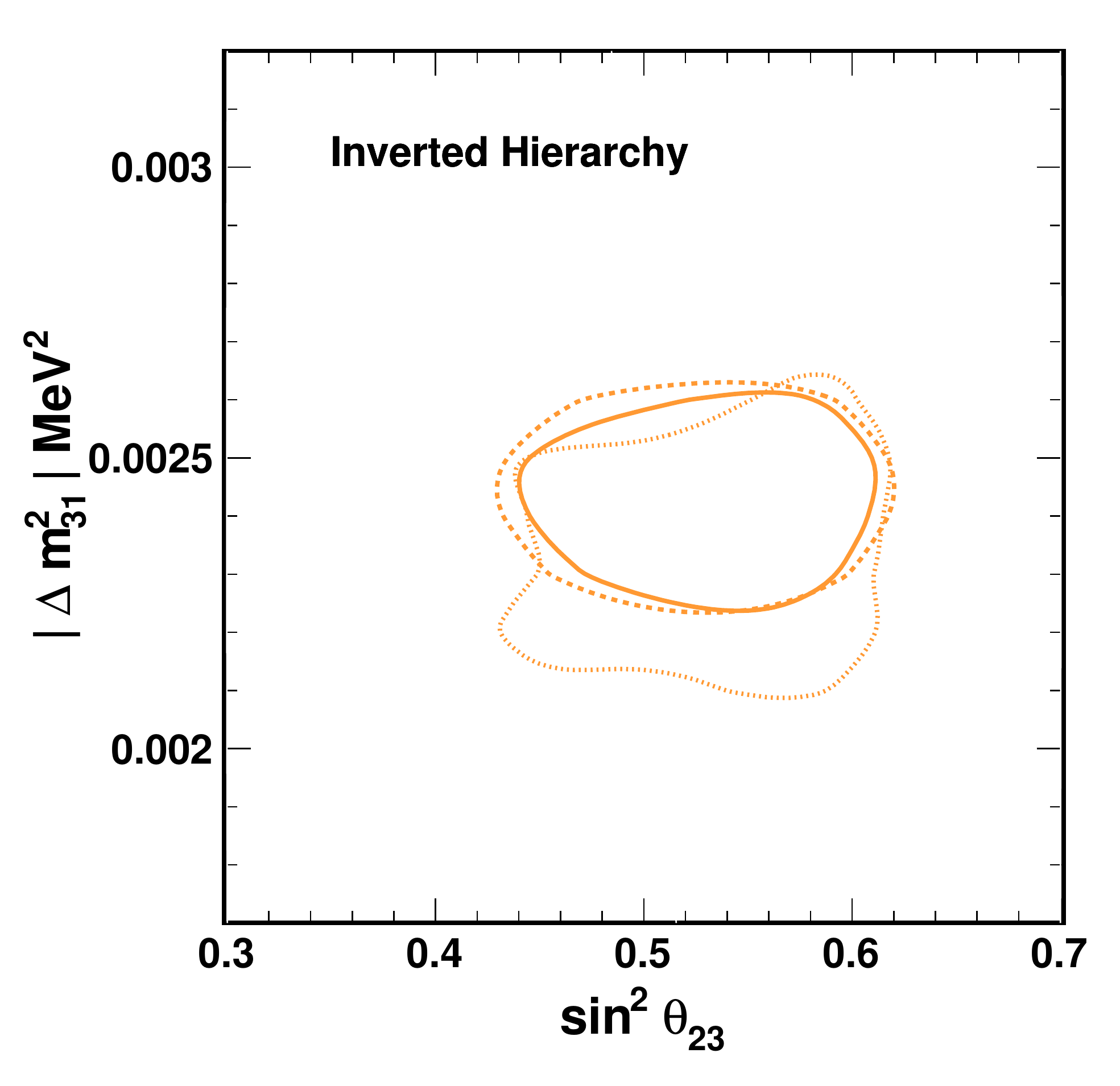}
  \caption{ Constraints on neutrino oscillation contours from a combined
            fit of Super-K atmospheric neutrino data and a model of the
            T2K experiment. The left figure shows 90\% C.L. 
            constraints from the atmospheric neutrino data (dotted),
            the T2K model (dashed), and their combination (solid) for the
            normal hierarchy. The right figure shows the same for the inverted hierarchy fit.
            In each contour $\mbox{sin}^{2} \theta_{13}$ is constrained to be $0.0219 \pm 0.0012$ .
            Normal and inverted hierarchy contours for each analysis are drawn relative 
            the best fit among the two.}
  \label{fig:twod}
\end{figure*}

In the second fit the addition of the T2K samples is expected to 
improve the constraint on the atmospheric mixing parameters due to T2K's 
more precise measurements.
The left two panels of Fig.~\ref{fig:skt2k_cont} show one-dimensional constraints 
on these parameters and two-dimensional contours appear in Fig.~\ref{fig:twod}. 
In the latter dotted lines denote the allowed region from the $\theta_{13}$-constrained fit 
to the atmospheric neutrino data only and dashed lines show the 
allowed regions from the T2K model fit by itself.
The combination of the two data sets, depicted as the solid line, shows that the fit to 
these parameters is dominated by the T2K model, with little improvement seen
in the contour when fit together with atmospheric neutrinos.

With less freedom to adjust the atmospheric mixing parameters, the combination of 
atmospheric neutrinos with the T2K model is expected to improve the mass hierarchy 
sensitivity on average (see Fig.~\ref{fig:hier_sens}). 
By itself, the T2K model favors the normal hierarchy by 
$\Delta \chi^{2} = -0.85 $~\cite{Abe:2015awa}. 
Though T2K has little mass hierarchy sensitivity on average, 
$\Delta \chi^{2} = -0.4$ at the Super-K best fit point, this result is driven by an excess of observed events in its appearance sample. 
When atmospheric neutrinos are combined with T2K, the hierarchy preference strengthens to
$\Delta \chi^{2} = -5.27$, with the majority of the expected sensitivity 
coming from the atmospheric samples appearing in Fig.~\ref{fig:ud_ratio}. 

Similar preferences in both samples for $\delta_{CP}$ near $3 \pi / 2$ result 
in a stronger constraint on this parameter when analyzed together. 
The right panel of Fig.~\ref{fig:skt2k_cont} shows the constraint for both 
hierarchy assumptions, with the offset in the two lines corresponding to the 
$\Delta \chi^{2}$ between the two.
Naturally, this preference is consistent with an increased 
$\nu_{e}$ (as opposed to $\bar \nu_{e}$) rate in T2K relative to the expectation from 
the measured value of $\theta_{13}$.
Though the constraint from the normal hierarchy fit disfavors the region 
around $\pi / 2$, the contour includes the CP-conserving value $\delta_{CP} = \pi$ 
at nearly $1{\sigma}$.

%
%
%
%
\begin{table*}
\begin{center}
{\renewcommand{\arraystretch}{1.4}
\begin{tabular}{ll|ccccc}
Fit & Hierarchy & $\chi^{2}$ & $\mbox{sin}^{2} \theta_{13}$ & $\mbox{sin}^{2} \theta_{23}$ & $|\Delta m^{2}_{32,31}|$ [$\times 10^{-3}$ eV$^2$] & $\delta_{CP}$ \\
\hline 
\hline 
SK    $\theta_{13}$ Free         & NH & 571.29  & $0.018^{+0.029}_{-0.013}$  &  $0.587^{+0.036}_{-0.069}$  & $2.50^{+0.13}_{-0.31} $ & $4.18^{+1.45}_{-1.66}$     \\  
                                 & IH & 574.77  & $0.008^{+0.017}_{-0.007}$  &  $0.551^{+0.044}_{-0.075}$  & $2.20^{+0.33}_{-0.13} $ & $3.84^{+2.38}_{-2.12}$     \\
\hline 
SK    $\theta_{13}$ Constrained  & NH & 571.33  & --  &  $0.588^{+0.031}_{-0.064}$  & $2.50^{+0.13}_{-0.20} $  & $4.18^{+1.41}_{-1.61}$     \\  
                                 & IH & 575.66  & --  &  $0.575^{+0.036}_{-0.073}$  & $2.50^{+0.08}_{-0.37} $  & $4.18^{+1.52}_{-1.66}$     \\
\hline 
SK+T2K $\theta_{13}$ Constrained & NH & 639.43  & --  &  $0.550^{+0.039}_{-0.057}$  & $2.50^{+0.05}_{-0.12} $  & $4.88^{+0.81}_{-1.48}$     \\  
                                 & IH & 644.70  & --  &  $0.550^{+0.035}_{-0.051}$  & $2.40^{+0.13}_{-0.05} $  & $4.54^{+1.05}_{-0.97}$     \\
\hline
\hline
\end{tabular}
}
\end{center}
\caption{Summary of parameter estimates for each analysis and hierarchy hypothesis considered. Here NH (IH) refers to the normal (inverted) hierarchy fit.
         The terms ``Free'' and ``Constrained'' refer to fits without and with a constraint on $\mbox{sin}^{2} \theta_{13}$, respectively, as described in the 
         text. The expected absolute $\chi^{2}$ value for the SK (SK+T2K) fits is 559.9 (636.2). The p-value for obtaining a smaller $\chi^{2}$ than the data is 0.439  (0.482) in the NH $\theta_{13}$-constrained fits. }
\label{tbl:bestfits}
\end{table*}

\section{Interpretation}
\label{sec:interp}

\begin{figure}[htbp]
  \includegraphics[width=0.50\textwidth]{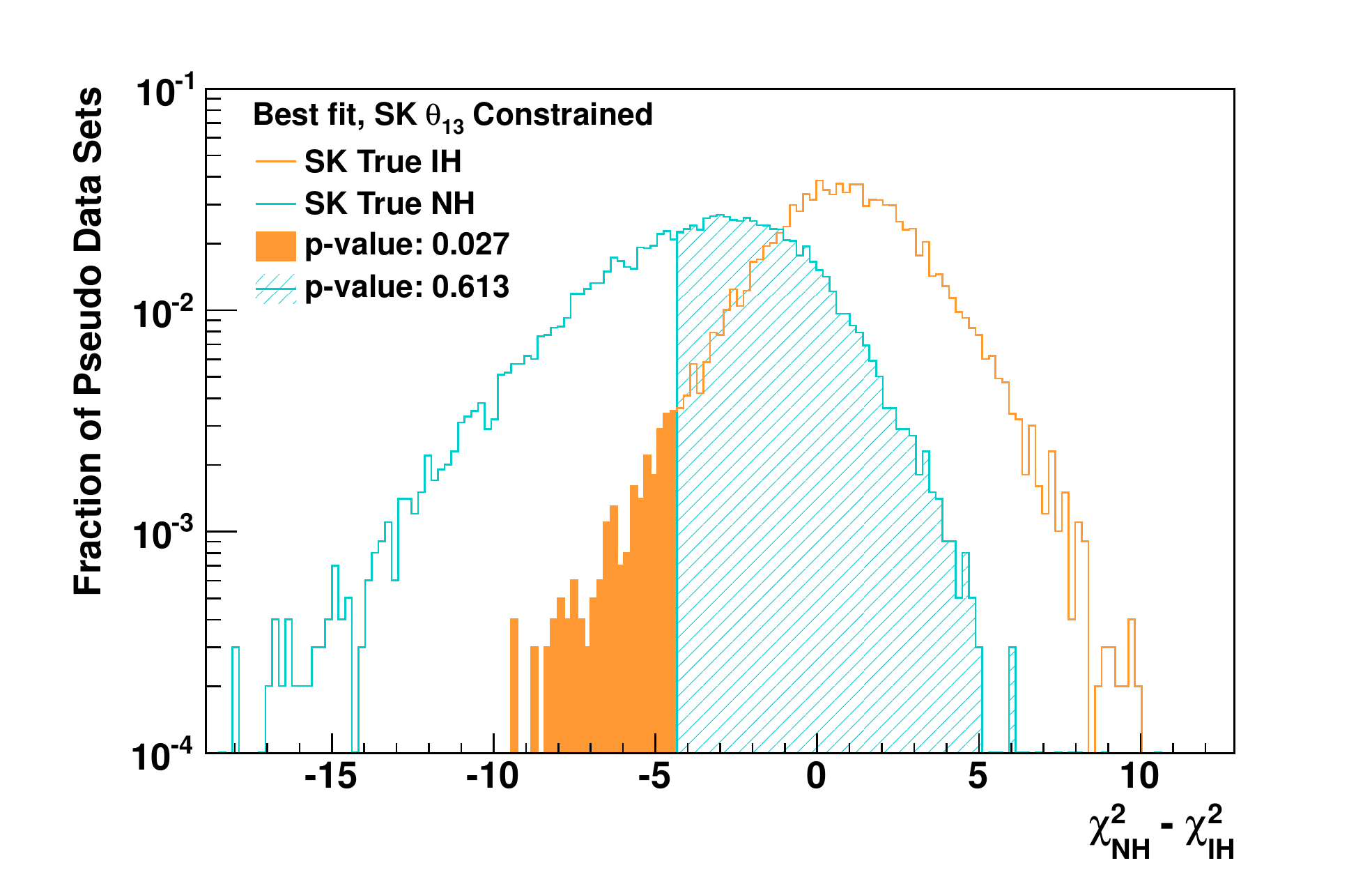}
  \caption{ Distributions of the difference in best fit $\chi^{2}$ 
            values between normal- and inverted-hierarchy fits 
            to pseudo data sets used in the generation of the $\mbox{CL}_{s}$ value 
            for the SK $\theta_{13}$ constrained analysis. 
            In the cyan (orange) histogram the pseudo data have been generated 
            assuming the normal (inverted) hierarchy at the analysis best fit 
            shown in Table~\ref{tbl:bestfits}. 
            Shaded portions of the histograms denote the fraction of 
            pseudo data sets with more extreme values than that observed in the data, 
             $\Delta \chi^{2}_{data} = -4.33$.}
  \label{fig:cls}
\end{figure}

It is known that the significance of a mass hierarchy determination
does not necessarily follow the expectation from a comparison of the
$\chi^2$ minima from each of the hierarchy hypotheses
(c.f. reference~\cite{Qian:2012zn}).  Indeed, the hierarchies do not
form a nested hypothesis and as a result Wilks'
theorem~\cite{Wilks:1938dza} is not applicable.  To address the issue
of the hierarchy significance in the present analysis, ensembles of
pseudo data sets generated from the atmospheric neutrino MC are used
to estimate $p$-values for obtaining a difference in $\chi^{2}$ between
the hierarchy hypotheses more extreme than that observed in data.
This condition is termed ``rejecting'' the alternative hierarchy
hypothesis for a given hierarchy assumption in what follows.

For the Super-K analysis, two important issues need to be considered.
First, as shown in Fig.~\ref{fig:hier_sens} the expected sensitivity
to the mass hierarchy is a strong function of the underlying
oscillation parameters and as such, $p$-value calculations are expected
to depend heavily on the parameters assumed in the generation of MC
ensembles.  Rather than attempting a Bayesian-like treatment of the
$p$-value calculation and marginalizing over the effect of each
parameter, a range of $p$-values has been computed using the $90\%$
C.L. intervals obtained from the present analysis to avoid ambiguities surrounding the
choice of parameter priors.

Second, it is also clear from the figure that at the current level of
statistics, Super-K has only modest sensitivity to reject either
hypothesis, making the interpretation of the $p$-value susceptible to
fluctuations of the background. While the $p$-value for rejecting the
inverted hierarchy (IH) hypothesis assuming the normal hierarchy (NH)
may be unlikely, the $p$-value in the reverse scenario may be equally
unlikely, leading to an overestimation of the significance when stated
in terms of the first $p$-value only.  Following the lead of the LHC
experiments, this issue is treated using the $\mbox{CL}_{s}$
method~\cite{Read:2002hq}, where
\begin{equation}
\mbox{CL}_{s} = \frac{ p_{0}(IH) }{ 1 - p_{0}(NH) }. 
\end{equation}
\noindent Here $p_{0}(IH)$ ($p_{0}(NH)$) represents the $p$-value for
obtaining a difference in the minimum $\chi^{2}$ values between both
hierarchy hypotheses, $\Delta \chi^{2} \equiv \chi^{2}_{NH} -
\chi^{2}_{IH}$ smaller (larger) than that from the data, $\Delta
\chi^{2}_{data}$, assuming the true hierarchy is the IH (NH).  While
$\mbox{CL}_{s}$ does not behave as a fully frequentist $p$-value, it is
a conservative method of preventing erroneous rejection of the null
hypothesis when the overall sensitivity is limited.

MC ensembles were generated 
assuming statistical fluctuations of the pseudo data sets
according to the current detector exposure, and Gaussian fluctuations
of the systematic errors.  
Figure~\ref{fig:cls} shows the distribution of MC ensembles 
used in the calculation of the $\mbox{CL}_{s}$ value for the 
SK $\theta_{13}$-constrained fit. 
Table~\ref{tbl:pval} shows the range of
$p$-values and $\mbox{CL}_{s}$ values based on ensembles generated 
with true oscillation parameters taken from the 90\%~C.L. bounds on $\theta_{23}$ and $\delta_{CP}$ and best fits from the analyses above.  
Since the data's preference for
the normal hierarchy is driven primarily by upward-going excesses seen
in hierarchy-sensitive $e$-like samples, smaller values of $p_{0}(IH)$ 
and larger $\mbox{CL}_{s}$ are obtained when assuming smaller values of
$\mbox{sin}^{2} \theta_{23}$ or when $\delta_{CP}$ is near $\pi/2$
since both of these regions predict the least amount of electron
neutrino appearance.
For $\mbox{sin}^{2} \theta_{23} > 0.60$ both metrics decrease 
as there is sufficient electron neutrino appearance to discriminate between the 
two hierarchy hypotheses at the level seen in the data.
In contrast, both metrics are found to vary only
slightly with $\Delta m^{2}_{32,31}$.
%
%
%
%
%
\begin{table*}[htbp]
\begin{center}
\begin{tabular}{l|ccc|ccc}
    & \multicolumn{3}{c}{ $p_{0}(IH)$ } & \multicolumn{3}{c}{ $\mbox{CL}_{s}$ }  \\ 
Fit & Lower 90\% C.L. & Best Fit & Upper 90\% C.L. & Lower 90\% C.L. & Best Fit & Upper 90\% C.L. \\
\hline 
\hline 
SK    $\theta_{13}$  Constrained  & 0.012 & 0.027 & 0.020 & 0.181 & 0.070 & 0.033 \\
\hline 
SK+T2K $\theta_{13}$ Constrained  & 0.004 & 0.023 & 0.024 & 0.081 & 0.075 & 0.056 \\ 
\hline 
\hline 
\end{tabular}
\end{center}
\caption{Normal hierarchy significance summarized in terms of the
  probability of observing a $\chi^{2}$ preference for the NH more
  extreme than that observed in data assuming an IH, $p_{0}(IH)$ , and
  $\mbox{CL}_{s}$ values for a range of assumed parameters.  The Best
  Fit column reports results assuming MC ensembles generated with
  oscillation parameters taken from the best fit point parameters from
  the NH fit in each analysis. Similarly, the Lower 90\% (Upper 90\%)
  column reports values assuming true parameters generated on the lower
  (upper) 90\% C.L. obtained from the fits. }
\label{tbl:pval}
\end{table*}
%
%
%
%
%

\section{Conclusion}
\label{sec:conclusion}

Analysis of Super-Kamiokande atmospheric neutrino data over a
328~kton-year exposure of the detector indicates a weak preference for the
normal mass hierarchy, disfavoring the inverted mass hierarchy at
93.0\% assuming oscillation parameters at the analysis best fit point 
and preferring matter over vacuum oscillations by $1.6{\sigma}$.
Assuming the normal mass hierarchy the constraints on the atmospheric
mixing parameters are $\mbox{sin}^{2} \theta_{23} =
0.588^{+0.031}_{-0.064}$ and $\Delta m^{2}_{32} = 2.50^{+0.13}_{-0.20}
$, with $\delta_{CP} = 4.18^{+1.41}_{-1.61}$.  Fitting in conjunction
with a model of the T2K experiment generally enhances these
constraints and the preference for the normal mass hierarchy.  Over
the range of parameters allowed at 90\% C.L. the inverted mass
hierarchy is disfavored by between 81.9\% and 96.7\% for SK by itself
and by between 91.9\% and 94.5\% when SK is combined with T2K for the 
$\theta_{13}$-constrained fits.
\section{Acknowledgments}
We gratefully acknowledge the cooperation of the Kamioka Mining and
Smelting Company. The Super-Kamiokande experiment has been built and
operated from funding by the Japanese Ministry of Education, Culture,
Sports, Science and Technology, the U.S. Department of Energy, and the
U.S. National Science Foundation. Some of us have been supported by
funds from the National Research Foundation of Korea NRF-2009-0083526
(KNRC) funded by the Ministry of Science, ICT, and Future Planning,
the European Union H2020 RISE-GA641540-SKPLUS, the Japan Society for
the Promotion of Science, the National Natural Science Foundation of
China under Grants No. 11235006, the National Science and Engineering
Research Council (NSERC) of Canada, the Scinet and Westgrid consortia
of Compute Canada, and the National Science Centre of Poland
(2015/17/N/ST2/04064, 2015/18/E/ST2/00758).

\appendix
\section{Systematic Uncertainties}\label{sec:systematics}

\renewcommand{\arraystretch}{1.0}
\newcolumntype{E}{D{.}{.}{-1}}

\begin{table*}
\centering
\begin{tabular}{lllEE}
\hline \hline
\multicolumn{3}{l}{Systematic Error} & \multicolumn{1}{c}{Fit Value (\%)} & \multicolumn{1}{c}{$\sigma$ (\%)} \\
\hline
Flux normalization                              & $E_\nu < \val{1}{GeV}$\footnote[1]{Uncertainty decreases linearly with $\log E_{\nu}$ from 25\,\%(0.1\,GeV) to 7\,\%(1\,GeV).} &                    
&   14.3 &    25 \\
                                                & $E_\nu > \val{1}{GeV}$\footnote[2]{Uncertainty is 7\,\% up to 10\,GeV, linearly increases with $\log E_{\nu}$ from 7\,\%(10\,GeV) to 12\,\%(100\,GeV) and then to 20\,\%(1\,TeV)} &                    
&    7.8 &    15 \\
$(\numu+\numubar)/(\nue+\nuebar)$               & \multicolumn{2}{l}{$E_\nu < \val{1}{GeV}$}            
&   0.08 &     2 \\
                                                & $1 < E_\nu < \val{10}{GeV}$      &                    
&  -1.1  &     3 \\
                                                & $E_\nu > \val{10}{GeV}$\footnote[3] {Uncertainty linearly increases with $\log E_{\nu}$ from 5\,\%(30\,GeV) to 30\,\%(1\,TeV).} &                    
&    1.6 &     5 \\
$\nuebar/\nue$                                  & \multicolumn{2}{l}{$E_\nu < \val{1}{GeV}$}            
&    1.6 &     5 \\
                                                & $1 < E_\nu < \val{10}{GeV}$      &                    
&    3.3 &     5 \\
                                                & $E_\nu > \val{10}{GeV}$\footnote[4] {Uncertainty linearly increases with $\log E_{\nu}$ from 8\,\%(100\,GeV) to 20\,\%(1\,TeV).} &                    
&   -1.6 &     8 \\
$\numubar/\numu$                                & \multicolumn{2}{l}{$E_\nu < \val{1}{GeV}$}            
&    0.24 &     2 \\
                                                & $1 < E_\nu < \val{10}{GeV}$      &                    
&    2.9 &     6 \\
                                                & $E_\nu > \val{10}{GeV}$\footnote[5] {Uncertainty linearly increases with $\log E_{\nu}$ from 6\,\%(50\,GeV) to 40\,\%(1\,TeV).} &                    
&   -2.9 &    15 \\
Up/down ratio                                   & $< \val{400}{MeV}$               & $e$-like           
& -0.026 &   0.1 \\
                                                &                                  & $\mu$-like         
& -0.078 &   0.3 \\
                                                &                                  & 0-decay $\mu$-like 
& -0.286 &   1.1 \\
                                                & $> \val{400}{MeV}$               & $e$-like           
& -0.208 &   0.8 \\
                                                &                                  & $\mu$-like         
& -0.130 &   0.5 \\
                                                &                                  & 0-decay $\mu$-like 
& -0.442 &   1.7 \\
                                                & Multi-GeV                        & $e$-like           
& -0.182 &   0.7 \\
                                                &                                  & $\mu$-like         
& -0.052 &   0.2 \\
                                                & Multi-ring Sub-GeV               & $e$-like           
& -0.104 &   0.4 \\
                                                &                                  & $\mu$-like         
& -0.052 &   0.2 \\
                                                & Multi-ring Multi-GeV             & $e$-like           
& -0.078 &   0.3 \\
                                                &                                  & $\mu$-like         
& -0.052 &   0.2 \\
                                                & PC                               &                    
& -0.052 &   0.2 \\
Horizontal/vertical ratio                       & $< \val{400}{MeV}$               & $e$-like           
&  0.018 &   0.1 \\
                                                &                                  & $\mu$-like         
&  0.018 &   0.1 \\
                                                &                                  & 0-decay $\mu$-like 
&  0.054 &   0.3 \\
                                                & $> \val{400}{MeV}$               & $e$-like           
&  0.252 &   1.4 \\
                                                &                                  & $\mu$-like         
&  0.341 &   1.9 \\
                                                &                                  & 0-decay $\mu$-like 
&  0.252 &   1.4 \\
                                                & Multi-GeV                        & $e$-like           
&  0.576  &   3.2 \\
                                                &                                  & $\mu$-like         
&  0.414 &   2.3 \\
                                                & Multi-ring Sub-GeV               & $e$-like           
&   0.252 &   1.4 \\
                                                &                                  & $\mu$-like         
&   0.234 &   1.3 \\
                                                & Multi-ring Multi-GeV             & $e$-like           
&   0.504 &   2.8 \\
                                                &                                  & $\mu$-like         
&   0.270 &   1.5 \\
                                                & PC                               &                    
&   0.306 &   1.7 \\
\multicolumn{3}{l}{K/$\pi$ ratio in flux calculation\footnote[6] {Uncertainty increases linearly from 5$\%$ to 20$\%$ between 100GeV and 1TeV.}} 
&   -9.3 &    10 \\
\multicolumn{3}{l}{Neutrino path length}                                                                
&  -2.13 &    10 \\
Sample-by-sample                                & \multicolumn{2}{l}{FC Multi-GeV}                      
&   -6.6 &     5 \\
                                                & PC + Stopping \UP                &                    
&   0.22 &     5 \\
\multicolumn{3}{l}{Matter effects}                                                                      
&    0.52 &   6.8 \\
\hline\hline
\end{tabular}
\caption{
Flux-related systematic errors that are common to all SK run periods. 
The second column shows the best fit value of the systematic error parameter, $\epsilon_j$, in percent and the third column shows the estimated $1{\sigma}$ error size in percent.}
\label{tab:sysa}
\end{table*}

\begin{table*}
\centering
\begin{tabular}{lllEE}
\hline \hline
Systematic Error & &  & \multicolumn{1}{c}{Fit Value (\%)} & \multicolumn{1}{c}{$\sigma$ (\%)} \\
\hline
$M_A$ in QE                                     &                                  &                    
&   -0.69 &    10 \\
Single $\pi$ Production, Axial Coupling         &                                   &
&   -4.4 &    10  \\
Single $\pi$ Production, $C_{A5}$               &                                   &
&   -3.1 &    10  \\
Single $\pi$ Production, BKG                    &                                   &
&   -8.7 &    10  \\
\multicolumn{3}{l}{CCQE cross section\footnote[1] {Difference from the Nieves~\cite{Nieves:2004wx} model is set to 1.0}} 
&    6.7 &    10 \\
\multicolumn{3}{l}{CCQE $\bar \nu/\nu$ ratio\footnotemark[1]}                                           
&    9.2 &    10 \\
\multicolumn{3}{l}{CCQE $\mu/e$ ratio\footnotemark[1]}                                                  
&   0.67 &    10 \\
\multicolumn{3}{l}{DIS cross section}                                                                   
&   -4.4 &     5 \\
\multicolumn{3}{l}{DIS model comparisons\footnote[2] {Difference from CKMT~\cite{CKMT94} parameterization is set to 1.0}} 
&    3.0 &    10 \\
\multicolumn{3}{l}{DIS $Q^2$ distribution (high W)\footnote[3] {Difference from GRV98~\cite{Gluck:1998xa} is set to 1.0}} 
&    8.2 &    10 \\
\multicolumn{3}{l}{DIS $Q^2$ distribution (low W)\footnotemark[3]}                                      
&   -5.8 &    10 \\
\multicolumn{3}{l}{Coherent $\pi$ production}                                                           
&   -10.0 &   100 \\
\multicolumn{3}{l}{NC/CC}                                                                               
&   12.1 &    20 \\
\multicolumn{3}{l}{\nutau cross section}                                                                
&  -13.8 &    25 \\
\multicolumn{3}{l}{Single $\pi$ production, $\pizero/\pi^\pm$}                                          
&   -20.3 &    40 \\
\multicolumn{3}{l}{Single $\pi$ production, $\bar \nu_{i} /\nu_{i}$ (i=$e,\mu $)\footnote[4] {Difference from the Hernandez\cite{Hernandez07} model is set to 1.0}} 
&    -11.0 &    10 \\
\multicolumn{3}{l}{NC fraction from hadron simulation}                                                  
&     -0.47 &    10 \\
$\pi^+$ decay uncertainty Sub-GeV 1-ring        & \multicolumn{2}{l}{$e$-like 0-decay}                  
&  -0.17  &   0.6 \\
                                                & $\mu$-like 0-decay               &                    
&  -0.22 &   0.8 \\
                                                & $e$-like 1-decay                 &                    
&   1.1  &   4.1 \\
                                                & $\mu$-like 1-decay               &                    
&   0.25 &   0.9 \\
                                                & $\mu$-like 2-decay               &                    
&   1.60 &   5.7 \\
\multicolumn{3}{l}{Final state and secondary interactions\footnote[5] {Error is set by the result of a fit to global data as presented in Ref.~\cite{Abe:2015awa}}}
&   -0.2 &    10 \\
\multicolumn{3}{l}{Meson exchange current\footnote[6] {Difference from NEUT without model from~\cite{Nieves:2004wx} is set to 1.0}} 
&   -1.8 &    10 \\
\multicolumn{3}{l}{\dmsq{21} \cite{Agashe:2014kda}}
&  0.022 &  2.4 \\
\multicolumn{3}{l}{\sn{12} \cite{Agashe:2014kda}}                                                           
&   0.32 &  4.6 \\
\multicolumn{3}{l}{\sn{13} \cite{Agashe:2014kda}}                                                                 
&   0.11 &  5.4 \\
\hline\hline
\end{tabular}
\caption{
Neutrino interaction, particle production, and PMNS oscillation parameter systematic errors that are common to all SK run periods.
The second column shows the best fit value of the systematic error parameter, $\epsilon_j$, in percent and the third column shows the estimated $1{\sigma}$ error size in percent.}
\label{tab:sysb}
\end{table*}

\renewcommand{\tabcolsep}{0pt}

\begin{table*}
\centering
\begin{tabular}{lllEEEEEEEE}
\hline \hline
 & & 
 & \multicolumn{2}{c}{SK-I} 
 & \multicolumn{2}{c}{SK-II} 
 & \multicolumn{2}{c}{SK-III} 
 & \multicolumn{2}{c}{SK-IV} 
\\
 \multicolumn{3}{l}{Systematic Error} 
 & \multicolumn{1}{c}{Fit Value} & \multicolumn{1}{c}{$\sigma$}
 & \multicolumn{1}{c}{Fit Value} & \multicolumn{1}{c}{$\sigma$}
 & \multicolumn{1}{c}{Fit Value} & \multicolumn{1}{c}{$\sigma$}
 & \multicolumn{1}{c}{Fit Value} & \multicolumn{1}{c}{$\sigma$} 
\\
\hline
FC reduction                                    &                                  &                    
&  -0.009 &   0.2 &  0.005 &   0.2 &  0.066 &   0.8 &  0.68 &   1.3 \\
\multicolumn{3}{l}{PC reduction}                                                                        
&   0.016 &   2.4 &  -3.43 &   4.8 & -0.012 &   0.5 &  -0.78 &     1 \\
\multicolumn{3}{l}{FC/PC separation}                                                                    
& -0.10 &   0.6 &  0.077 &   0.5 &  -0.13 &   0.9 & 0.0004 &  0.02 \\
\multicolumn{3}{l}{PC stopping/through-going separation (bottom)}                                       
&  -15.8 &    23 &   -2.4 &    13 &   -0.32 &    12 &  -1.5 &   6.8 \\
\multicolumn{3}{l}{PC stopping/through-going separation (barrel)}                                       
&   3.8 &     7 &  -5.7 &   9.4 &   -13.9 &    29 &  -0.40 &   8.5 \\
\multicolumn{3}{l}{PC stopping/through-going separation (top)}                                          
&   8.5 &    46 &   -3.0 &    19 &   -12.6 &    87 &   -24.1 &    40 \\
Non-$\nu$ background                            & \multicolumn{2}{l}{Sub-GeV $\mu$-like}                
&  0.010 &   0.1 &  0.065 &   0.4 &  0.105 &   0.5 & -0.011 &   0.02 \\
                                                & Multi-GeV $\mu$-like             &                    
&  0.040 &   0.4 &  0.065 &   0.4 &  0.105 &   0.5 & -0.011 &   0.02 \\
                                                & Sub-GeV 1-ring 0-decay $\mu$-like &                    
&  0.010 &   0.1 &  0.049 &   0.3 &  0.084 &   0.4 & -0.052 &   0.09 \\
                                                & PC                               &                    
&  0.020 &   0.2 &  0.115 &   0.7 &  0.381 &   1.8 & -0.282 &  0.49 \\
                                                & Sub-GeV $e$-like (flasher event)                 &                    
&  0.068 &   0.5 &  0.000 &   0.2 & -0.004 &   0.2 & -0.000 &   0.02 \\
                                                & Multi-GeV $e$-like (flasher event)               &                    
&  0.014 &   0.1 &  0.000 &   0.3 & -0.014 &   0.7 & -0.000 &   0.08 \\
                                                & Multi-GeV 1-ring $e$-like        &                    
&    3.6 &    13 &  -5.2 &    38 &    -1.0 &    27 &    2.6 &    18 \\
                                                & Multi-GeV Multi-ring $e$-like    &                    
&    3.7 &    12 &    3.8 &    11 &   0.75 &    11 &    0.34 &    12 \\
\multicolumn{3}{l}{Fiducial Volume}                                                                     
&  -0.85 &     2 &  -0.11 &     2 &   0.22 &     2 &   -1.5 &     2 \\
Ring separation                                 & $< \val{400}{MeV}$               & $e$-like           
&   0.45 &   2.3 &  -1.07 &   1.3 &   0.80 &   2.3 &   0.96 &   1.6 \\
                                                &                                  & $\mu$-like         
&  0.14  &   0.7 &  -1.91 &   2.3 &   1.04 &     3 &   1.79 &     3 \\
                                                & $> \val{400}{MeV}$               & $e$-like           
&  0.078 &   0.4 &  -1.40 &   1.7 &   0.45 &   1.3 &  -0.60 &     1 \\
                                                &                                  & $\mu$-like         
&  0.14  &   0.7 & -0.576 &   0.7 &  0.208 &   0.6 & -0.36 &   0.6 \\
                                                & Multi-GeV                        & $e$-like           
&   0.72 &   3.7 &  -2.14 &   2.6 &   0.45 &   1.3 &  -0.60 &     1 \\
                                                &                                  & $\mu$-like         
&   0.33 &   1.7 &  -1.41 &   1.7 &   0.35 &     1 &   0.72 &   1.2 \\
                                                & Multi-ring Sub-GeV               & $e$-like           
&  -0.68 &   3.5 &   3.13 &   3.8 &   0.45 &   1.3 &   1.14 &   1.9 \\
                                                &                                  & $\mu$-like         
&  -0.88 &   4.5 &   6.75 &   8.2 &  -0.90 &   2.6 &   1.37 &   2.3 \\
                                                & Multi-ring Multi-GeV             & $e$-like           
&  -0.61 &   3.1 &   1.56 &   1.9 &  -0.38 &   1.1 &  0.54 &   0.9 \\
                                                &                                  & $\mu$-like         
&  -0.80 &   4.1 &  0.658 &   0.8 &  -0.73 &   2.1 &  -1.43 &   2.4 \\
Particle identification (1 ring)                & Sub-GeV                          & $e$-like           
&  0.039 &  0.23 &  0.227 &  0.66 &  0.053&  0.26 & -0.123 &  0.28 \\
                                                &                                  & $\mu$-like         
& -0.030 &  0.18 & -0.172 &   0.5 & -0.038 &  0.19 &  0.097&  0.22 \\
                                                & Multi-GeV                        & $e$-like           
&  0.032 &  0.19 &  0.082 &  0.24 &  0.062 &  0.31 & -0.154 &  0.35 \\
                                                &                                  & $\mu$-like         
& -0.032 &  0.19 & -0.089 &  0.26 & -0.060 &   0.3 &  0.154 &  0.35 \\
Particle identification (multi-ring)            & Sub-GeV                          & $e$-like           
&  -0.23 &   3.1 &  -3.44 &     6 &   3.49 &   9.5 &   -2.24 &   4.2 \\
                                                &                                  & $\mu$-like         
&  0.049 &  0.66 &   1.38 &   2.5 &  -1.91 &   5.2 &   0.85  &   1.6 \\
                                                & Multi-GeV                        & $e$-like           
&   0.48 &   6.5 &   5.57 &   9.7 &  -1.80 &   4.9 &   -1.76 &   3.3 \\
                                                &                                  & $\mu$-like         
&  -0.21 &   2.9 &  -2.24 &   3.9 &   0.99 &   2.7 &   0.85 &   1.6 \\
Multi-ring likelihood selection                 &  Multi-ring $e$-like             &   $\nu_{e}$,$\bar{\nu}_{e}$                    
&  -6.5  &   6.0 &  -1.3  &   3.8 &   -5.3 &   5.3 &   -2.3  &   3.0 \\
                                                &  Multi-ring Other                            &                    
&   6.2  &   5.7 &   1.4  &   4.1 &    4.7 &   4.9 &    2.7  &   3.4 \\
\multicolumn{3}{l}{Energy calibration}                                                                  
&   -0.75 &   3.3 &  -0.90 &   2.8 &   0.06 &   2.4 &  0.08 &   2.1 \\
\multicolumn{3}{l}{Up/down asymmetry energy calibration}                                                
&  0.26 &   0.6 & 0.24 &   0.6 &   0.74 &   1.3 & -0.15 &   0.4 \\
\UP reduction                                   & \multicolumn{2}{l}{Stopping}                          
& -0.091 &   0.7 & -0.090 &   0.7 &  0.162 &   0.7 &  0.087 &   0.5 \\
                                                & Through-going                    &                    
& -0.065 &   0.5 & -0.064 &   0.5 &  0.115 &   0.5 &  0.052 &   0.3 \\
\multicolumn{3}{l}{\UP stopping/through-going separation}                                               
&  0.003 &   0.4 &  -0.004 &   0.6 &  0.030 &   0.4 & -0.102 &   0.6 \\
\multicolumn{3}{l}{Energy cut for stopping \UP}                                                         
&  -0.043 &   0.9 &   -0.122 &   1.3 &   0.957 &     2 &   -0.122 &   1.7 \\
\multicolumn{3}{l}{Path length cut for through-going \UP}                                               
&   -0.416 &   1.5 &   -0.826 &   2.3 &  0.993 &   2.8 &  1.47 &   1.5 \\
\multicolumn{3}{l}{Through-going \UP showering separation}                                              
&   7.53 &   3.4 &  -4.68 &   4.4 &   2.90 &   2.4 &  -3.30 &     3 \\
Background subtraction for \UP                  & \multicolumn{2}{l}{Stopping\footnote[1]{The uncertainties in BG subtraction for upward-going muons are only for the most horizontal bin,  $-0.1 < \cos\theta < 0$.}} 
&   10.0 &    16 &   -3.1 &    21 &   -4.9 &    20 &   -6.7 &    17 \\
                                                & Non-showering\footnotemark[1]    &                    
&   -3.6 &    18 &    -3.6 &    14 &    1.4 &    24 &    2.1 &    17 \\
                                                & Showering\footnotemark[1]        &                    
&   -12.3 &    18 &  -15.7 &    14 &    0.1 &    24 &   -0.9 &    24 \\
\multicolumn{3}{l}{$\nue/\nuebar$ Separation}                                                           
&  -0.98 &   7.2 &   6.96 &   7.9 &  0.45 &   7.7 &  2.46 &   6.8 \\
Sub-GeV 1-ring \pizero selection                & \multicolumn{2}{l}{$100 < P_e < \val{250}{MeV/c}$}    
&   1.7 &     9 &    7.0 &    10 &   0.98 &   6.3 &   5.2 &   4.6 \\
                                                & $250 < P_e < \val{400}{MeV/c}$   &                    
&   1.7 &   9.2 &    9.8 &    14 &    0.76 &   4.9 &   3.4 &     3 \\
                                                & $400 < P_e < \val{630}{MeV/c}$   &                    
&    3.0 &    16 &    7.7 &    11 &   3.7 &    24 &   14.8 &    13 \\
                                                & $630 < P_e < \val{1000}{MeV/c}$  &                    
&    2.6 &    14 &   11.2 &    16 &   1.3 &   8.2 &   19.4 &    17 \\
                                                & $1000 < P_e < \val{1330}{MeV/c}$ &                    
&    2.2 &    12 &   6.8  &   9.8 &    1.7 &    11 &    27.4 &    24 \\
\multicolumn{3}{l}{Sub-GeV 2-ring \pizero}                                                              
&   1.3 &   5.6 &  -2.7 &   4.4 &  1.6 &   5.9 &   -0.72 &   5.6 \\
\multicolumn{3}{l}{Decay-e tagging}                                                                     
&   -3.2 &    10 &   -1.0 &    10 &    0.9 &    10 &    1.3 &    10 \\
\multicolumn{3}{l}{Solar Activity}                                                                      
&    -1.8 &    20 &   20.0 &    50 &    2.7 &    20 &    0.6 &    10 \\
\hline\hline
\end{tabular}
\caption{
Systematic errors that are independent in SK-I, SK-II, SK-III, and SK-IV. 
Columns labeled `fit' show the best fit value of the systematic error parameter, $\epsilon_j$, in percent and columns labeled $\sigma$ shows the estimated $1{\sigma}$ error size in percent.}
\label{tab:sysc}
\end{table*}

\bibliography{bibliography}

\end{document}

%% file: authors.tex

\newcommand{\AFFicrr}{\affiliation{Kamioka Observatory, Institute for Cosmic Ray Research, University of Tokyo, Kamioka, Gifu 506-1205, Japan}}
\newcommand{\AFFkashiwa}{\affiliation{Research Center for Cosmic Neutrinos, Institute for Cosmic Ray Research, University of Tokyo, Kashiwa, Chiba 277-8582, Japan}}
\newcommand{\AFFipmu}{\affiliation{Kavli Institute for the Physics and
Mathematics of the Universe (WPI), The University of Tokyo Institutes for Advanced Study,
University of Tokyo, Kashiwa, Chiba 277-8583, Japan }}
\newcommand{\AFFmad}{\affiliation{Department of Theoretical Physics, University Autonoma Madrid, 28049 Madrid, Spain}}
\newcommand{\AFFubc}{\affiliation{Department of Physics and Astronomy, University of British Columbia, Vancouver, BC, V6T1Z4, Canada}}
\newcommand{\AFFbu}{\affiliation{Department of Physics, Boston University, Boston, MA 02215, USA}}
\newcommand{\AFFbnl}{\affiliation{Physics Department, Brookhaven National Laboratory, Upton, NY 11973, USA}}
\newcommand{\AFFuci}{\affiliation{Department of Physics and Astronomy, University of California, Irvine, Irvine, CA 92697-4575, USA }}
\newcommand{\AFFcsu}{\affiliation{Department of Physics, California State University, Dominguez Hills, Carson, CA 90747, USA}}
\newcommand{\AFFcnm}{\affiliation{Department of Physics, Chonnam National University, Kwangju 500-757, Korea}}
\newcommand{\AFFduke}{\affiliation{Department of Physics, Duke University, Durham NC 27708, USA}}
\newcommand{\AFFfukuoka}{\affiliation{Junior College, Fukuoka Institute of Technology, Fukuoka, Fukuoka 811-0295, Japan}}
\newcommand{\AFFgifu}{\affiliation{Department of Physics, Gifu University, Gifu, Gifu 501-1193, Japan}}
\newcommand{\AFFgist}{\affiliation{GIST College, Gwangju Institute of Science and Technology, Gwangju 500-712, Korea}}
\newcommand{\AFFuh}{\affiliation{Department of Physics and Astronomy, University of Hawaii, Honolulu, HI 96822, USA}}
\newcommand{\AFFicl}{\affiliation{Department of Physics, Imperial College London , London, SW7 2AZ, United Kingdom }}
\newcommand{\AFFkek}{\affiliation{High Energy Accelerator Research Organization (KEK), Tsukuba, Ibaraki 305-0801, Japan }}
\newcommand{\AFFkobe}{\affiliation{Department of Physics, Kobe University, Kobe, Hyogo 657-8501, Japan}}
\newcommand{\AFFkyoto}{\affiliation{Department of Physics, Kyoto University, Kyoto, Kyoto 606-8502, Japan}}
\newcommand{\AFFliv}{\affiliation{Department of Physics, University of Liverpool, Liverpool, L69 7ZE, United Kingdom}}
\newcommand{\AFFmiyagi}{\affiliation{Department of Physics, Miyagi University of Education, Sendai, Miyagi 980-0845, Japan}}
\newcommand{\AFFnagoya}{\affiliation{Institute for Space-Earth Enviromental Research, Nagoya University, Nagoya, Aichi 464-8602, Japan}}
\newcommand{\AFFkmi}{\affiliation{Kobayashi-Maskawa Institute for the Origin of Particles and the Universe, Nagoya University, Nagoya, Aichi 464-8602, Japan}}
\newcommand{\AFFpol}{\affiliation{National Centre For Nuclear Research, 00-681 Warsaw, Poland}}
\newcommand{\AFFsuny}{\affiliation{Department of Physics and Astronomy, State University of New York at Stony Brook, NY 11794-3800, USA}}
\newcommand{\AFFokayama}{\affiliation{Department of Physics, Okayama University, Okayama, Okayama 700-8530, Japan }}
\newcommand{\AFFosaka}{\affiliation{Department of Physics, Osaka University, Toyonaka, Osaka 560-0043, Japan}}
\newcommand{\AFFox}{\affiliation{Department of Physics, Oxford University, Oxford, OX1 3PU, United Kingdom}}
\newcommand{\AFFqmul}{\affiliation{School of Physics and Astronomy, Queen Mary University of London, London, E1 4NS, United Kingdom}}
\newcommand{\AFFregina}{\affiliation{Department of Physics, University of Regina, 3737 Wascana Parkway, Regina, SK, S4SOA2, Canada}}
\newcommand{\AFFseoul}{\affiliation{Department of Physics, Seoul National University, Seoul 151-742, Korea}}
\newcommand{\AFFsheff}{\affiliation{Department of Physics and Astronomy, University of Sheffield, S10 2TN, Sheffield, United Kingdom}}
\newcommand{\AFFshizuokasc}{\affiliation{Department of Informatics in
Social Welfare, Shizuoka University of Welfare, Yaizu, Shizuoka, 425-8611, Japan}}
\newcommand{\AFFstfc}{\affiliation{STFC, Rutherford Appleton Laboratory, Harwell Oxford, and Daresbury Laboratory, Warrington, OX11 0QX, United Kingdom}}
\newcommand{\AFFskk}{\affiliation{Department of Physics, Sungkyunkwan University, Suwon 440-746, Korea}}
\newcommand{\AFFtokyo}{\affiliation{The University of Tokyo, Bunkyo, Tokyo 113-0033, Japan }}
\newcommand{\AFFtodai}{\affiliation{Department of Physics, University of Tokyo, Bunkyo, Tokyo 113-0033, Japan }}
\newcommand{\AFFtit}{\affiliation{Department of Physics,Tokyo Institute of Technology, Meguro, Tokyo 152-8551, Japan }}
\newcommand{\AFFtus}{\affiliation{Department of Physics, Faculty of Science and Technology, Tokyo University of Science, Noda, Chiba 278-8510, Japan }}
\newcommand{\AFFtoronto}{\affiliation{Department of Physics, University of Toronto, ON, M5S 1A7, Canada }}
\newcommand{\AFFtriumf}{\affiliation{TRIUMF, 4004 Wesbrook Mall, Vancouver, BC, V6T2A3, Canada }}
\newcommand{\AFFtokai}{\affiliation{Department of Physics, Tokai University, Hiratsuka, Kanagawa 259-1292, Japan}}
\newcommand{\AFFtsinghua}{\affiliation{Department of Engineering Physics, Tsinghua University, Beijing, 100084, China}}
\newcommand{\AFFuw}{\affiliation{Department of Physics, University of Washington, Seattle, WA 98195-1560, USA}}
\newcommand{\AFFynu}{\affiliation{Faculty of Engineering, Yokohama National University, Yokohama, 240-8501, Japan}}

\AFFicrr
\AFFkashiwa
\AFFmad
\AFFbu
\AFFubc
\AFFbnl
\AFFuci
\AFFcsu
\AFFcnm
\AFFduke
\AFFfukuoka
\AFFgifu
\AFFgist
\AFFuh
\AFFicl
\AFFkek
\AFFkobe
\AFFkyoto
\AFFliv
\AFFmiyagi
\AFFnagoya
\AFFkmi
\AFFpol
\AFFsuny
\AFFokayama
\AFFosaka
\AFFox
\AFFqmul
\AFFregina
\AFFseoul
\AFFsheff
\AFFshizuokasc
\AFFstfc
\AFFskk
\AFFtokai
\AFFtokyo
\AFFtodai
\AFFipmu
\AFFtit
\AFFtus
\AFFtoronto
\AFFtriumf
\AFFtsinghua
\AFFuw
\AFFynu

\author{K.~Abe}
\AFFicrr
\AFFipmu
\author{C.~Bronner}
\AFFicrr
\author{Y.~Haga}
\AFFicrr
\author{Y.~Hayato}
\AFFicrr
\AFFipmu
\author{M.~Ikeda}
\AFFicrr
\author{K.~Iyogi}
\AFFicrr 
\author{J.~Kameda}
\AFFicrr
\AFFipmu 
\author{Y.~Kato}
\AFFicrr
\author{Y.~Kishimoto}
\AFFicrr
\AFFipmu 
\author{Ll.~Marti}
\AFFicrr
% ...

\collaboration{The Super-Kamiokande Collaboration}
\noaffiliation